\documentclass[a4paper,11pt]{article}
\pdfoutput=1 
\usepackage{jheppub} 

\usepackage[T1]{fontenc} 
\usepackage[table]{xcolor}
\usepackage[english]{babel}
\usepackage[utf8]{inputenc}
\usepackage{amsmath}
\usepackage{graphicx}
\usepackage[colorinlistoftodos]{todonotes}
\usepackage{amsmath}
\usepackage{indentfirst}
\usepackage{fancyhdr}
\usepackage{color}
\usepackage{amssymb}
\usepackage{array} 
\usepackage{colortbl}
\usepackage{multirow}
\newcolumntype{M}[1]{>{\centering\arraybackslash}m{#1}}
\newcolumntype{N}{@{}m{0pt}@{}}
\usepackage{subcaption}
\usepackage{caption}
\usepackage{soul}
\usepackage{mathtools}
\usepackage{ulem}

\def\ls{\mathrel{\lower4pt\vbox{\lineskip=0pt\baselineskip=0pt
			\hbox{$<$}\hbox{$\sim$}}}}
\def\gs{\mathrel{\lower4pt\vbox{\lineskip=0pt\baselineskip=0pt
			\hbox{$>$}\hbox{$\sim$}}}}
\def\drawbox#1#2{\hrule height#2pt
	\hbox{\vrule width#2pt height#1pt \kern#1pt
		\vrule width#2pt}
	\hrule height#2pt}

\def\Asym#1#2{\vcenter{\vbox{\drawbox{#1}{#2}
			\kern-#2pt       
			\drawbox{#1}{#2}}}}

\newcommand{\be}{\begin{equation}}
	\newcommand{\ee}{\end{equation}}
\newcommand{\bea}{\begin{eqnarray}}
	\newcommand{\eea}{\end{eqnarray}}

\newcommand{\gsim}{\lower.7ex\hbox{$\;\stackrel{\textstyle>}{\sim}\;$}}
\newcommand{\lsim}{\lower.7ex\hbox{$\;\stackrel{\textstyle<}{\sim}\;$}}

\newcommand{\vo}{\mathcal{V}}

\newcommand{\ben}{\begin{enumerate}}
	\newcommand{\een}{\end{enumerate}}
\newcommand{\bei}{\begin{itemize}}
	\newcommand{\eei}{\end{itemize}}
\newcommand{\mc}{\mathcal}

\newcommand{\ol}{\overline}
\newcommand{\coma}{\, , \quad}
\newcommand{\fstop}{\, .}

\title{\boldmath Fuzzy Dark Matter Candidates from String Theory}

\author[a,b]{Michele Cicoli}
\author[c]{Veronica Guidetti}
\author[c]{Nicole Righi}
\author[c]{Alexander Westphal}

\preprint{DESY 21-153}

\affiliation[a]{\small Dipartimento di Fisica e Astronomia, Universit\`a di Bologna, \\ via Irnerio 46, 40126 Bologna, Italy}
\affiliation[b]{\small INFN, Sezione di Bologna, viale Berti Pichat 6/2, 40127 Bologna, Italy}
\affiliation[c]{\small Deutches Electronen-Synchrotron, DESY,\\ Notkestra\ss e 85, 22607 Hamburg, Germany}

\emailAdd{michele.cicoli@unibo.it}
\emailAdd{veronica.guidetti@desy.de}
\emailAdd{nicole.righi@desy.de}
\emailAdd{alexander.westphal@desy.de}

\abstract{String theory has been claimed to give rise to natural fuzzy dark matter candidates in the form of ultralight axions. In this paper we revisit this claim by a detailed study of how moduli stabilisation affects the masses and decay constants of different axion fields which arise in type IIB flux compactifications. We find that obtaining a considerable contribution to the observed dark matter abundance without tuning the axion initial misalignment angle is not a generic feature of 4D string models since it requires a mild violation of the $S f\lesssim M_P$ bound, where $S$ is the instanton action and $f$ the axion decay constant. Our analysis singles out $C_4$-axions, $C_2$-axions and thraxions as the best candidates to realise fuzzy dark matter in string theory. For all these ultralight axions we provide predictions which can be confronted with present and forthcoming observations.}


\begin{document} 
	\maketitle
	\flushbottom
	
	\section{Introduction}
	\label{sec:introduction}
	
	Despite long model building efforts, the origin and nature of dark matter remains one of the biggest puzzles in Physics and astronomy. In recent years, Cold Dark Matter (CDM) has been pointed out as the best class of models that is able to reproduce large scale structure formation of the universe. In these models, dark matter is made out of weakly interacting non-relativistic particles with a small initial velocity dispersion relation inherited from interactions in the early universe that do not erase structures on galactic and sub-galactic scales. Among the various models, the combination of cosmic acceleration measurement and the CMB evidence for a flat universe led to the choice of $\Lambda$CDM model which is nowadays considered as the `Standard Model' of cosmology.
	Despite its success in explaining the large scale structure of the universe, $\Lambda$CDM was believed to suffer from some problems related to galaxy formation~\cite{Bullock_2017} that may be actually explained with unaccounted baryonic feedback mechanisms or to new exotic dark matter physics on small scales~\cite{Brooks_2013,Spergel:1999mh,Read:2018fxs,Colin:2000dn,Salucci:2018hqu} but a final and exhaustive solution is still lacking. Regardless of the veracity of small-scale problems, Weakly Interacting Massive Particles (WIMPs) having mass $\sim \mc{O}(100)$ GeV that were considered the most promising CDM candidates have continuously eluded whatever kind of experimental measurement as collider searches and direct/indirect detection experiments.
	
	 These concerns about $\Lambda$CDM and WIMPs led to the study of alternative DM models. Among those, in recent years the idea of bosonic ultralight CDM, also called Fuzzy Dark Matter (FDM), has been proposed~\cite{Hu:2000ke,Schive:2014dra,Hui:2016ltb,Hui:2021tkt}. In one of its prominent versions, DM is made of ultralight axion-like particles that form halos as Bose-Einstein condensates.
	In this theory each axionic particle can develop structures on the scale of de Broglie wavelength thanks to gravitational interactions. This is an ensemble effect which is given by the mean properties of every single axion field. A prominent soliton, i.e. a state where self-gravity is balanced by the effective pressure arising from the uncertainty principle, develops at the centre of every bound halo. The soliton properties depend on the axion mass but usually its extension is assumed to be much smaller than the galaxy or galaxy cluster size. In the original proposal, an axion having mass around $10^{-22}$ eV and decay constant $f\sim 10^{16\div 17}$ GeV was pointed out as the best candidate to represent the dominant part of CDM in the universe since the wave nature of such a particle can suppress kpc scale cusps in DM halos and reduce the abundance of low mass halos~\cite{Schive:2014hza,Schive:2014dra,Hui:2016ltb}.
	
	Recent studies put severe constraints on the vanilla FDM model without self-interactions where the usual cosine axionic potential is approximated as $1-\cos(\phi/f)\sim \frac{\phi^2}{2f^2}$. Various analyses of Lyman-$\alpha$ forest, satellite galaxies formation, dwarf galaxies, the Milky Way core and Black Hole superradiance~\cite{Marsh:2018zyw,Chan:2021ukg,Jones:2021mrs,Nadler:2020prv,Zu:2020whs,Nebrin_2019,Maleki:2020sqn,Marsh:2021lqg} leave as the only viable mass windows $m_\phi \sim 10^{-24}$ eV and $m_\phi \sim 10^{-15}$ eV, although certain of these bounds could be relaxed and open a window near $10^{-21}$ eV also.
	These experimental bounds imply that FDM cannot solve the alleged small-scale problems affecting $\Lambda$CDM as the Jeans mass (representing the lower bound on DM halos mass production) rapidly decreases at increasing ultralight boson masses~\cite{Nebrin_2019}. Nevertheless, even in this case, these problems can be solved by baryonic physics and a better understanding of galaxy formations may allow us to discriminate between standard CDM and FDM models. Indeed, it was proven that small-mass halos suppression in the FDM model causes a delay in the onset of Cosmic Dawn and the Epoch of Reionization. Future experiments, such as the HERA survey, will measure the neutral hydrogen (HI) 21cm line power spectrum at high statistical significance across a broad range of redshifts~\cite{Jones:2021mrs,Nebrin_2019} and their findings may be able to discriminate between standard WIMP and FDM scenarios. 
	Since experimental bounds and simulations strongly constrain the original FDM model with negligible self-interaction, many extensions of it have been studied. It was shown that for large initial misalignment angles ALPs self-interactions can affect the baryonic structure and accelerate star formation in the early universe or induce oscillon formation that can give rise to detectable low frequency stochastic gravitational waves~\cite{Arvanitaki:2019rax}. Other authors suggest that FDM may not represent the entirety of DM~\cite{Schwabe:2020eac} or that FDM may not be given by a single component, being made out of multiple ultralight ALPs~\cite{Broadhurst:2018fei}.

    The extremely high value of the decay constant together with the possible multiple axionic nature of FDM have been claimed to be a possible sign in favour of the string axiverse~\cite{Hui:2016ltb,Visinelli:2018utg}, where a plenitude of axion-like particles (ALPs) naturally emerge from 4D effective theories. However, in this paper we point out that obtaining a FDM axion with the correct mass and decay constant is not automatic in string theory. Indeed, even if one would naively think that ultralight axions generally emerge from string theory equipped with naturally high decay constants, reproducing the right relic abundance turns out to be hard and provides sharp predictions for fundamental microscopical parameters. We carry out a detailed analysis, studying the general features of closed and open string ALPs coming from type IIB string theory. Focusing on simple extra-dimensions geometries and using the most common moduli stabilisation prescriptions, for each class of ALPs we provide general predictions for the expected mass, decay constant and dark matter abundance. We discuss the settings of the microscopical parameters that lead to ultralight axions representing non-negligible fractions of DM, and we estimate how these requirements put stringent predictions for the relevant energy scales of the 4D effective field theory, such as the Kaluza-Klein (KK) scale, the gravitino mass and the scale of inflation. Finally, we compare our predictions for FDM ALPs with current observational constraints and we highlight which stringy FDM candidates occupy a region of the parameter space that will be probed by next generation experiments. We would like to stress that in this work we only consider the simplest setups, thus neglecting the effects that may arise from considering a large number of axionic fields. Indeed, we assume that the axionic potential does not create local minima, and that there are no turns in the field dynamics when they start to oscillate at times where $m\sim H$. We also neglect the possibility of having axion alignment~\cite{Kim:2004rp} as this is not the most common situation and its implementation often involves a considerable amount of any parameter tuning. Despite our simple assumptions, we believe the results presented in this work remain generally true also for more general extradimensional geometries. Indeed, we find that among closed string axions only those related to large cycles can be good FDM candidates. Although it is not possible to write the most generic volume of a CY, the number of moduli entering the volume with a positive sign must be finite.
	
	This work is organised as follows: in Section~\ref{sec:StringOrigin} we introduce our notation and we briefly sum up how ALPs naturally arise from type IIB string theory as closed and open string axions. Moreover, we discuss the non-trivial theoretical implication hiding behind the requirements of matching the right mass, decay constant and abundance. In Section~\ref{sec:closedALPs} we focus on closed string axion FDM models. We will work with type IIB string theory compactified to 4D on six dimensional Calabi-Yau (CY) orientifolds. Considering the two most prominent moduli stabilisation prescriptions for this setting, i.e. Large Volume Scenario (LVS)~\cite{Balasubramanian:2005zx} and KKLT~\cite{Kachru:2003aw}, we scan over the different axion classes that can represent significant fractions of DM, i.e $C_4$, $C_2$ axions and thraxions. We find that both moduli stabilisation prescriptions allow for having a considerable amount of ultralight axionic dark matter, but only LVS predicts masses in the FDM range. In Section~\ref{sec:PredExpConstr} we discuss our findings and compare them with state of the art experimental bounds also considering how future experiments will be able to constrain the allowed ultralight axions parameter space. We also provide some intuition about the probabilistic distribution of these particles in the string landscape and we try to figure out how our results would be affected by considering more complex extra-dimension geometries.

	\section{String origin of ultralight axionic DM candidates}
	\label{sec:StringOrigin}
	The 4D effective field theory coming from string compactification  contains many scalar fields, named moduli, which parametrise the size and the shape of the extra dimensions. Moduli appear at tree-level as massless and uncharged scalar fields which, thanks to their effective gravitational coupling to all ordinary particles, would mediate some undetected long-range fifth forces. For this reason, it is necessary to develop a potential for these particles in order to give them a mass. This problem goes under the name of moduli stabilisation.
	
	Since the number of ALPs is related to the number of moduli, which can easily reach the value of several hundreds, we can have many ultralight axion candidates which create the so called axiverse~\cite{Cicoli:2012sz}. On the other hand, it is essential to notice that, although string compactifications carry plenty of candidates for axion and axion-like weakly interacting particles, there are several known mechanisms by which they can be removed from the low energy spectrum.
	
	The low energy spectrum below the compactification scale generically contains many axion-like particles which arise either as closed string axions, which are the KK zero modes of 10D antisymmetric tensor fields, or as the phase of open string modes. While the number of closed string axions is related to the topology of the internal manifold, the number of open string axions is more model dependent since their existence relies upon the brane setup. In the next section, we will briefly describe the main properties of both closed and open string axions, trying to understand what conditions are required in order to reproduce viable FDM particles. 
	
	Let us now focus on the most relevant features that our fields need to satisfy in order to be good FDM candidates. Considering for simplicity a single axion field, a commonly used set of axion conventions is
	\be
	{\cal L}=\frac{1}{2}f^2(\partial \theta)^2-Ae^{-S}\cos(\theta)\,,
	\label{eq:AxionLagr}
	\ee
	where $f$ is the axion decay constant and $S$ represents the instanton action that gives rise to the axion potential. From the above expression, where we set the instanton charge to one for simplicity, we see that the axion mass is given by 
	\be\label{eq:axionmassgen} 
	m_\phi^2=A M_P^4 e^{-S}/f^2\fstop
	\ee 
	Using this notation the axion periodicity is $2\pi/f$ and the value for $Sf$ corresponding to (half of) a Giddings-Strominger wormhole (for a review see~\cite{Hebecker:2018ofv}) is
	\be
	Sf=\frac{\sqrt{6}\pi}{8}\simeq 0.96\fstop
	\ee
	Given that FDM particles have to be produced through the misalignment mechanism and that a GUT scale decay constant implies that the Peccei-Quinn symmetry is broken before the inflationary stage, the DM abundance of the physical ALP particle, $\phi=f\theta$, can be expressed as~\cite{Cicoli:2012sz}:
	\be
	\label{eq:DMabundance}
	\frac{\Omega_{\phi}h^2}{0.112}\simeq 2.2 \times \left(\frac{m_\phi}{10^{-22} \mbox{eV}}\right)^{1/2}\left(\frac{f}{10^{17}\mbox{GeV}}\right)^2 \theta_{mi}^2\sim1 \coma
	\ee
	where $\theta_{mi}\in [0,2\pi]$ is the initial misalignment angle with respect to the minimum of the potential.\footnote{Given that $\frac{\Omega_{\phi}h^2}{0.112}\propto e^{-S/4}f^{3/2}$, we see that the representation fraction of every axion in the DM halo changes depending on its value of $S$ and/or $f$. In general, for the same value of $f$, axions with smaller $S$ are more represented, hence the DM abundance is dominated by the heavier axions (cf.~\eqref{eq:axionmassgen}). For axions with the same $S$, those with larger $f$ have a larger DM abundance and (cf.~\eqref{eq:axionmassgen}) are also lighter. This last case is less generic.} In Eq.~\eqref{eq:DMabundance} we are considering small field initial displacement, large misalignment will be briefly treated in Appendix~\ref{sec:anharm}.  
	
	Therefore, assuming an initial misalignment angle $\theta_{mi}\sim\mathcal{O}(1)$, a prefactor $A\sim\mathcal{O}(1)$, and imposing the right value for the axion mass and decay constant, $m_\phi\sim 10^{-22}$ eV and $f\sim 10^{-2}  M_P$, we have that
	\begin{equation}
		\label{eq:naivebound}
		Sf=-f\ln\left(\frac{m_\phi^2f^2}{A M_P^4} \right)\gtrsim 1\fstop
	\end{equation}
	This means that the existence of a FDM candidate tends to slightly violate the Weak Gravity Conjecture (WGC)\cite{Alonso:2017avz,Hebecker:2018ofv}. Hence, in this paper we are going to check the most generic closed string axion candidates in terms of their ability to reach a regime where they acquire their mass from an instanton with $Sf={\cal O}$(a few) as indicated by Eq.~\eqref{eq:naivebound}. What we find is summarised in Table~\ref{tab:closedaxions}, showing that only few candidates, $C_2$ axions and thraxions, and to some extent also $C_4$ in certain limits, can violate the bound $Sf\lesssim 1$, thus potentially allowing for all dark matter to be FDM.

	The WGC \cite{ArkaniHamed:2006dz,Palti:2019pca} suggests that there must exist (some) charged states whose charge-to-mass ratio is larger than that of an extremal black hole in the theory, implying that gravity should be the weakest force. Since axions can be seen as $0$-form gauge fields, the WGC should hold for them as well. The axionic version of the WGC states that there must be an instanton whose action satisfies
	\begin{equation}\label{eq:WGC}
		S f \lesssim  \alpha M_P\coma
	\end{equation} 
	where $\alpha$ is an $\mathcal{O}(1)$ constant depending on the extremality bound entering the formulation of the conjecture. However, general extremal solutions for instantons have not been found yet, therefore the precise value of $\alpha$ is known only for special cases (see e.g.~\cite{Rudelius:2015xta,Brown:2015iha,Hebecker:2015zss,Demirtas:2019lfi}). 
	Let us mention here that in the literature many different versions of the WGC were proposed up to date (see e.g.~\cite{Harlow:2022gzl} for a recent review). In this work we mainly distinguish between `strong' and `mild' forms of the WGC. By strong WGC we mean that \textit{all} the axions present in a given model will acquire their dominant instanton potential from instantons satisfying the WGC bound. Instead, with mild WGC we refer to the statement that the WGC-satisfying instantons may give subleading contributions to the non-perturbative axion potential. This means that the mild WGC allows for some axions to acquire the leading potential from instantons with an effective $Sf>1$.
	
	Nevertheless, we can refine the statement below Eq.~\eqref{eq:naivebound} in the following way. Since the axion mass has an exponential dependence on the instanton action $S$, the accordance with or the violation of the WGC crucially depends on the precise extremality bound, i.e. on the value of $\alpha$ entering the formulation of the WGC. It appears indeed quite interesting that experiments constraining the parameter space of FDM ALPs may be able to probe the upper limit of the axionic WGC, thus shedding some light on the underlying theory of quantum gravity.

	\subsection{Closed string axions}
	\label{sec:csALPs}
	In String Theory, axion-like particles coming from closed string modes arise from the integration of $p$-form gauge field potentials over $p$-cycles of the compact space. In what follows we consider type IIB string compactifications where axions arise from the integration of the NS-NS 2-form $B_2$ and R-R 2-form $C_2$ over 2-cycles, $\Sigma_2^I$, or from the integration of the R-R 4-form $C_4$ over 4-cycles, $\Sigma_4^I$. Another axion is given by the R-R 0-form $C_0$. In order to understand where these axionic particles come from, we define the set of harmonic $(1,1)$-forms $\omega_I$, $I=\{1,\dots,h^{1,1}\}$ which are representatives of the Dolbeault cohomology group $H_{\bar{\partial}}^{1,1}(X_6,\mathbb{C})$ and the dual basis $\tilde{\omega}_I$ of $H^{2,2}$ that satisfy the following normalisation condition~\cite{Baumann2015}
	
	\begin{equation}
		\displaystyle\int_{\Sigma_2^I} \omega^J=\alpha'\delta_I^J\coma \qquad \displaystyle\int_{\Sigma_4^I}\tilde{ \omega}^J=(\alpha')^2\delta_I^J \coma
	\end{equation}

	\begin{equation}
		b_I=\frac{1}{(2\pi)^2\alpha'}\displaystyle\int_{\Sigma_2^I}B_2\coma \; c_I=\frac{1}{(2\pi)^2\alpha'}\displaystyle\int_{\Sigma_2^I}C_2\coma\; d_I=\frac{1}{(2\pi)^4(\alpha')^2}\displaystyle\int_{\Sigma_4^I}C_4\coma \;
	\end{equation}
	
	\begin{equation}
		B_2=B_2(x)+b^I(x)\omega_I\coma \quad C_2=C_2(x)+c^I(x)\omega_I\coma \quad C_4=d^I(x)\tilde{\omega}_I\fstop
	\end{equation}
	Here $B_2(x)=B_{\mu\nu}dx^\mu\wedge dx^\nu$ and $C_2(x)=C_{\mu\nu}dx^\mu\wedge dx^\nu$ are 4-dimensional 2-forms and $\alpha'$ is the inverse string tension.
	After orientifold involution the cohomology group $H^{1,1}$ splits into a direct sum of orientifold even and orientifold odd 2-forms cohomology. Therefore $\omega^I$ decomposes into  $\omega^i$  (even) and $\omega^\alpha$ (odd) respectively, where $i=1,\dots,h^{1,1}_+$, $\alpha=1,\dots,h^{1,1}_-$ and $h^{1,1}_++h^{1,1}_-=h^{1,1}$. In addition $B_2(x)$, $C_2(x)$ are projected out and we are left with the following invariant 2- and 4-form fields:
	\begin{equation}
		B_2=b^\alpha(x)\omega_\alpha\coma C_2=c^\alpha(x)\omega_\alpha\coma C_4=d^i(x)\tilde{\omega}_i\fstop
	\end{equation}
	The K\"ahler form can be written as $J=t^i(x)\omega_i$, where $t^i(x)$ are orientifold invariant real scalar fields which parametrise the volume of internal 2-cycles that are even under orientifold involution. 
	The invariant complex structure moduli are given by $\zeta^a$, $a=1,\dots,h^{2,1}_-$ while the dilaton $\phi$ and $C_0$ are automatically invariant under orientifold involution.
	After we determine the invariant scalar degrees of freedom, we need to rearrange them into the bosonic components of chiral multiplets of $\mc N=1$ supersymmetry.
	The proper coordinates of the moduli space turn out to be $h_-^{1,1}$ 2-form fields $G_\alpha$, $h_+^{1,1}$ K\"ahler moduli $T_i$, $h_-^{2,1}$ complex structure moduli $\zeta^a$ and the axio-dilaton $S$~\cite{Grimm:2004uq}:
	\begin{equation} 
	\label{eq:field_def}
		S=C_0+i\, e^{-\phi}\coma  G_\alpha=c_\alpha-S b_\alpha\coma  T_i=\tau_i + i\,d_i+\frac{i\,\kappa_{i\alpha\beta}}{2 \left(S-\bar{S}\right)}G^\alpha\left(G-\bar{G}\right)^{\beta}\fstop
	\end{equation}
	where $\tau_i=\frac{1}{2}k_{ijk}t^{j}t^{k}$ while $\kappa_{ijk}$ and $\kappa_{i\alpha\beta}$ are intersection numbers.
	We immediately see that the axionic content of the theory coming from closed string modes is given by the fields $C_0$, $c_\alpha$, $b_\alpha$, $d_i$, whose number depends on the geometrical structure of the extra dimensions.
	
		\begin{figure}
		\begin{center}
			\includegraphics[width=0.25\textwidth, angle=0]{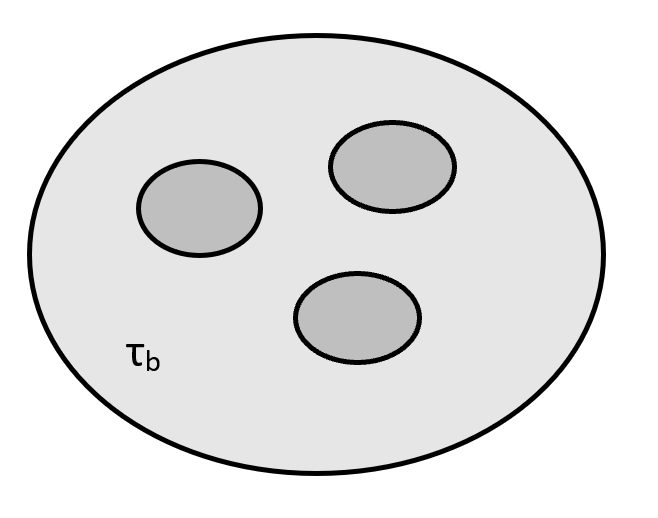}  \includegraphics[width=0.35\textwidth, angle=0]{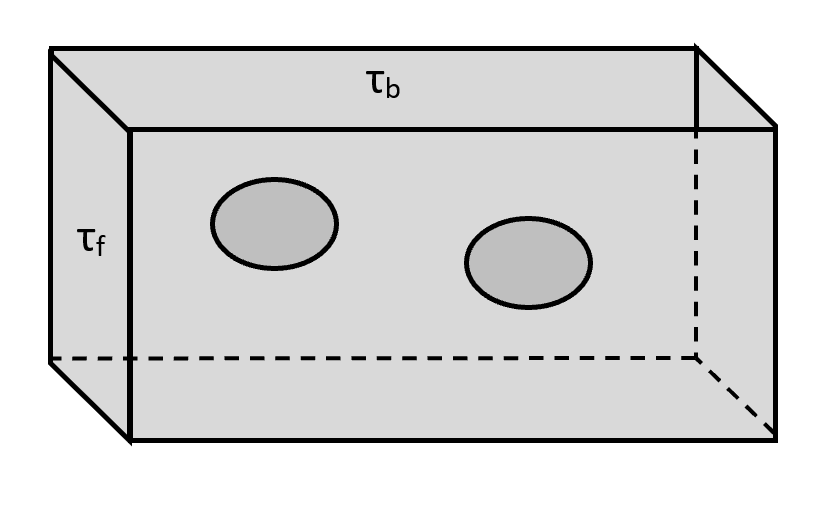}
			\includegraphics[width=0.35\textwidth, angle=0]{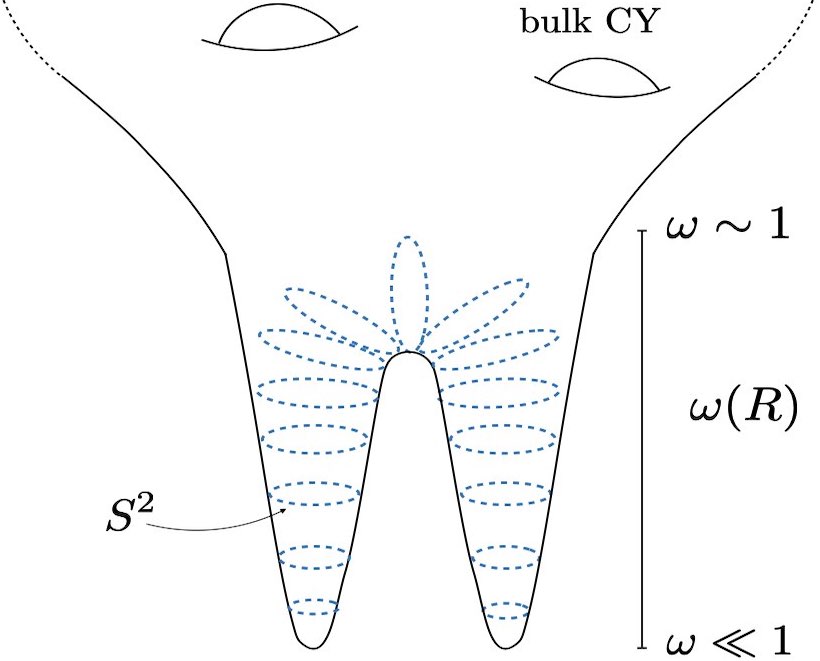}
			\caption{Pictorial representation of Swiss-cheese (left), K3 fibred geometry (centre) and double-throats (right) in Calabi-Yau (CY) threefolds.} \label{fig:c4picture}
		\end{center}
	\end{figure}
	
	Moreover, a new class of ultralight axions coming from flux compactification of type IIB string theory was recently discovered~\cite{Hebecker:2018yxs}. These so-called \textit{thraxions} are axionic modes living at the tip of warped multi-throats of the compact manifold, near a conifold transition locus in complex structure moduli space. As shown in~\cite{Hebecker:2018yxs}, at the tip of such throats there exists a 4D mode $c$ that can be thought of as the integral of the two-form $C_2$ over the $S^2$ collapsing at the conifold point, as measured far away from that point. Although so far no study has been carried out on the phenomenology of such axions, it was shown in \cite{Carta:2020ohw} that they do exist in a quite interesting fraction of orientifolds of the known compact manifolds realised as complete intersections of polynomial equations in products of projective spaces, also known as CICYs \cite{Candelas:1987kf}. More in general, it is expected that Klebanov-Strassler throats with tiny warp factor are widely present in type IIB CY orientifolds or F-theory models \cite{Ashok:2003gk,Denef:2004ze,Hebecker:2006bn}. Therefore, in this work we study how they behave as possible FDM candidates, as they are known theoretically to be ultralight and they possess a flux-enhanced decay constant. 
	
	 \begin{table}[t!]
		\centering
		\begin{tabular}{l|lcc}
			Axion &    $Sf$          \\[5pt]
			\hline\\[-5pt]
			$C_0$ 		& $ 1/\sqrt{2}\,M_P$      \\[7pt]
			$B_2$	 	& $<\,M_P$  \\[7pt]
			$C_2$ 	    & $\left\{\begin{array}{ll}
				S_{ED1}f\lesssim  \,M_P \\
				S_{ED3}f\lesssim \sqrt{g_s}\,{\cal V}^{1/3} M_P 
			\end{array}\right.$	  \\[15pt]
			$C_4\; (1\; \mbox{dof})\qquad$ 	    &  $\lesssim \sqrt{3/2}\,M_P$ 	  \\[5pt]
			$C_4\; (2\; \mbox{dof})$ 	    &  $\lesssim M_P$	  \\[5pt]
			$C_{2,\text{thrax}}$ 	    & 
        $S_{\text{eff}}f_{\text{eff}} \lesssim \frac{3\pi M^3\sqrt{g_s} }{\mathcal{V}^{1/3}} M_P
		$	  \\[6pt]
			\hline
		\end{tabular}
		\caption{Bounds on $Sf$ for different classes of closed string axions. These results arise from the study of simple extra-dimensions geometries. Further details are contained in Appendix~\ref{sec:closed_examples}. Our simple explicit constructions here saturate these bounds (`$\sim$'), while we expect more general compactifications to satisfy them (`$<$'). $C_0$ and $B_2$ are listed for completeness, but generically they cannot be FDM candidates as they get very high masses from flux stabilisation.		\label{tab:closedaxions}}
	\end{table}

	Being interested in axions that can nearly saturate the WGC bound, we analyse some simple setups that allow us to estimate the maximum value of $Sf$. These results are summed up in Table~\ref{tab:closedaxions} and further details can be found in Appendix~\ref{sec:closed_examples}. From our analysis it turns out that $C_2$, $C_4$ axions and thraxions are the best candidates to satisfy the constraint of Eq.~\eqref{eq:naivebound}. To study the behaviour of $C_4$ axions having $Sf\sim\mc{O}(1)$, we consider two different CY geometries: the Swiss-cheese case and the fibred case where the overall volume of the extra dimensions is parametrised by a single or by two degrees of freedom (dof) respectively. A pictorial view of these geometries is given in Figure~\ref{fig:c4picture}.
	Then, we study $C_2$ axions in the Swiss-cheese geometry. These fields can be viable FDM candidates in case they get a mass through non-perturbative effects coming from pure ED1 and ED3/ED1 instanton corrections. In the former case the bound on $Sf$ is similar to the $C_4$-axion case but these axions tend to be naturally lighter. In presence of ED3/ED1 corrections it seems that the strong version of WGC can be slightly violated, as in LVS $Sf\sim \mc{V}^{1/3}/\sqrt{\ln(\mc{V})}> 1$. Nevertheless, it may be not appropriate to apply the WGC in this case as this is a hybrid setup where we are effectively comparing the $C_2$ decay constant with the $C_4$ ED3 instanton action.
	These results are in agreement with what previously stated in the literature about the construction of explicit models and in works where a full mathematical analysis has been carried out for specific axion classes~\cite{Demirtas:2019lfi}. Indeed, no cases have been reported for $C_4$ and $C_2$ axions, where it was possible to clearly violate the constraint of the WGC even in its weak form while keeping the theory under control. 
	
	Concerning thraxions, we analyse both the case in which their mass is independent of the stabilisation of K\"ahler moduli and also when it gets lifted by their presence. Since their existence relies only on the presence of multi-throats and fluxes inside such throats, we do not have to specify any type of geometry for the compact manifold as thraxion features only depend on in its volume size $\mathcal{V}$. As shown in Table~\ref{tab:closedaxions}, we are also concerned with $K, \,M$, the flux numbers coming from the integral of the $H_3,\, F_3$ field strengths over $\mathcal{B}$-type and $\mathcal{A}$-type 3-cycles respectively, and the string coupling $g_s$. Thraxions will acquire a potential due to the constituting warped-down 3-form flux energy density at the IR end of a throat as well as from ED1 instanton contributions. As we discuss in Section~\ref{sec:thraxions}, the former case is more appropriate for our purpose, and we will show how to rewrite the intrinsic flux-generated thraxion potential in terms of an effective `instanton' action $S_{\text{eff}}$ which we can arrange to be dominant compared to ED1 effects.

	Besides looking at the constraint on $Sf$, we also need to consider that a good FDM axion must be extremely light. The current techniques developed to perform moduli stabilisation in type IIB are able to exclude already some possible candidates. The axio-dilaton, together with complex structure moduli, are stabilised at high energies by background fluxes, so that they are naturally too heavy to represent FDM. The same conclusion is true for the orientifold-odd axions $B_2$ which are usually much heavier than the overall volume modulus \cite{Gao_2014,Hristov_2009}. The remaining candidates are given by $C_2$, $C_4$ axions and thraxions that we analyse in the following sections.

	\subsection{Open string axions}
	
	\noindent If we are dealing with CY manifolds which contain collapsed cycles carrying a $U(1)$ charge, we might work with open string axions which come from anomalous $U(1)$ symmetries belonging to the gauge theory located at the singularity. 
	Anomalous $U(1)$ factors derive from D7-branes wrapping 4-cycles in the geometric regime or from D3-branes at singularities. The anomalous $U(1)$ gauge boson acquires a mass in the process of anomaly cancellation eating up the open string axion for D7-branes or the closed string axion for D3-branes, when the 4-cycle saxion is collapsed at singularity \cite{Green:1984sg,Allahverdi:2014ppa}. At energies below the gauge boson mass, the theory features a global $U(1)$ symmetry. In the presence of 4-cycles that are collapsed at singularity, some complex scalar matter field $C=|C|e^{i\sigma}$ can be charged under the global $U(1)$ symmetry and its phase $\sigma$ may represent an open string axion. Indeed, the global symmetry can be broken by subdominant supersymmetry breaking contributions coming from background fluxes~\cite{Cicoli:2013cha}, making $\sigma$ the Nambu-Goldstone boson of the broken $U(1)$. Under these conditions, the open string axion decay constant becomes
	\be
	\begin{array}{ll}
		f\propto \frac{1}{\mathcal{V}^\alpha}\coma\quad \alpha=1,2
	\end{array}
	\ee
	where the values of $\alpha$ are related to sequestered ($\alpha=1$)  and to super-sequestered ($\alpha=2$) scenario respectively~\cite{Cicoli:2013cha}.
	This particle acquires a mass through hidden sector strong dynamics instanton effects. The scale of strong dynamics in the hidden sector is given by
	\be
	\Lambda_{hid}=M_P e^{-c/g_s^2}\coma
	\ee
	where $c$ is fixed by the 1-loop $\beta$ function
	\begin{equation}
		\frac{1}{g_s^2}=\frac{1}{g_{s,0}^2}+\frac{\beta}{4\pi}\ln(\dots)\fstop
	\end{equation}
	These quantities fix the open string axion mass scale to be
	\begin{equation}
		m_{\sigma}^2=\Lambda_{hid}^4/f_{seq}^2\fstop
	\end{equation}
	Being interested in ultralight axions, we will need an extremely low scale of strong dynamics in the hidden sector. The only parameter choice that may lead to a high decay constant is given by $\alpha=1$, i.e. the sequestered scenario, where
	\begin{equation}
		\begin{array}{ll}
			f_{\sigma}= p\, \frac{M_P}{\mathcal{V}}\fstop
		\end{array}
	\end{equation}
	Plugging this result inside Eq.~\eqref{eq:DMabundance} and assuming $m_\sigma= 10^{-22}$ eV, we get 
	$\mc{V}\simeq 2 \times 10^{2}$, which is consistent with the sequestered assumptions described in Appendix~\ref{sec:open_example}.
	
	On the other hand, matching the right mass requires
	\be
	\langle S \rangle=\frac{1}{c}\ln{\left(\frac{M_P}{\Lambda_{hid}}\right)}\simeq \frac{59}{c}\fstop
	\ee 
	For $c\sim \mc{O}(1)$, this is consistent with the use of a perturbative approach to string theory, being $g_s\simeq 0.13$ and implies that $\Lambda_{hid}\simeq  70 \; \mbox{eV}$. Therefore, we see that if we want an open string axion to be the FDM particle we need to deal with small extra dimension volumes and extremely low scales for the hidden sector strong dynamics.
	
	Despite fine-tuning of parameters being quite reduced in this context, the required setup is not as general as for closed string axions and it is not easy to give these axions a precise upper bound on $Sf$. In addition, one should take into account that strong dynamics may induce a non-negligible production of glueballs that may represent a non-vanishing contribution to DM. In order to give a precise estimate of the amount of glueballs production it would be necessary to focus on some explicit models but this is far beyond the aim of this paper. Outside string theory, a concrete example where FDM candidates arise from infrared confining dynamics can be found in \cite{Davoudiasl:2017jke}.
	
	Given that closed string axions represent a model-independent feature of string compactifications, the forthcoming sections will be devoted to the general constraints and the predictions coming from explicit FDM constructions in this context. 
	
	\section{FDM from closed string axions}
	\label{sec:closedALPs}
	Let us start by reviewing how to compute $C_2$ and $C_4$ axions decay constant and mass, which are the two relevant quantities in FDM models.
	The axion fields $d_i$ and $c^\alpha$ arise as harmonic zero modes of $p$-forms gauge potentials on $p$-cycles of the compactification space. Hence, at the perturbative level, the 10D gauge invariances of the $p$-form gauge fields descend to continuous shift symmetries of their associated $p$-form axions in 4D $\Phi_i\sim \Phi_i+c$, $c\in \mathbb{R}$.
	The kinetic part of the 4D Lagrangian contains the following terms associated to the axions:
	\be
	\mc{L}\supset \frac{g_{ij}}{2} \partial_\mu \Phi^i \partial^\mu \Phi^j\coma
	\ee
	where $g_{ij}=2\frac{\partial^2 \mathcal{K}}{\partial T^i \partial \bar{T}^j}$ for $C_4$ axions, $g_{ij}=2\frac{\partial^2 \mathcal{K}}{\partial G^i \partial \bar{G}^j}$ for $C_2$ axions and thraxions, and $\mathcal{K}$ is the K\"ahler potential of the theory. 
	In order to work with canonically normalised fields, we need to diagonalise the K\"ahler metric and find the axion metric eigenvalues $\lambda_i$ and eigenvectors $\tilde{\Phi}_i$. After that, we define the canonically normalised axion fields as $\phi_i =\sqrt{\lambda_i} \tilde{\Phi}_i M_P$ (restoring proper powers of $M_P$) where \cite{Cicoli:2012sz}
	\be
	\mc{L}_{\rm kin}\supset \frac{\lambda_i M_P^2}{2} \partial_\mu \tilde{\Phi}_i \partial^\mu \tilde{\Phi}_i=
	\frac12 \partial_\mu \phi_i \partial^\mu \phi_i \fstop
	\ee
	In the case of massless axions, it is quite common to refer to $\hat{f}_i = \sqrt{\lambda_i}\,M_P$ as the axion decay constant. This derives from the fact that the couplings of the physical axions with all other fields scale as $\propto 1/\hat{f}$. So far we have only considered massless axions but, as with the rest of the moduli, these fields need to be stabilised. 
	
	Axions acquire a mass through non-perturbative quantum corrections (instantons coming from branes wrapping internal cycles) that break their continuous shift symmetry down to their discrete subgroup. The typical form of the potential arising from a single non-perturbative correction reads
	\be
	V(\phi_i)=\Lambda_i \cos(a_i \Phi_i)\coma
	\ee
	where $a_i=2\pi/N_i$, with $N_i\in \mathbb N^+$ being the rank of the gauge group living on the branes. In general, to work with physical fields we need to find the field basis that diagonalises both the mass matrix and the field space metric. In the simplest case where the K\"ahler metric is approximately diagonal ($\Phi_i\sim \tilde{\Phi}_i$) and we have a single non-perturbative correction, computing the decay constant becomes rather simple. Noticing that the field periodicity corresponds to that of the potential, the stabilised axion decay constant, $f_i$, derives from
	\be
	\begin{array}{lll}
		a_i \,\Phi_i\quad &\rightarrow \quad a_i\,\Phi_i+ 2\pi\;k \qquad\; \mbox{implying that}\\[10pt]
		\phi_i \quad &\rightarrow \quad \phi_i + 2\pi f_i\;k \qquad\mbox{where}\qquad f_i = \sqrt{\lambda_i}\,\frac{M_P}{a_i}=\,\frac{\hat{f}_i}{a_i}   \fstop
	\end{array}
	\ee
	For a complete and exhaustive treatment about dealing with a non-diagonal field space metric, multiple instanton corrections, and non-trivial instanton charge matrix, see \cite{Bachlechner:2014gfa}.

	The thraxion potential comes instead from corrections to the superpotential $W$ governed by powers of the warp factor $\omega$, which tends to zero when approaching the conifold limit. In the pure ISD solution of~\cite{Giddings:2001yu}, the effective potential for the thraxion $c$ takes the form 
	$
	V\sim \omega^6\cos\left(c/M\right)
	$,
	where $M$ is the flux quantum coming from the presence of a 3-form flux  integrated over the 3-cycle that is shrinking at the bottom of the warped throat. The corrections to $W$ break the continuous shift symmetry but they preserve a set of discrete ones. Using the same notation as in the $C_4$ case above, we get that the effective decay constant is enhanced by a factor $M$, namely $f_{\mbox{\tiny{eff}}}\simeq M \hat{f}$.
	However, in the thraxion case this computation is quite model dependent. It is better to derive the decay constant in its general form from the 10D perspective, by dimensionally reducing to 4D the $|F_3|^3$ term and plugging the expansion of $C_2$ in harmonic forms. In this way, one can show that $f$ depends explicitly on inverse powers of the warp factor $\omega$ coming from the Klebanov-Tseytlin throat metric~\cite{Klebanov:2000nc}.
	In order to estimate mass and decay constant values, we have to analyse how these depend on the microscopic parameters of the theory through moduli stabilisation. Two prominent prescriptions to perform moduli stabilisation in type IIB string compactification are given by LVS~\cite{Balasubramanian:2005zx} and KKLT~\cite{Kachru:2003aw}. These rely on different constructions and give rise to different mass spectra for the moduli fields. For these reasons we analyse them separately. To clear the physical meaning and the values of the parameters used in the following sections, we summarize them in Table~\ref{tab:all_parameters}. 
    \begin{table}[t!]
		\centering
		\begin{tabular}{l|lcc}
			               & description                    & LVS range             & KKLT range       \\[3pt]
			\hline
			$W_0$          & tree-level superpotential 	&  $1\div10^2$          &  $\exp{\left(-\frac{2\pi}{N}\, \mc{V}^{2/3}\right)}$ \\[3pt]
			$g_s$           & string coupling	            &  $(\ln\mc{V})^{-1}$  & $10^{-2}\div 0.2$ \\[3pt]
			$A$	        	& non-perturbative correction prefactor &  $10^{-4}\div 10^{4}$ & $10^{-4}\div 10^{4}$ \\[3pt]
			$N$ 	        & number of D7-branes           &  $1\div 10$           & $30\div 60$  \\[3pt]   
			$M,\, K$ 	        & flux numbers from $F_3,\, H_3$           &  $\geq 10$           & $\geq 10$   \\[3pt] 
			\hline
		\end{tabular}
		\caption{Description of microscopic parameters and their associated ranges considered in the study of closed string axions in LVS and KKLT moduli stabilisation. Where we provide a functional form, i.e. $g_s$ in LVS or $W_0$ in KKLT, no a priori range can be given. For instance, $W_0$ cannot be interpreted as a parameter in KKLT as its value fixes the whole stabilisation. The same reasoning applies to $g_s$ in LVS. For simplicity, we do not make any assumptions about the distribution of parameters. The parameter ranges are chosen according to model building literature \cite{Kachru:2003aw,Balasubramanian:2005zx}.  }		\label{tab:all_parameters}
	\end{table}	
	
	\subsection{LVS: FDM from $C_4$ axions}
	\label{sec:LVSC4}
	As its name suggests, LVS moduli stabilisation allows the volume of the extra dimensions to be stabilised at exponentially large values. This creates a natural hierarchy between energy scales that can be parametrised by inverse powers of the overall volume. This is particularly convenient for phenomenology, since it allows us to perform moduli stabilisation step by step, at different energies. After flux stabilisation, the K\"ahler moduli are still flat directions thanks to the so called `no-scale structure'. They can be stabilised using perturbative and non-perturbative corrections to the K\"ahler potential and the superpotential. In this section, we will assume for simplicity that $h^{1,1}_-=0$.
	LVS describes a way to stabilise K\"ahler moduli using the interplay between non-perturbative corrections to the superpotential coming from euclidean ED3 instantons or gaugino condensation and leading order $\alpha'$ corrections to the K\"ahler potential of the form
	\begin{equation}\label{eq:LVS_contr}
		\begin{cases}
			\mathcal{K}&=\mathcal{K}_0 -2\ln \left(\mc{V}+\frac{\hat{\xi}}{2}\right)  \\
			W&= W_0 + \sum_{i} A_ie^{-a_iT_i} \coma
		\end{cases}
	\end{equation}
	where $\hat{\xi}=\xi/g_s^{3/2}$,  $\xi=-\frac{\zeta(3)\chi(Y_6)}{2(2\pi)^3}$, $\chi(Y_6)$ is the effective Euler characteristic of the CY manifold \cite{Becker:2002nn}, $W_0$ is the tree-level superpotential coming from background fluxes stabilisation, $A_i$ depends on the VEVs of complex structure moduli and the dilaton, and $a=2\pi/N$ where $N=1$ for euclidean ED3 instantons or $N>1$ for gaugino condensation. The moduli stabilisation prescription of LVS holds if the number of 3-cycles is larger than the number of 4-cycles, i.e. $h^{2,1}>h^{1,1}>1$ and in presence of at least one shrinkable 4-cycle.	
	Being interested in large 4-cycles parametrising the overall volume of extra dimensions, let us consider a simplified version of the so-called weak Swiss-cheese volume form, namely
	\be
	\label{eq:CY_vol}
	\mathcal{V}=\left(f_{3/2}(\tau_i)-\tau_s^{3/2}\right)\qquad i=1\dots N \coma
	\ee
	where $f_{3/2}$ is a function of degree $3/2$ in $\tau_i$ that we assume to be given by a single term for simplicity and $\tau_s$ is a diagonal contractible blow-up cycle.
	Given this simplifying assumptions and considering non-perturbative corrections to $W$ only related to the small cycle $T_s$, LVS stabilisation is able to fix three directions in the K\"ahler moduli space, namely the overall volume $\mc{V}$, the small cycle $\tau_s$ and the $C_4$ axion $d_s$ at
	\be 
	\langle \tau_s\rangle^{3/2}\simeq \frac{\hat{\xi}}{2}\,, \qquad e^{-a_s\langle \tau_s\rangle}\simeq\frac{\sqrt{\tau_s}|W_0|}{a_s|A_s| \mc{V}}\,, \qquad a_s d_s=(1 + 2k)\pi\coma
	\ee
	where $k\in\mathbb{Z}$.
	From the previous equations we see that the LVS minimum lies at exponentially large volume $\vo\sim e^{a_s\tau_s}\gg 1$ and does not require any fine-tuning on the tree-level superpotential $W_0\sim 1\div 100$. Non-perturbative effects do not destabilise the flux-stabilised complex structure moduli and the dilaton. Moreover, supersymmetry is mostly broken by the F-terms of the K\"ahler moduli and the gravitino mass is exponentially suppressed with respect to $M_P$, allowing to get low-energy supersymmetry in a natural way. These models are characterised by a non-supersymmetric anti de Sitter minimum of the scalar potential at exponentially large volume. Since the value of the scalar potential in its minimum gives the value of the cosmological constant, we must find a way to uplift this negative minimum to a de Sitter vacuum. This can be done by switching on magnetic fluxes on D7-branes~\cite{Burgess:2003ic}, adding anti D3-branes~\cite{Kachru:2003aw, Kallosh:2014wsa, Bergshoeff:2015jxa,Kallosh:2015nia,Aparicio:2015psl,Garcia-Etxebarria:2015lif,GarciadelMoral:2017vnz,Moritz:2017xto,Crino:2020qwk}, hidden sector T-branes~\cite{Cicoli:2015ylx}, non-perturbative effects at singularities~\cite{Cicoli:2012fh}, non-zero F-terms of the complex structure moduli~\cite{Gallego:2017dvd} or via the winding mechanism coming from a flat direction in the complex-structure moduli space~\cite{Hebecker:2020ejb,Carta:2021sms}. Note that the uplift to de Sitter does not change the axion potential, as \textit{generically} the fields responsible for the uplift are the real part of the K\"ahler moduli, $\tau$, appearing in Eq.(\ref{eq:field_def}).
	
	If the CY volume of Eq.~\eqref{eq:CY_vol} is parametrised by a single K\"ahler modulus, i.e. ($f_{3/2}=\tau_1^3$) LVS is able to stabilise all the real part of K\"ahler moduli. If this is not the case, i.e. ($f_{3/2}=\tau_1\sqrt{\tau_2}$ or $f_{3/2}=\sqrt{\tau_1\tau_2\tau_3}$) we will be left with some flat directions in the K\"ahler moduli space. A potential for these fields can be generated at lower energies by e.g. higher order $\alpha'$ and $g_s$-loop corrections. Once these fields get stabilised, the scalar potential for the axions associated to volume cycles is just induced by non-perturbative terms as in Eq.~\eqref{eq:LVS_contr}.
	
	The field dependence of the decay constant associated to $C_4$ axions is given by \cite{Cicoli:2012sz}
	\be
	\label{eq:dc}
	f\sim \left\{
	\begin{array}{ll}
		\frac{M_P}{\tau}&\qquad \mbox{volume closed string axion,}  \\[10pt]
		\frac{M_P}{\sqrt{\vo}} &\qquad \mbox{blow-up closed string axion.} \\[10pt]
	\end{array}\right.
	\ee
	Moreover, in this setup the instanton action appearing in the axion potential of Eq.~\eqref{eq:AxionLagr} is given by $S=a\tau$. 
	Looking for a particle having a high decay constant and an extremely small mass, we immediately see from Eq.~\eqref{eq:dc} that a FDM particle is more likely represented by axions related to large cycles parametrising the overall volume. In fact, while blow-up cycles seem to have a higher decay constant ($\sim \mc{V}^{-1/2}$) compared to volume cycles ($\sim \mc{V}^{-2/3}$), LVS stabilisation requires that $\mc{V}\sim e^{a_s\tau_s}$. This implies that matching the right FDM mass value tunes the overall volume too large ($\mc{V}\sim e^{220}$) making the match between $m$ and $f$ unfeasible and, above all, this would cause the string scale to be much lower than eV where the theory is no longer under control.  In addition, looking at the $\tau$ dependence of the decay constant, the axion mass and the total amount of FDM, Eq.~\eqref{eq:DMabundance}, we can easily conclude that in presence of multiple volume axions, the heavier particles will represent a higher percentage of dark matter. Indeed, assuming that all the other parameters and the initial misalignment angle are the same for every axion, we have that $\frac{\Omega_{\theta}}{0.112}\sim e^{-S/4}\propto m_{\theta}^{1/2}$. 
	
	In what follows, we are going to analyse two simple examples of concrete 4D effective models coming from type IIB string theory: Swiss-cheese and fibred CY threefolds.

	\paragraph{Swiss-cheese geometry}
	\noindent This model is based on a CY having the typical Swiss-cheese shape
	\be
	\vo=\alpha \left(\tau_\vo^{3/2} -\lambda_s \tau_{s}^{3/2}\right)\coma
	\ee
	where $\alpha$ and $\lambda_s$ are positive real coefficients of order one.
	After LVS stabilisation, all K\"ahler moduli but the overall volume axion $d_{\mc{V}}$ have been stabilised. This will represent our FDM candidate whose mass is given by
	\be
	m_{d_\vo}^2=\frac{8\kappa S_\vo^3 A_\vo\,W_0\, e^{-S_\vo}}{3 \vo^2} M_P^2 \coma\, S_\vo=a_\vo \tau_\vo\coma
	\ee
	while its decay constant is
	\be
	f_\vo = \sqrt{\frac32}\frac{M_P}{ S_\mathcal{V}}\fstop
	\ee
	The previous relations are based on the assumption that both the kinetic Lagrangian and the mass matrix associated to $C_4$ axions are diagonal. Working in the large volume limit, this can be safely assumed, as the off-diagonal terms of the K\"ahler matrix are suppressed by powers of $\tau_s/\tau_b\ll1$ while the off-diagonal terms of the mass matrix are exponentially suppressed. 
	In what follows, we try to understand which requirements are needed to match the FDM prescriptions. Assuming to have no prior knowledge on the cosmological history of the universe, we assume a constant axion field distribution. Given a uniform probability density on the range $[0;2\pi]$, the mean value of $d_{\mc{V}}$ is given by $\pi$ and its standard deviation is $\sigma_d^2=\pi^2/3$, therefore we consider a misalignment angle $d_{mi}=\pi/\sqrt{3}$ as it represents the most likely value. This assumption is supported by~\cite{Graham:2018jyp} where it was shown that for any inflationary scale $H\gtrsim 1$ keV, the misalignment angle distribution becomes flat through stochastic diffusion. The most stringent constraint on inflationary model building in FDM models comes from isocurvature perturbation bounds as we briefly describe in Appendix~\ref{sec:anharm}.
	
	\begin{table}[t!]
	\begin{center}
		\begin{tabular}{l|l|l|lc}
			&    $N=1$   & $N=2$ & $N=10$       \\[3pt]
			\hline
			$\vo$ &  $200\div 300$  & $500\div 800$  & $6000 \div 9000$\\[3pt]
			\hline
		\end{tabular}
		\caption{Predicted overall volumes for different values of $N$ imposing 100\% of FDM.	}\label{tab:cvalues}
	\end{center}
	\end{table}
	\begin{figure}[t!]
		\begin{center}
			\includegraphics[width=0.47\textwidth]{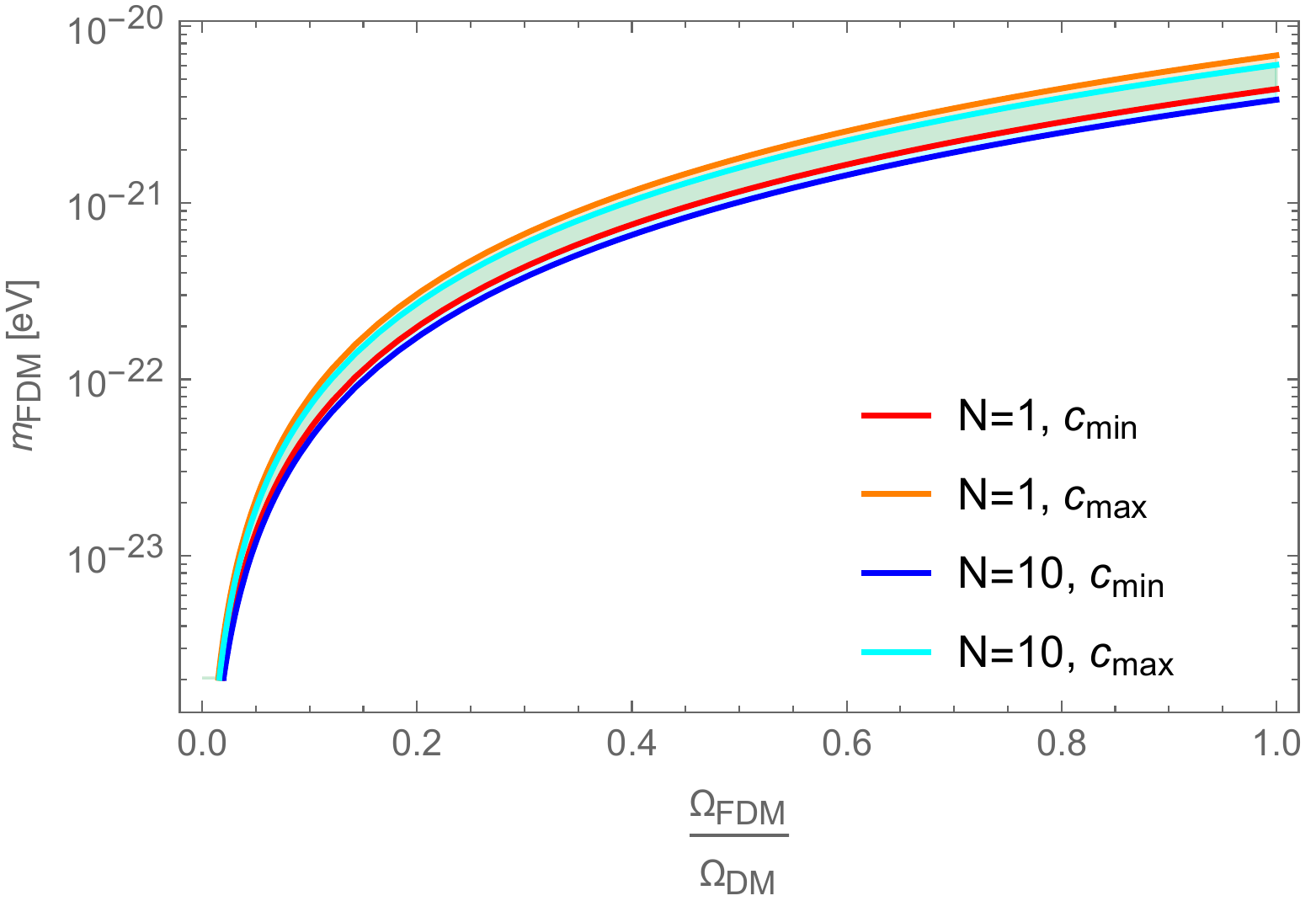}\hspace{6pt}\includegraphics[width=0.49\textwidth]{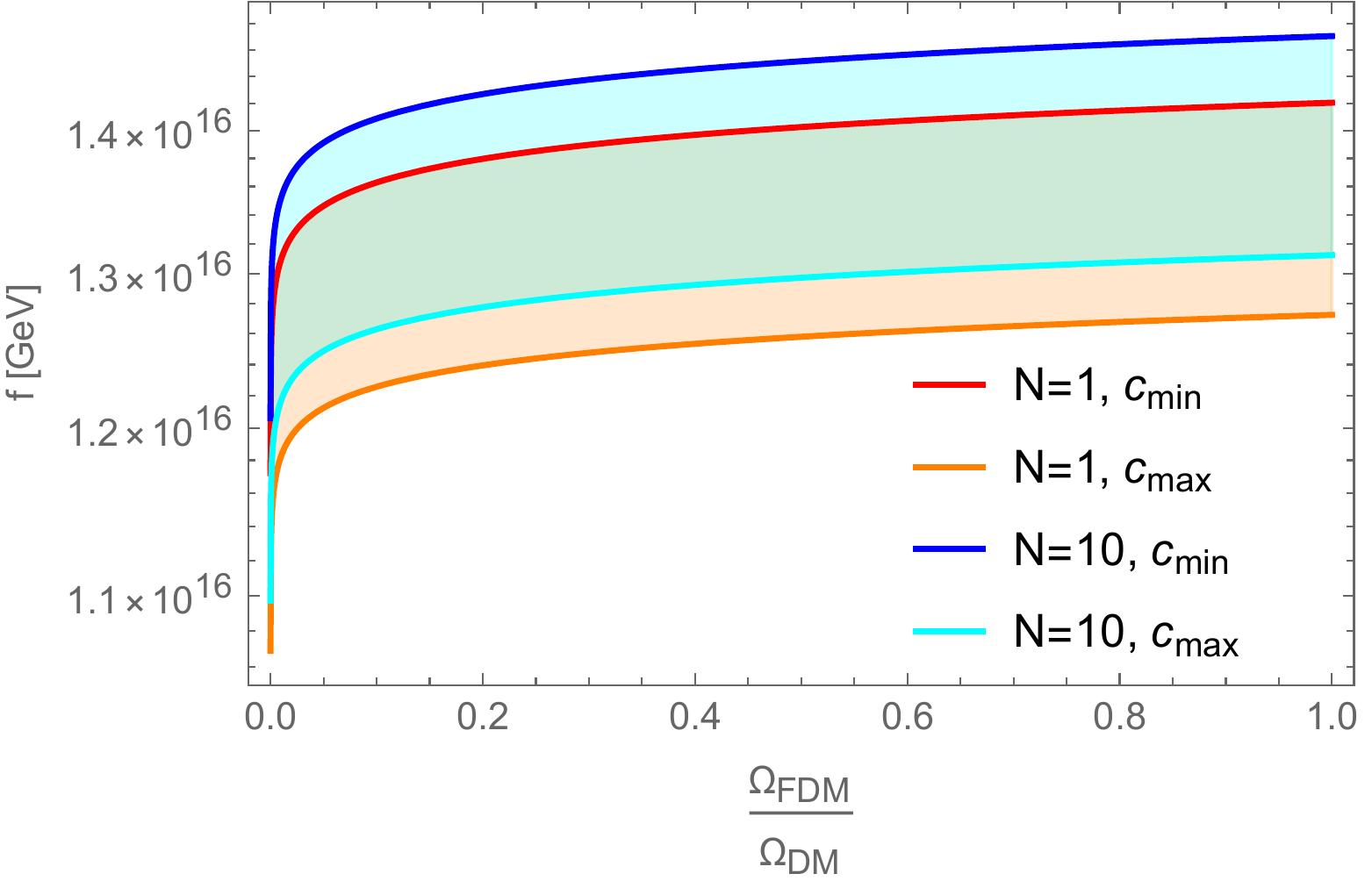}
			
			\caption{Predictions for axion mass (left) and decay constant (right) varying the percentage of axionic FDM and the non-perturbative effects giving rise to the ALP potential.  }\label{Fig:MassFPercentage} 
		\end{center}
	\end{figure} 
	
	In the Swiss-cheese geometry, the amount of DM depends only on the instanton action $S_\vo$. This implies that once we fix the required amount of DM, we can immediately compute the natural value of mass and decay constant the FDM axion candidate needs to have.
	Knowing the shape of the instanton action, we can write $\vo\simeq (S_\vo/a_\vo)^{3/2}$, being $a_\vo=2\pi/N$, so that the formula for the DM abundance, Eq.~\eqref{eq:DMabundance}, becomes:
	\begin{equation}
		\label{eq:thetaVamount}
		\frac{\Omega_{\theta}h^2}{0.112}\simeq 6.36\times 10^{27}  a_\vo^{3/4} (c\, g_s)^{1/4}   \frac{e^{-S_\vo/4}}{S_\vo^2 }\coma
	\end{equation}
	where $c=W_0 A_\vo$.
	Given that the value of the parameters may vary across different models, we decide to fix the maximum and minimum values that they may acquire and we choose different values of $N=\{1,2,10\}$. Moreover, we use the LVS relation between the string coupling and the overall volume: $\mc{V}\sim e^{g_s^{-1}}$ to reduce the amount of fine-tuning. The extrema of the values we consider are listed in Table~\ref{tab:all_parameters}.

	Looking at the previous formula it is clear that once we fix the upper and lower bounds for $W_0$, $A_\vo$ and the fraction of FDM $\frac{\Omega_{FDM}}{\Omega_{DM}}$, we can easily determine the mass and decay constant values of our axion candidate.
	The natural amount of axionic DM with the right mass and decay constant range can be found in Fig.~\ref{Fig:MassFPercentage}. While the predictions for the decay constant are not significantly influenced by changing parameters, the particle mass can vary across different setups. As shown in Table~\ref{tab:cvalues}, these setups put strong constraints on the predicted overall volume $\vo$. The lightest DM particles representing a considerable fraction of FDM have $m\lesssim 10^{-20}$ eV. It is worth noticing that neither the mass nor the decay constant value seem to be sensitive to the gauge theory on the brane stack.  
	
	Concerning the implications related to this FDM model, let us now estimate what the relevant energy scales are going to be. The KK scales, i.e. the maximum energies at which a 4D treatment of the theory is allowed, that are associated with bulk KK modes and KK replicas of open string modes living on D7-branes wrapping 4-cycles are given by
	\be
	\label{eq:KKmassC4}
	M_{KK}^{(i)}=\frac{\sqrt{\pi}M_P}{\sqrt{\mc{V}}\tau_i^{1/4}}\fstop
	\ee 
	This implies that for the Swiss-cheese geometry $M_{KK}=\frac{\sqrt{\pi}M_P}{\mc{V}^{2/3}}\sim 10^{15}\div 10^{16}$ GeV. Moreover, we have that the blow-up moduli which are stabilised through LVS prescription receive masses comparable to the gravitino mass, $m_{3/2}=M_P W_0/\mc{V}\sim 10^{14}\div 10^{16}$ GeV. The last relevant energy scale is given by the inflationary scale.
	Looking at the ALP decay constant and mass, we can estimate what are the predictions for inflation that would arise from the ultralight $C_4$ axion detection. These are mainly due to isocurvature perturbations constraint and imply that the Hubble parameter during inflation, $H_I$, needs to be low, $H_I<5\cdot 10^{11}$ GeV, giving rise to undetectable stochastic gravitational waves, being the tensor-to-scalar ratio $r< 10^{-6}$. An extended derivation of these results can be found in Appendix~\ref{sec:anharm}. We conclude this paragraph by stressing that since FDM needs to be the dominant DM component, the mass spectrum of the theory between the inflationary scale and the FDM scale should be nearly empty. In particular, as we already stressed, since heavier axions naturally represent higher DM fractions, the axion spectrum in the aforementioned range needs to be exactly empty.
	
	\paragraph{Fibred geometry} Consider a fibred CY, whose volume can be written as
	\be
	\vo =\alpha\left( \tau_b \sqrt{\tau_f}-\lambda_s \tau_s^{3/2} \right)\coma
	\ee
	where $\tau_f$ parametrises the volume of a K3 fibre over a $\Bbb P^1$ base whose volume is controlled by $\tau_b$, and $\tau_s$ represents the volume of a rigid del Pezzo divisor. Again, $\alpha$ and $\lambda_s$ are positive real coefficients of order one. After LVS stabilisation, the fibre modulus is still a flat direction and requires additional corrections to be stabilised. These are usually taken to be $\alpha'$ corrections or KK and winding $g_s$ loop corrections~\cite{Berg:2005ja,Berg:2004ek,vonGersdorff:2005bf,Berg:2007wt,Cicoli:2007xp,Cicoli:2008va}. In this setup, the two good FDM candidates are the closed string axions related to the base and the fibre modulus. The shape of the decay constants in case of ED3 brane instantons or purely gauge theories on the D7-branes wrapping the 4-cycles, are given by
	\be
	\label{eq:decayconstants}
	\left\{
	\begin{array}{ll}
		f_{d_b}=\frac{M_P}{a_b\tau_b}=\frac{M_P}{S_b}\\[10pt]
		f_{d_f}=\frac{M_P}{\sqrt{2} a_f \tau_f}=\frac{1}{\sqrt{2}}\frac{M_P}{S_f}\\[10pt]
	\end{array}\right.
	\ee
	while their masses are
	\be
	\label{eq:massesbf}
	\begin{array}{lll}
		m_{d_f}^2\simeq \frac{8\kappa S_f^3 A_f \,W_0\, e^{-S_f}}{\mathcal{V}^2} M_P^2\coma \\[10pt]
		m_{d_b}^2\simeq\frac{4\kappa S_b^3 A_b \,W_0\, e^{-S_b}}{\mathcal{V}^2} M_P^2 \fstop\\
	\end{array}
	\ee
	Again, these relations are based on the assumption that both the kinetic Lagrangian and the mass matrix associated to $C_4$ axions are diagonal. As the field-space metric related to $\tau_f$ and $\tau_b$ is exactly diagonal, the same considerations provided in the Swiss-cheese geometry apply. 
	Without loss of generality we can consider the case where $\alpha=1$ so that
	\begin{equation}
		\mc{V}=\tau_b \sqrt{\tau_f}=\frac{S_b \sqrt{S_f}}{a_b \sqrt{a_f}} \fstop
	\end{equation}
	The masses of the two axions become
	\be
	\begin{array}{lll}
		m_{d_f}^2\simeq c_f a_b^2a_f \frac{S_f^2\,}{S_b^2}  e^{-S_f} M_P^2\coma  \qquad c_f=2 g A_f  \\[10pt]
		m_{d_b}^2\simeq c_b a_b^2a_f \frac{S_b\,}{S_f}  e^{-S_b} M_P^2\coma \qquad c_b= g A_b\\
	\end{array}
	\ee
	where $g=4\kappa W_0$.
	Fixing the ratio between the two decay constants to be $q=f_{d_b}/f_{d_f}$ we immediately see that the ratio between the abundance of DM components is given by
	\begin{equation}
		\frac{\Omega_{b}}{\Omega_f}\simeq 1.09\left(\frac{c_b}{c_f}\right)^{1/4}q^{5/4}e^{-\frac{M_P}{4f_b}\left(1-\frac{q}{\sqrt{2}}\right)}\fstop
	\end{equation}
	\begin{figure}[!t]
		\begin{center}
			\includegraphics[width=0.7\textwidth]{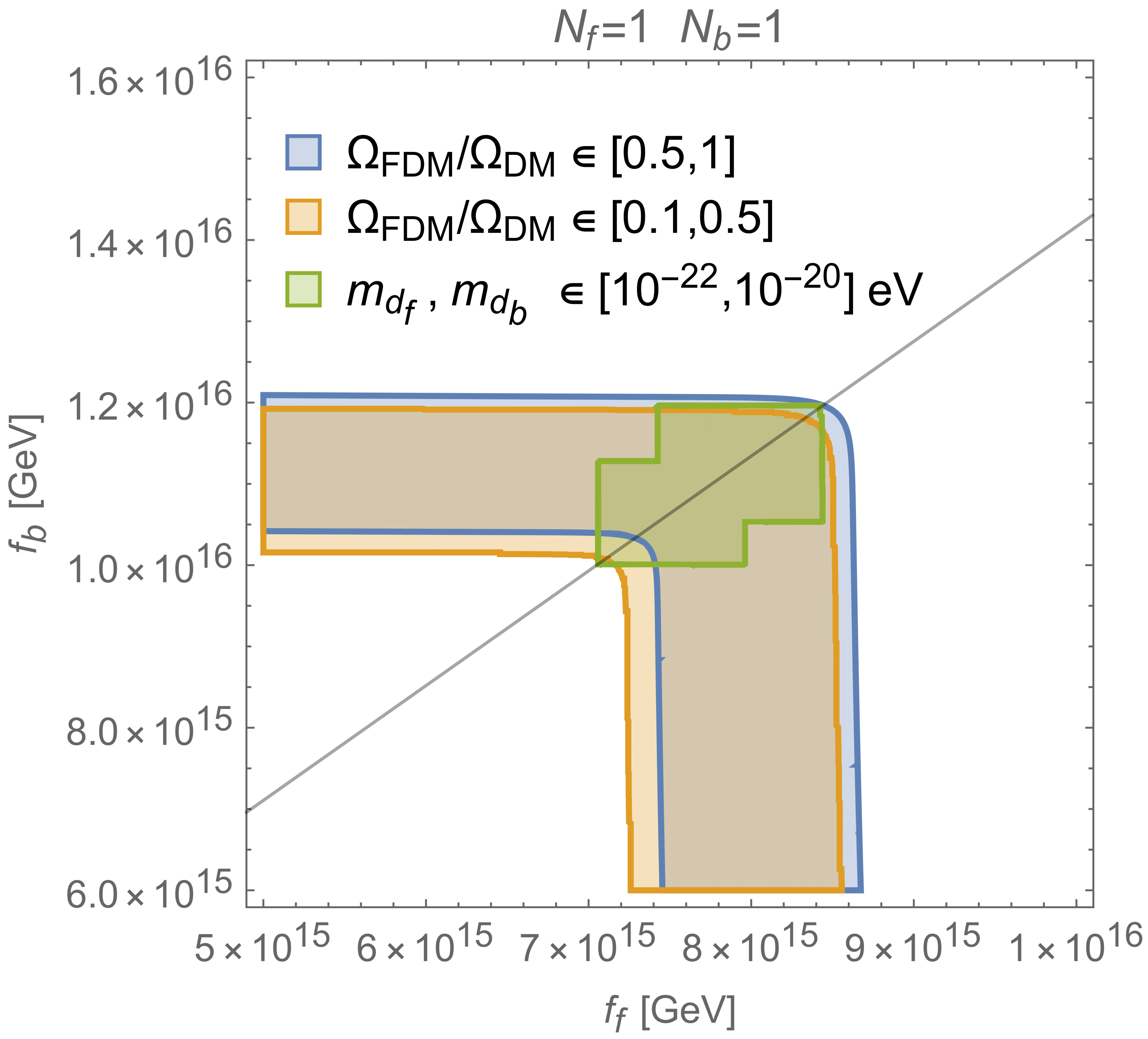}
		\end{center}
		\caption{\label{fig:FDMfibrePercentages} Allowed percentages of axionic DM as a function of the axion decay constants. The coloured areas satisfy the constraint $\frac{\Omega_{b}h^2}{0.112}+\frac{\Omega_{f}h^2}{0.112}\leq 1$. The blue and yellow areas refer to regions where ultralight axionic DM represents different percentages of the total amount of DM of the universe. The green area identifies the region where we have two FDM axions. The black line is given by $q=f_b/f_f=\sqrt{2}$. }
	\end{figure}
	
	\begin{figure}
		\begin{center}
			\includegraphics[width=0.49\textwidth]{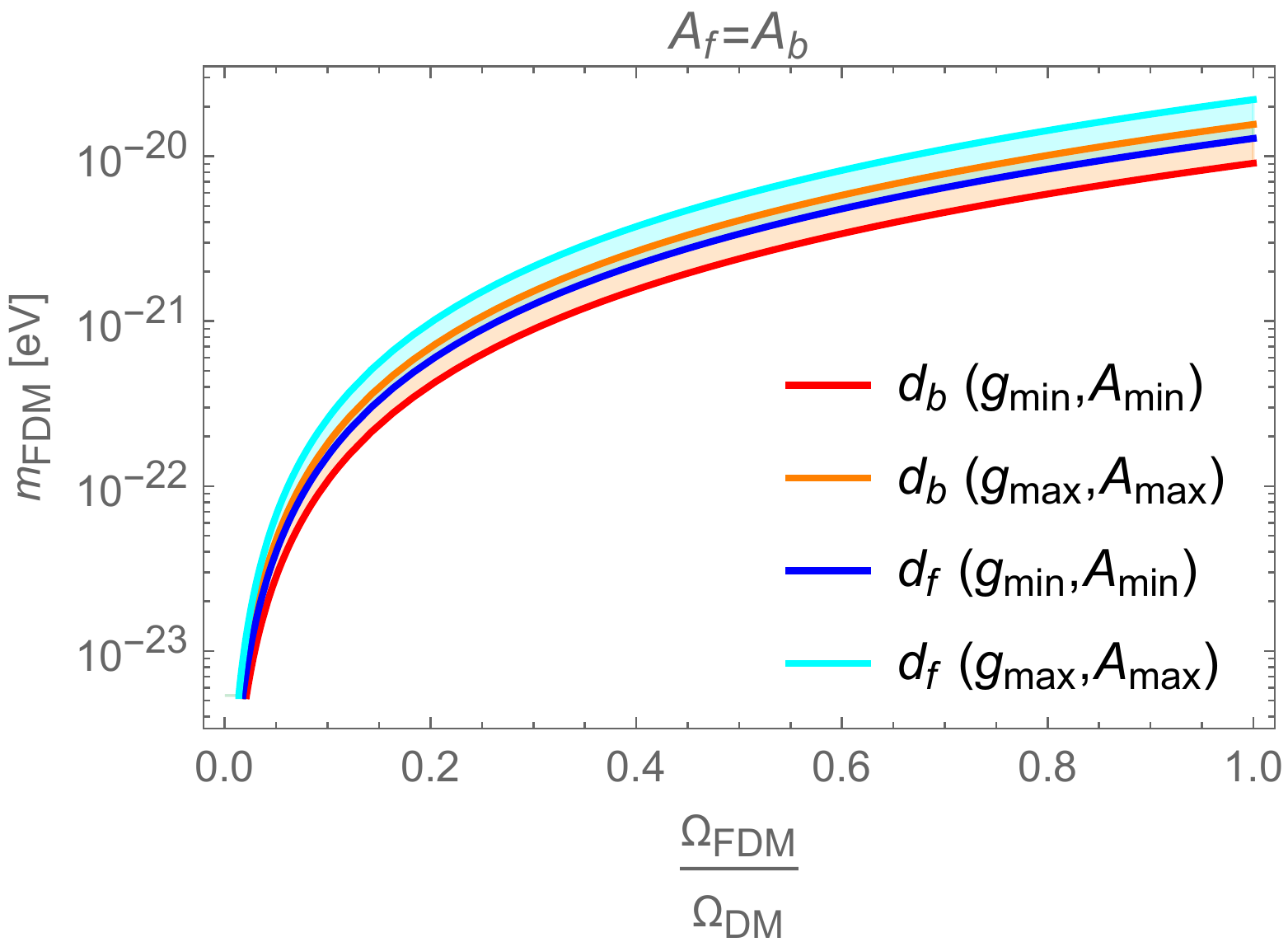}\includegraphics[width=0.49\textwidth]{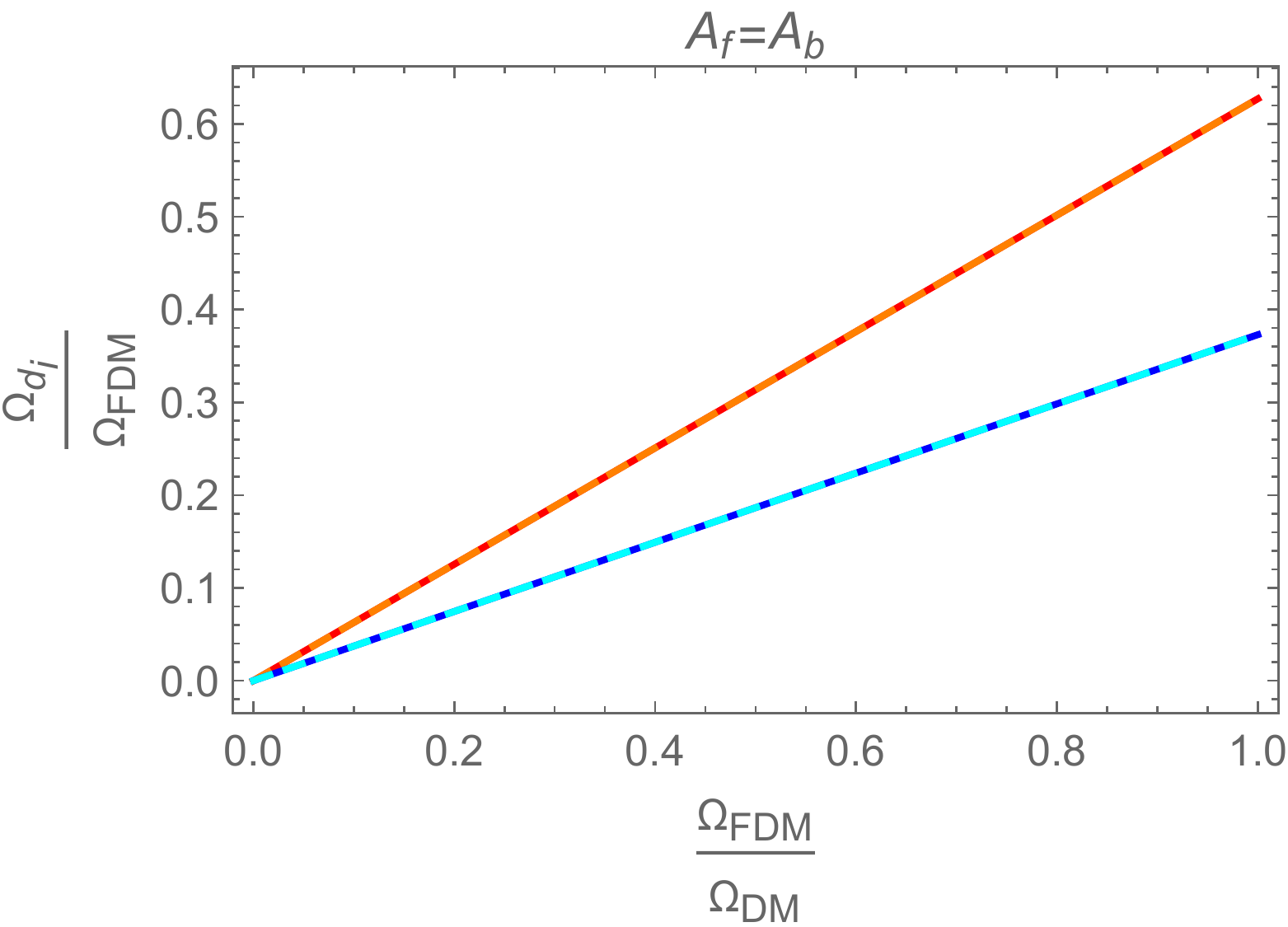}
			\vspace{5pt}
			\includegraphics[width=0.49\textwidth]{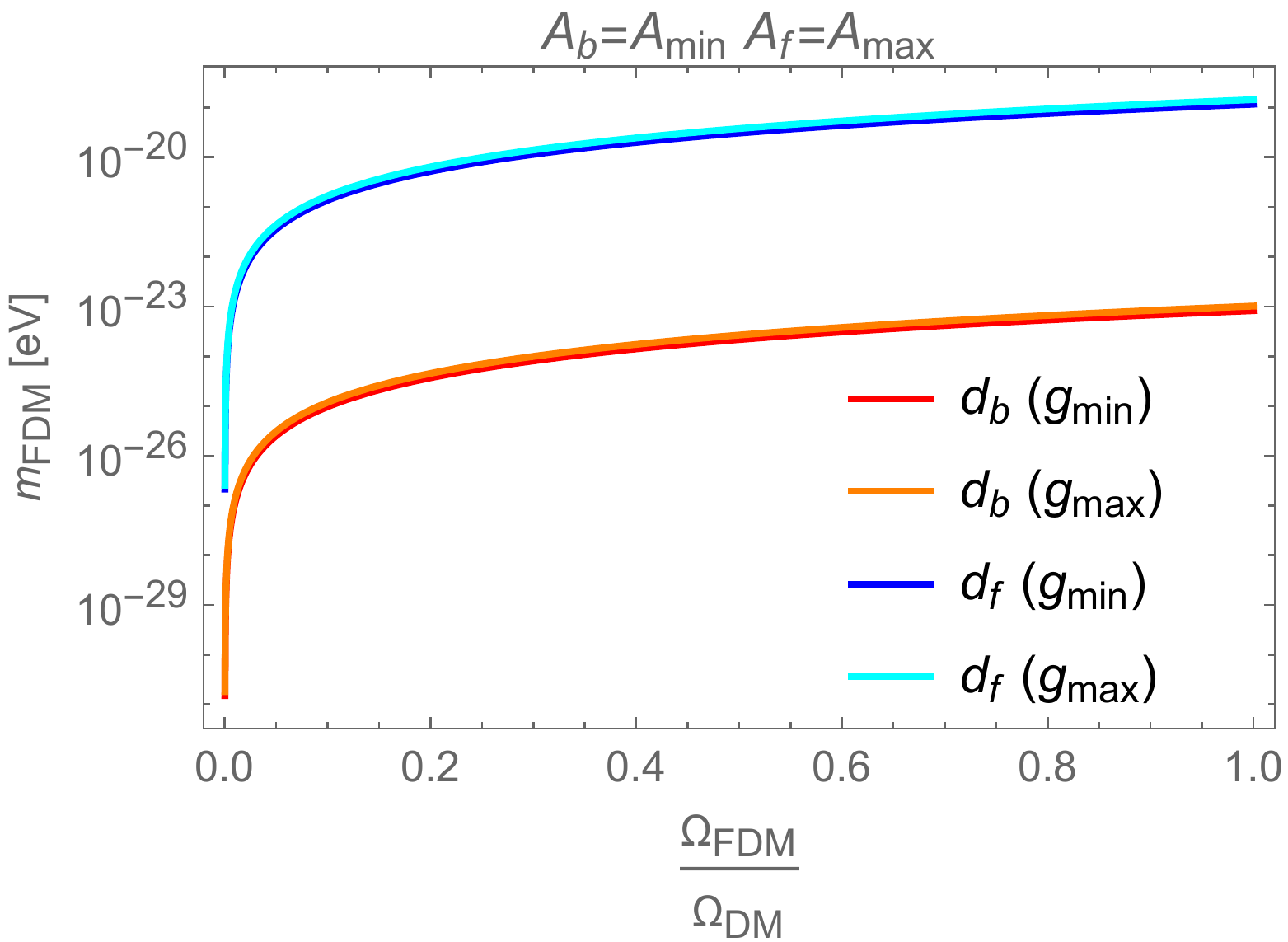}\includegraphics[width=0.49\textwidth]{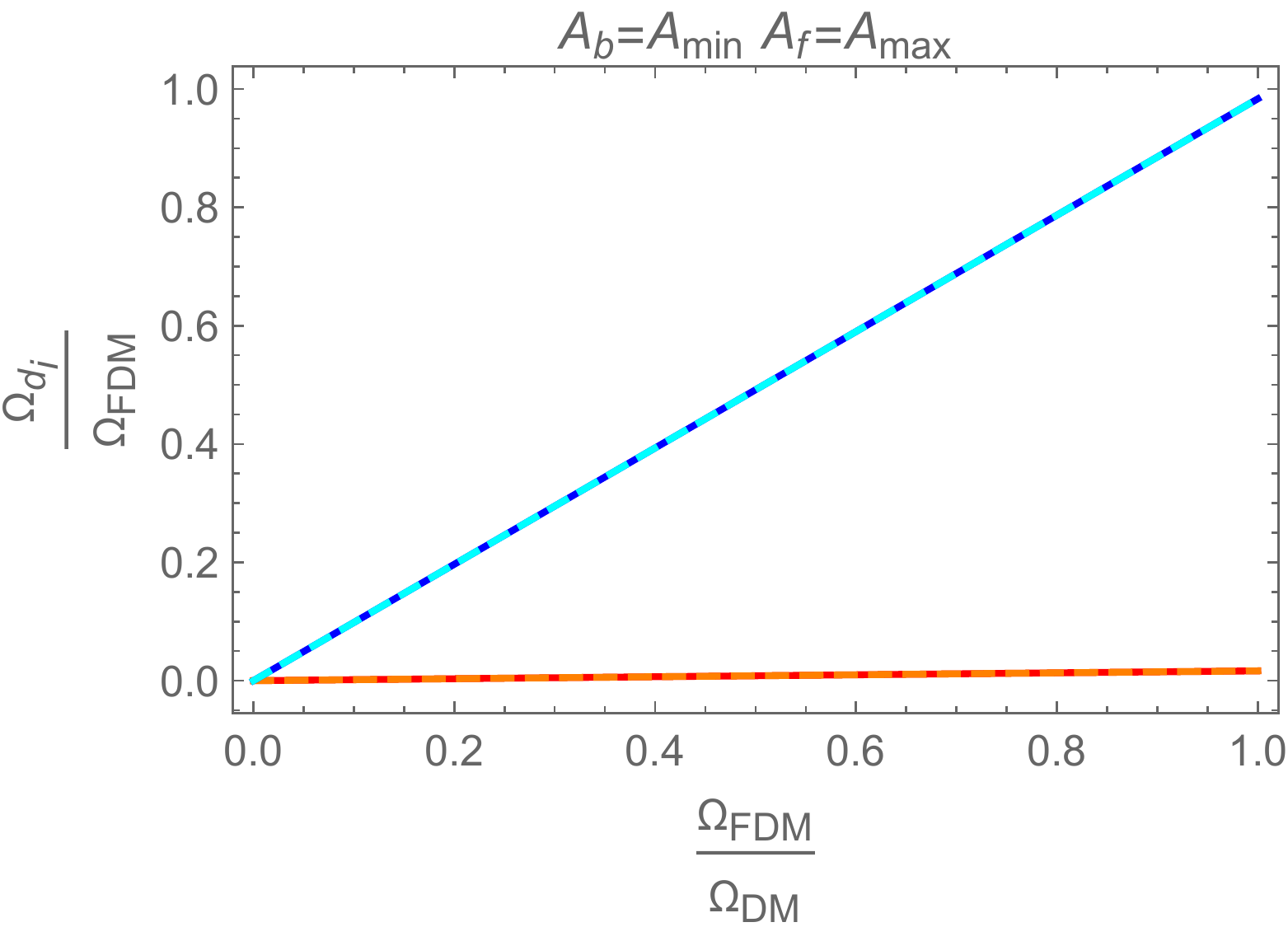}
			\vspace{5pt}
			\includegraphics[width=0.49\textwidth]{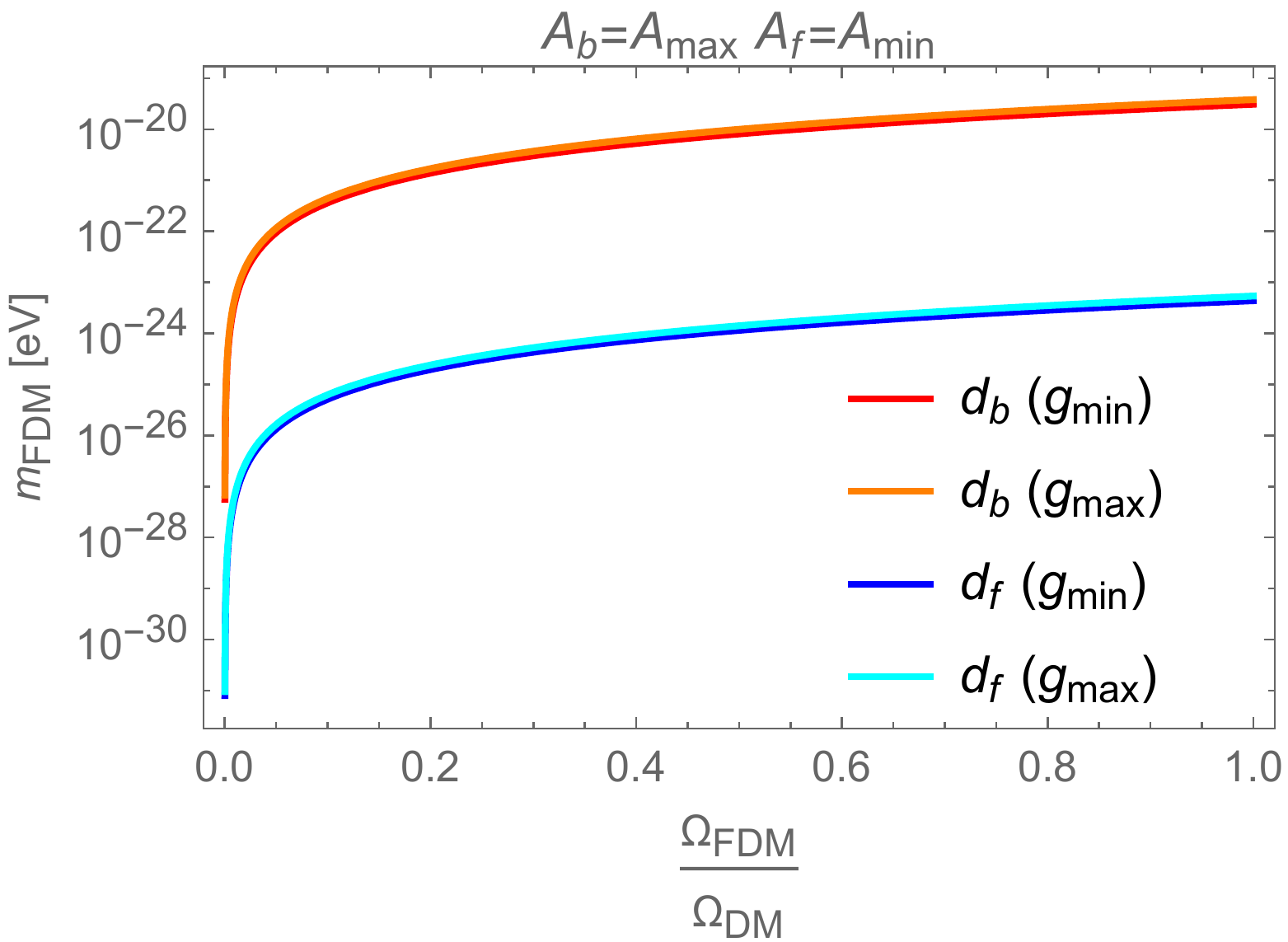}\includegraphics[width=0.49\textwidth]{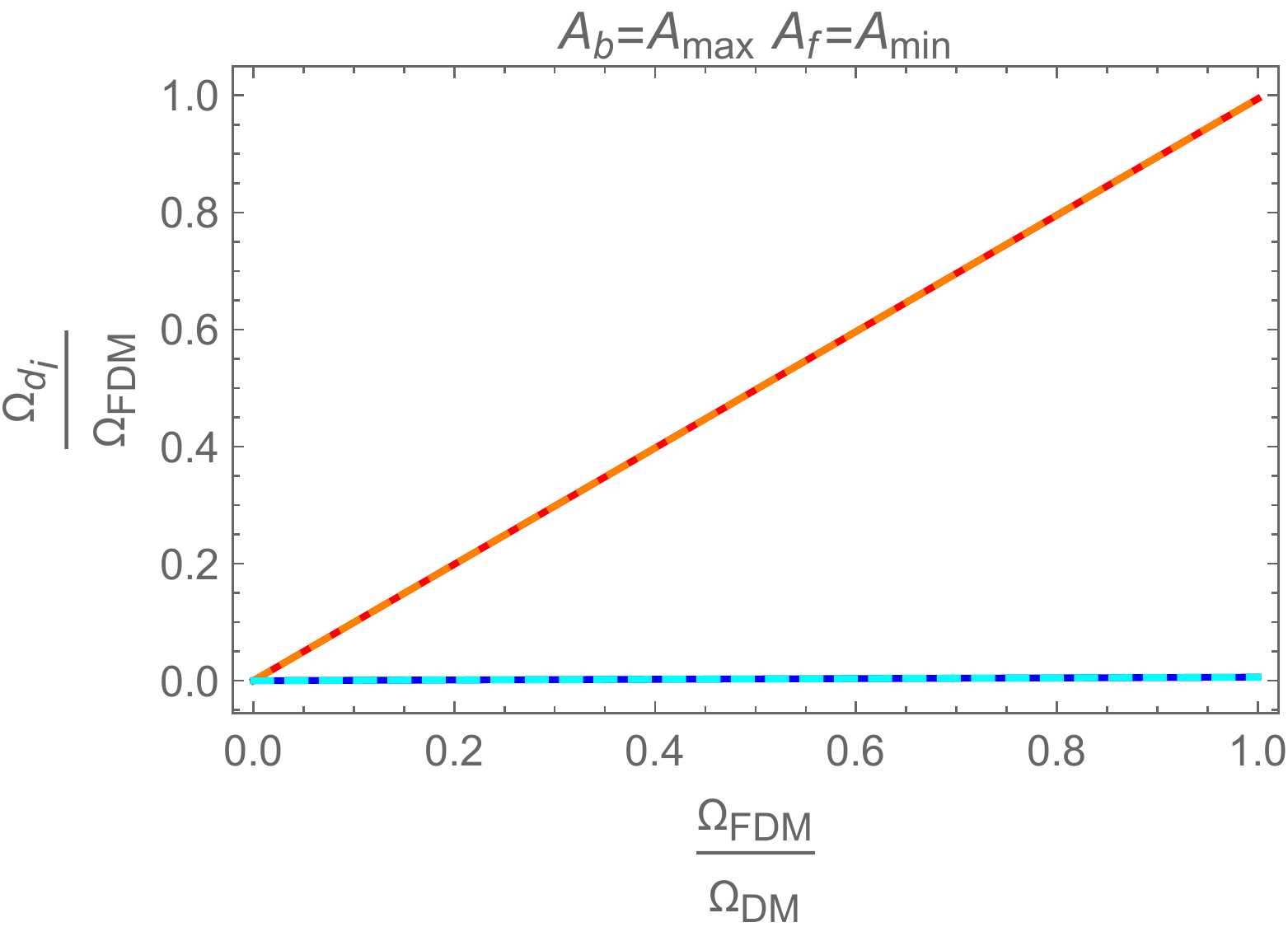}
		\end{center}
		\caption{\label{fig:massisotropic} Left: $d_f$ and $d_b$ axion masses as a function of the total ultralight axion fraction of  dark matter. Right: Relative contributions to $\Omega_{DM}$ coming from $d_f$ and $d_b$ axions. Top panels are referred to equal values of the perturbative corrections prefactors $A_i$, $i=f,b$. Central and bottom panels contain the results related to the cases where  $A_f=A_{max}\gg A_b=A_{min}$ and $A_f=A_{min}\ll A_b=A_{max}$ respectively. The extreme values $A_{min}$ and $A_{max}$ can be read from Table~\ref{tab:all_parameters}.}
	\end{figure}
	
	This result highlights that we can face two opposite scenarios. Isotropic compactification ($q\simeq \sqrt{2}$) implies that the two axions have similar masses and represent similar percentages of DM. On the other hand, given the exponential sensitivity of $\frac{\Omega_{b}}{\Omega_f}$ on the parameter $q$, in anisotropic compactifications ($q\ll 1$ or $q\gg 1$) just one axion can play the r\^ole of the FDM particle. As already mentioned in the previous sections, also in case of nearly isotropic compactifications, the heavier axion will naturally represent the higher fraction of DM. Let us consider for $W_0$, and $A_i$, $i=b,f$, the same parameter range as described in Table~\ref{tab:all_parameters}.
	Moreover, given that the mass range of the two particles will follow the same behaviour as in the Swiss-cheese geometry, we us focus on the case where $N_f=N_b=1$. Also considering the whole parameter space, we can already dramatically restrict the predictions for the allowed decay constants. The results of this analysis are represented in Fig.~\ref{fig:FDMfibrePercentages}. 
	
	From this plot, we can identify the narrow region where we have two suitable FDM candidates. Let us now fix the decay constant ratio $q$ in order to inspect the green central area and understand what will be the composition and the mass of the two axions. The results obtained fixing $q=\sqrt{2}$ are represented in Fig.~\ref{fig:massisotropic}. If we fix $A_b=A_f$, we find two different axions having mass $\sim 10^{-20}$ representing similar percentages of DM. If $A_f$ and $A_b$ get different values, one of the axions becomes much lighter than $ 10^{-22}$ eV representing a negligible fraction of DM.

	\begin{table}[t!]
		\centering
		\begin{tabular}{l|lc}
			&   $\vo$  \\
			\hline
			$q=0.01$     &   $(2.4\div 3.6)\cdot 10^4$  \\
			$q=0.1$      &   $(2.5\div 3.8)\cdot 10^3$ \\
			$q=\sqrt{2}$ &   $(1.9\div 2.9)\cdot 10^2$ \\
			$q=10$       &   $(4.9\div 7.3)\cdot 10^2 $ \\
			$q=100$      &   $(1.5\div 2.3)\cdot 10^3$ \\
			\hline
		\end{tabular}
		\caption{Overall volume of the extra-dimensions for different values of $q$. The range of $q$ values was chosen so that both $\tau_b$ and $\tau_f$ get stabilised at values $\gg 1$, in accordance with instanton expansion and EFT prescriptions.}
		\label{tab:qvalues}
	\end{table}

	\begin{figure}
		\begin{center}
			\includegraphics[width=0.7\textwidth]{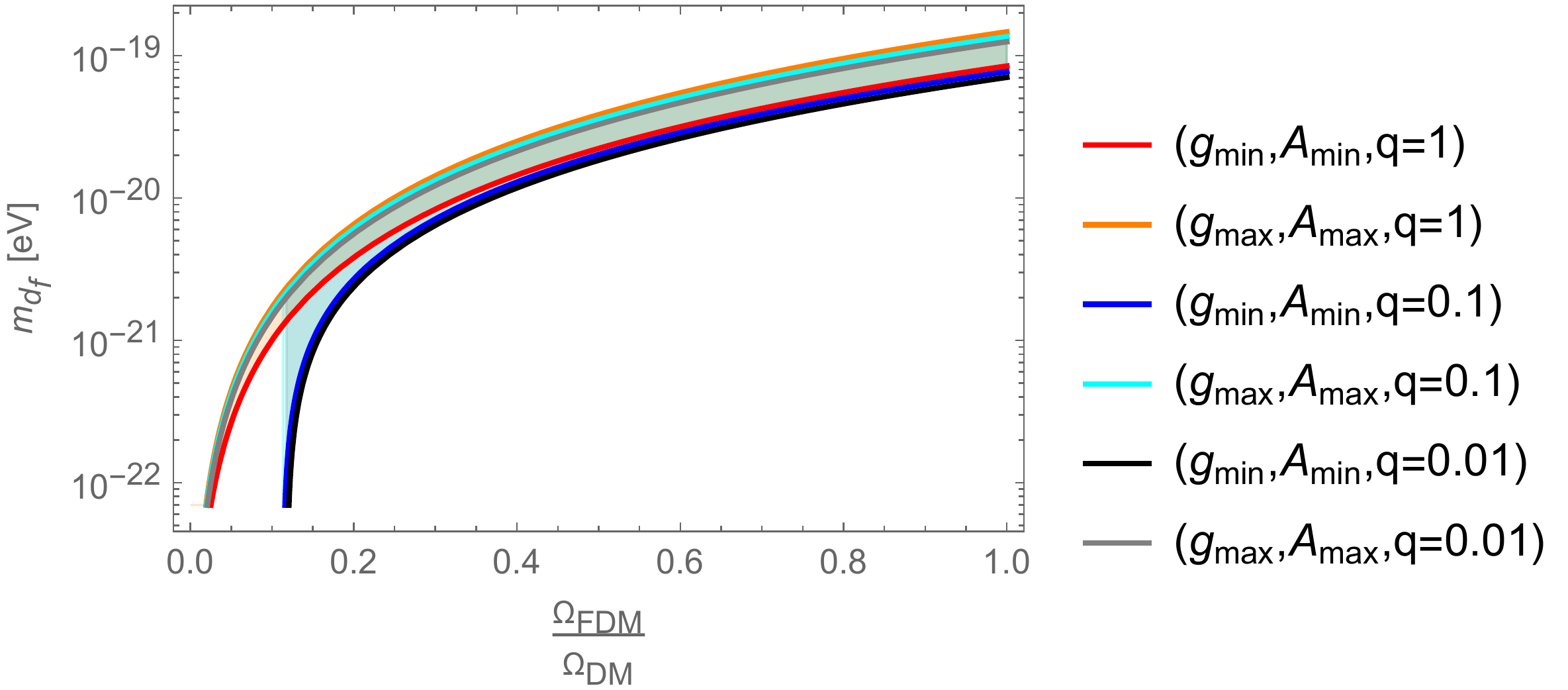}\vspace{5pt}
			\includegraphics[width=0.7\textwidth]{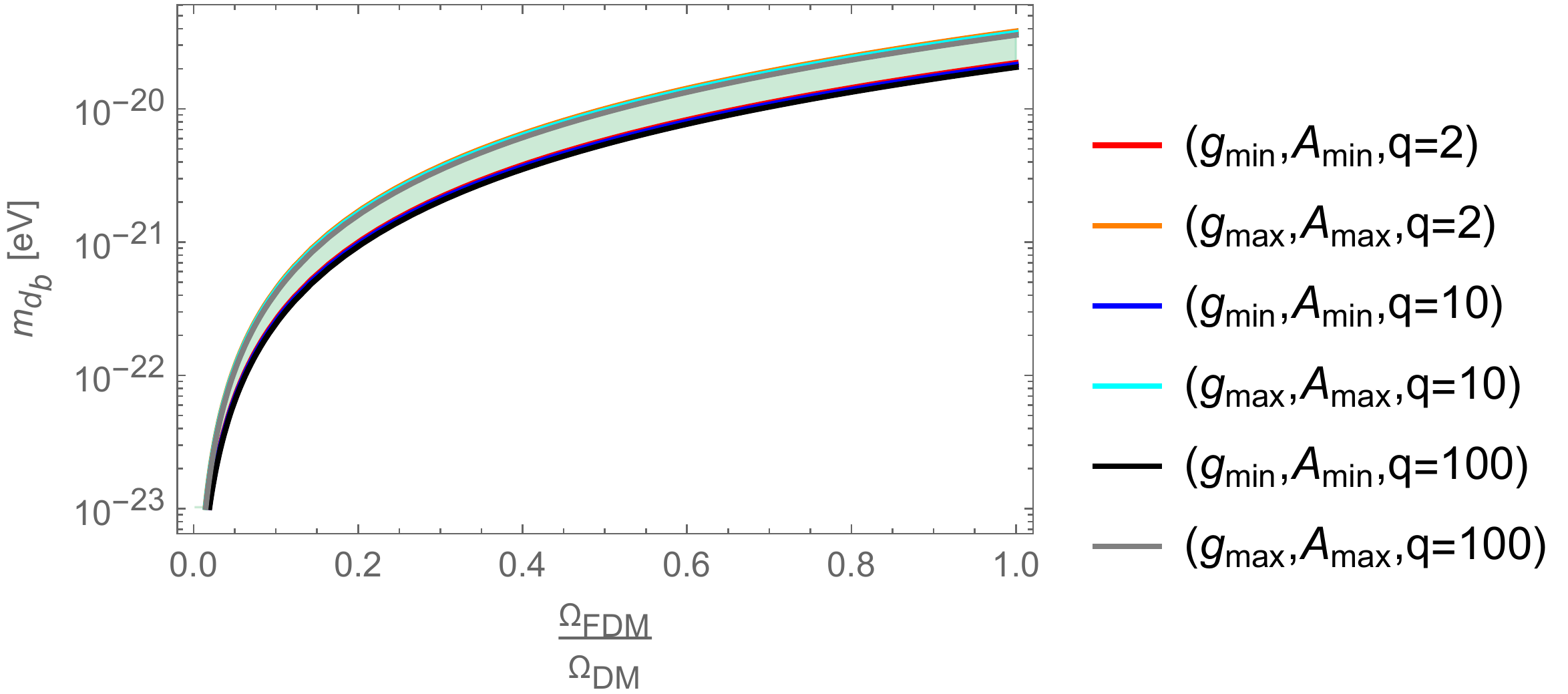}
		\end{center}
		\caption{\label{fig:massanisotropic}We show $m_{d_b}$ and $m_{d_f}$ masses as a function of the axion DM fraction varying the ratio between the decay constants $q=f_b/f_f$. Top: When $q\leq 1$ the contributions coming from $d_b$ are negligible and its mass becomes $\ll 10^{-40}$ eV. Bottom: When $q\geq 2$ the contributions coming from $d_f$ are negligible and its mass becomes $\ll 10^{-40}$ eV. The maximum and minimum values of the parameters used to compute the allowed mass range can be found in Table \ref{tab:all_parameters}. In these plots, we assumed for simplicity that $A_f=A_b$ given that, considering small and large $q$ values, a relative variation of these parameters does not have any impact on the predictions.}
	\end{figure}
	
	We now consider the effect coming from an anisotropic compactification. Also in this case, the predictions for the mass of the two candidates are quite robust. Indeed, as we show in Fig.~\ref{fig:massanisotropic}, choosing different values of $q=\{0.01,0.1,1,2,10,100\}$ the results about the mass and DM fraction do not change. For large values of $q$, $d_b$ is the FDM axion which represents a significant fraction of DM when it acquires a mass $m\gtrsim 10^{-20}$ eV, while $d_f$ is much lighter ($\ll 10^{-44}$ eV) and has a negligible impact on DM abundance. On the other hand, when $q$ is fixed to small values, $d_f$ is the right FDM candidate representing a large amount of DM when its mass is given by $m\sim 10^{-19}$ eV, while the contribution coming from $d_b$ is negligible. The predictions for the overall volume $\vo$ for different values of $q$ and varying parameters are listed in Table~\ref{tab:qvalues}.
	
	For what concerns the relevant energy scales of the model, i.e. KK masses, Eq.~\eqref{eq:KKmassC4} and the gravitino mass, the results found for the Swiss-cheese geometry are still valid in presence of CY fibrations in the isotropic compactification limit. Anisotropic compactifications may lead to different results depending on the overall volume considered. The ratio between the KK masses related to $\tau_f$ and $\tau_b$ 4-cycles scales as $M_{KK}^{(b)}/M_{KK}^{(f)}\sim q^{1/4}$. In this setup, the inflationary scale and the tensor-to-scalar ratio are suppressed compared to the Swiss-cheese geometry. For both isotropic and anisotropic compactifications and for any value of initial misalignment angles, we have that the inflationary scale $H_I< 10^{11}$ GeV and the tensor-to-scalar ratio $r<10^{-7}$. Further details can be found in Appendix~\ref{sec:anharm}.

	\subsection{LVS: FDM from $C_2$ axions}
	
	Our discussion at the beginning of Sec.~\ref{sec:csALPs} made it clear that it is the $C_2$-axions which can lay claim to be the arguably best axion candidates of the type IIB O3/O7 orientifold closed string axion sector. This is so because their shift symmetry remains protected even under orientifolding, and they acquire a potential from non-perturbative effects less easily than $C_4$ axions, as we now summarise (see e.g.~\cite{Gao:2013pra,Cicoli:2021tzt}).
	In the absence of brane~\cite{McAllister:2008hb} or flux monodromy~\cite{Kaloper:2008fb,Dong:2010in,Kaloper:2011jz}, scalar potentials for $C_2$ axions arise either via ED1-brane instantons, via bound states of ED3/ED1-brane instantons or via gaugino condensation on stacks of 4-cycle wrapping D7-branes with gauge flux.
	\begin{itemize}
	    \item{The $C_2$ axion potential can be generated by ED1 branes wrapped on 2-cycles. Such effects induce non-perturbative contributions to the metric of R-R two-forms axions themselves, but cannot contribute to the superpotential in our setup~\cite{Grimm:2007hs, McAllister:2008hb}. In the following, we will use that both KKLT and LVS can be arranged to stabilize the $B_2$ axion at vanishing VEV at high mass scale. In this case, K\"ahler potential corrections scale like $e^{-2\pi t_+/\sqrt{g_s}}$ where $t_+$ represents the Einstein frame volume of the orientifold-even 2-cycle, $\Sigma_2^+$ wrapped by the ED1 brane. These are easily suppressed by considering modest $t_+$ volumes potentially giving rise to light $C_2$ fields. }
		\item{The structure of the ED3/ED1-bound state instanton contribution to the superpotential is given by a modular theta function. For a large enough real argument, this becomes exponentially damped. In our cases, the total scalar potential results in stabilising $b=0$ so no extra damping from a finite $b$-VEV arises in the exponential in $W$. The suppression of the $C_2$-cosine potential comes from $e^{-T}$ dependence of the ED3-parent instanton. Hence in total, if you have an ED3 that has a dissolved ED1 \cite{Grimm:2007hs} this gives a non-perturbative correction to $W$ like $e^{-T -G}$. Formally, the $G$-dependence of the ED1 dissolved inside the ED3 arises as an ED3 magnetised by 2-form gauge flux threading 2-cycles in the ED3-wrapped 4-cycles. As the ED3-brane itself is a purely Euclidean instanton effect, the path integral enforces summation over the unmagnetised ED3 and all magnetised ED3/ED1-bound states, mandating the appearance of the $G$-dependence in $W$ for ED3-contributions on 4-cycles intersecting with orientifold-odd 2-cycle combinations.} 
		\item{If you use instead a D7-brane stack to stabilise the $T$ moduli, magnetisation of the D7-brane stack is a choice of compactification data (no path integral forces you to sum over magnetised D7-brane states, since unlike a purely euclidean instanton the full D7-brane fills 4D space--time as well). Thus, by avoiding putting gauge fluxes on the D7-branes you prevent single-suppressed $e^{-T -G}$ terms in $W$ from arising~\cite{Long:2014dta,Jockers:2004yj,Jockers:2005pn,Jockers:2005zy,Grimm:2011dj}. However, the path integral will generate contributions from ED3/ED1-bound state instantons to the gauge kinetic function. Such a correction to the gauge kinetic function of the 7-brane stack scaling like $e^{-T -G}$ in turn induces a superpotential correction of order $e^{-2T-G}$~\cite{McAllister:2008hb}. Compared to the scale of the superpotential terms $e^{-T}$ stabilising the $T$ moduli, this leads to a double suppression of the potential for the $C_2$ axion.}
	\end{itemize}
	
	We shall now summarise the scaling of the scalar potential for the $C_2$ axion arising from these non-perturbative effects in the concrete scenarios of KKLT and LVS stabilisation of the volume moduli on Swiss-cheese CY orientifolds with two volume moduli.
	\begin{itemize}
		\item{We first look at KKLT: if a harmonic zero-mode $C_2$ axion counted by $h^{1,1}_-$ acquires a single suppressed non-perturbative scalar potential from ED3/ED1-bound state instantons, then in KKLT it is too heavy to form FDM. Even if its potential comes from the double suppressed contribution of an unmagnetised 7-brane stack, the $C_2$ axion remains too heavy to constitute FDM. The reason is that in KKLT the lowest volume moduli masses $\sim e^{-T}$ are always around the gravitino mass scale. Since this in turn is bounded from below by ${\cal O}({\rm TeV})$ the resulting $C_2$ mass scale $\sim e^{-2T}$ is still too heavy. 
		On the other hand, if this axion receives a mass through pure ED1 contributions appearing in the K\"ahler potential, it may represent a good FDM candidate. Indeed, in this setup, the $C_2$ axion can become much lighter than the $C_4$ one and its mass would scale as $e^{-2T-\sqrt{T/g_s}}$.}
		\item{In LVS we should always have a CY manifold with a volume form such that it has at least two volume moduli appearing in the Swiss-cheese form. For a $C_2$ axion we can now consider intersection couplings with either the small LVS blow-up or the CY volume-carrying big cycle. We begin by looking at the case of $C_2$ intersecting with the small cycle. If you have a double suppressed term in $W$ from an unmagnetised 7-brane stack on a small cycle, the term in the exponent would scale as $2 (2\pi/N) \tau_s$. For $N=2$ this scales like a single ED3/ED1-bound state instanton wrapping the small cycle. Moreover, in this case $N=2$ is the most favourable setup for a potential FDM role as the volume needs to be $\mc{V}\sim e^{100}$, while it would be even large for $N>2$ driving the EFT out of the controlled regime. Also the case of ED1 branes wrapped around a blow-up cycle does not lead to good FDM candidates. Indeed, being $\tau_s\sim g_s^{-1}$, K\"ahler potential corrections scale like $ e^{-2\pi/g_s}$ and matching the right mass value requires $\mc{V}\gtrsim 10^{20}$ making the axion decay constant, $f\sim M_P/\sqrt{\mc{V}}\sim 10^8$ GeV, way too small.
		Hence, $C_2$-FDM cannot arise in LVS from the LVS blow-up cycle or similarly small blow-up cycles.}
		\item{Conversely, in LVS a D7-brane stack wrapped around the large volume cycle induces a double suppressed mass term that would imply either a too light axion or volume too small for control of the $\alpha'$-expansion.
			
		What thus remain are the cases of an ED3/ED1-bound state instanton in LVS or an ED1 instanton wrapping the volume cycle in both KKLT and LVS. In the first case, the resulting single-suppressed cosine potential for $C_2$ on the big cycle leads to a borderline situation and the relation between the $C_4$ and $C_2$ masses, the decay constants and the FDM abundances requires further investigation. Also the second case of a pure ED1 instanton may lead to interesting results as the K\"ahler potential corrections scaling as $e^{-\mc{V}^{1/3}/\sqrt{g_s}}$ can give rise to sufficiently light $C_2$ fields for both stabilisation prescriptions under study.}
	\end{itemize}   
	In what follows we consider the simple setup where we have a single orientifold-odd modulus $G$, the extradimensional geometry is Swiss-cheese and there is are non-vanishing intersection number between the pair of 2-cycles projected by the O7-action onto $t_+$ and a harmonic $C_2$ axion, and the large volume 4-cycle. While extensions to multiple odd moduli lead so similar results, moving towards more complex geometries is highly non-trivial. Given that we are only interested in the overall scaling of the mass and the decay constant, we leave this analysis for future work. We separately study the cases where the $C_2$ axion gets a mass from pure ED1 (in $\mathcal{K}$) or ED3/ED1 instanton effects (in $W$). In both cases, the K\"ahler potential has the following form:
	\be 
	\label{eq:KC2}
	\mathcal{K}=-2\ln\left(\mc{V}+\frac{\hat{\xi}}{2}\right) \coma
	\ee
	where $\mc V=\mc V(T_i,G,S)$ is the extra-dimensions volume. 
	
	\paragraph{Pure ED1 effects in LVS} Let us consider the case where the ED1 wraps a 2-cycle $t_b$ parametrizing the overall volume. For simplicity, we assume that the volume dependence on $t_b$ is given by $\mc{V}\supset \kappa_{bbb}\,t_b^3/6$, where $\kappa_{bbb}$ is the big cycle self-intersection number.\footnote{Similar results hold for more complex intersection polynomials once we go to the LVS limit, where one of the 2-cycles dominates over the other ones.} If this condition is satisfied, the 2-cycle volume can be written as $t_b=\sqrt{(T_b+\bar T_b)/\kappa_{bbb}}$. In the simplest Swiss-cheese setup, the CY volume is given by~\cite{McAllister:2008hb}:
    \be
    \label{eq:VC2ED1}
    \mc{V}/\alpha\simeq \left[T_b+\bar{T}_b-\kappa_{b}(G - \bar{G})^2 +C e^{-\frac{2\pi}{\sqrt{g_s}}\sqrt{\frac{(T_b+\bar T_b)}{\kappa_{bbb}}}} \mbox{Re}\left[ e^{i\pi G}\right]\right]^{3/2}- (T_s+\bar{T}_s)^{3/2}\coma
	\ee
	where $\alpha=\frac{1}{3}\sqrt{\frac{2}{\kappa_{bbb}}}$, $\kappa_b=\kappa_{b--}g_s/4$ being $\kappa_{b--}$ the intersection number of the big divisor with the odd cycle and $\mbox{Re}[e^{iG}]=e^{i\pi(G-\bar G)}\cos\left[\pi (G+\bar G )\right]$. We further assume the small blow-up 4-cycle $\Sigma_s$ to be wrapped by an ED3 instanton or a small D7-brane stack generating the following non-perturbative correction to the superpotential:
    \be
    \label{eq:WC2ED1}
        W=W_0 + A_s e^{-a_s T_s}\fstop
    \ee
       \begin{figure}[t!]
		\begin{center}
			\includegraphics[width=0.47\textwidth]{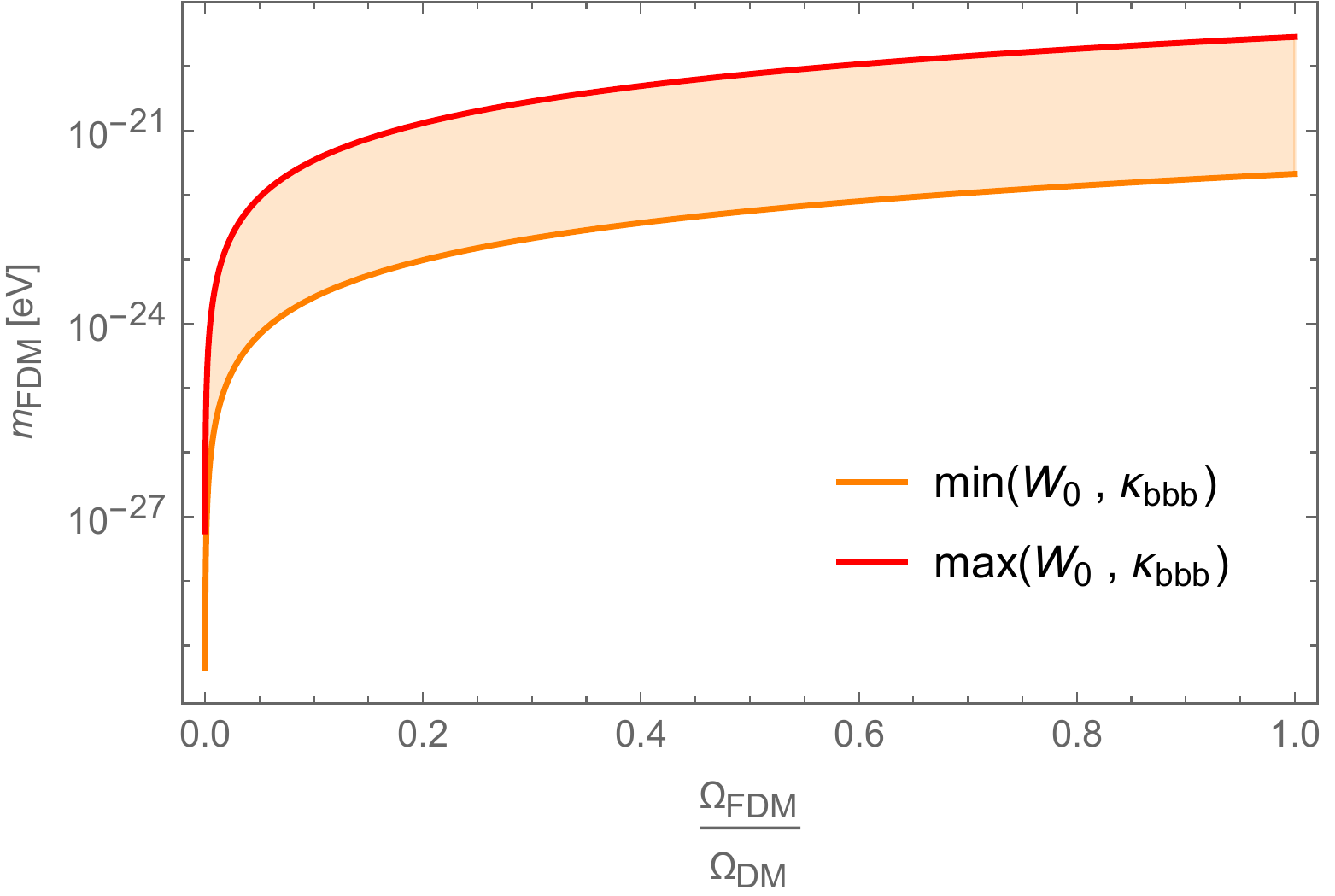}\hspace{6pt}\includegraphics[width=0.49\textwidth]{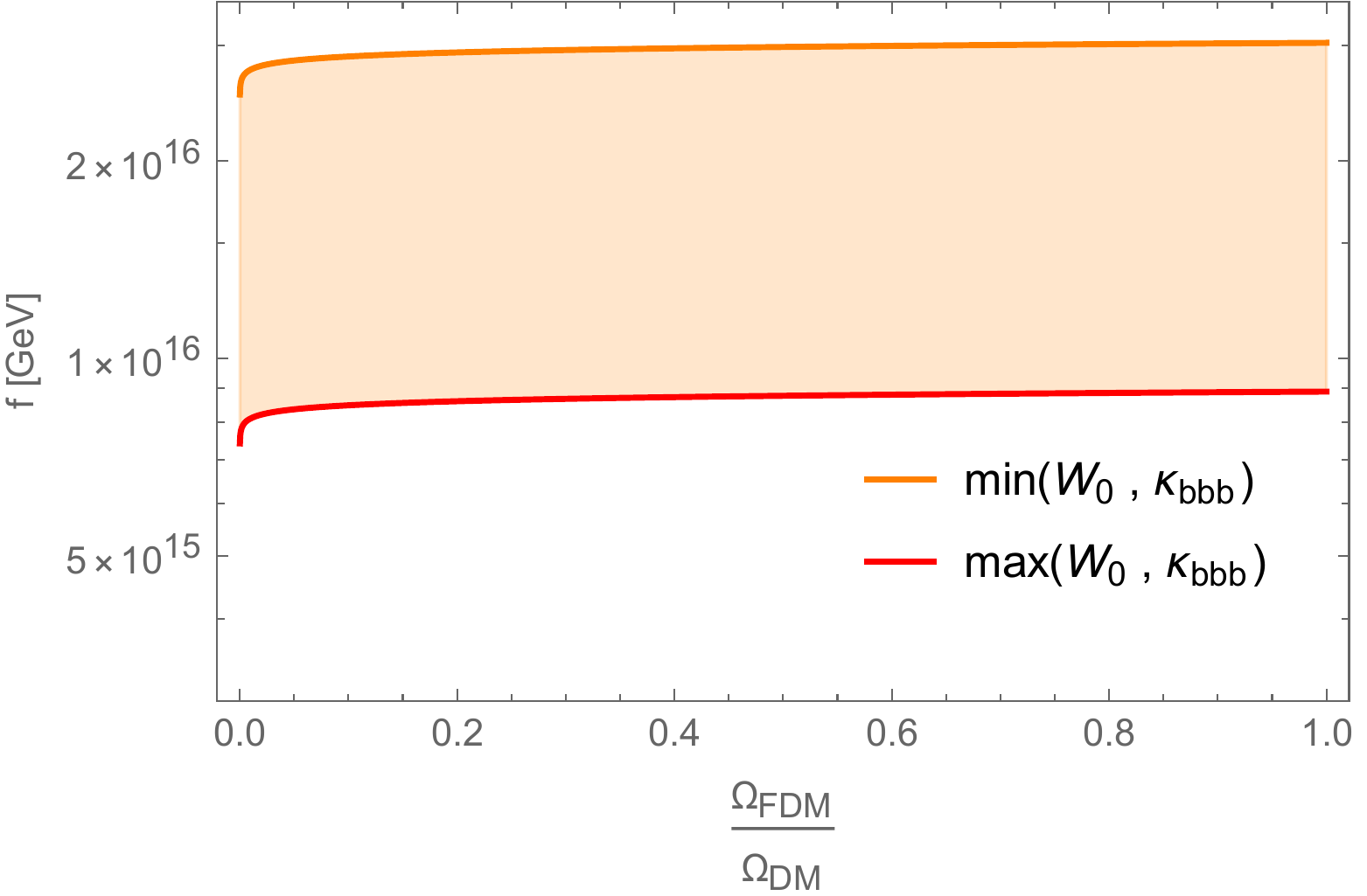}
			
			\includegraphics[width=0.47\textwidth]{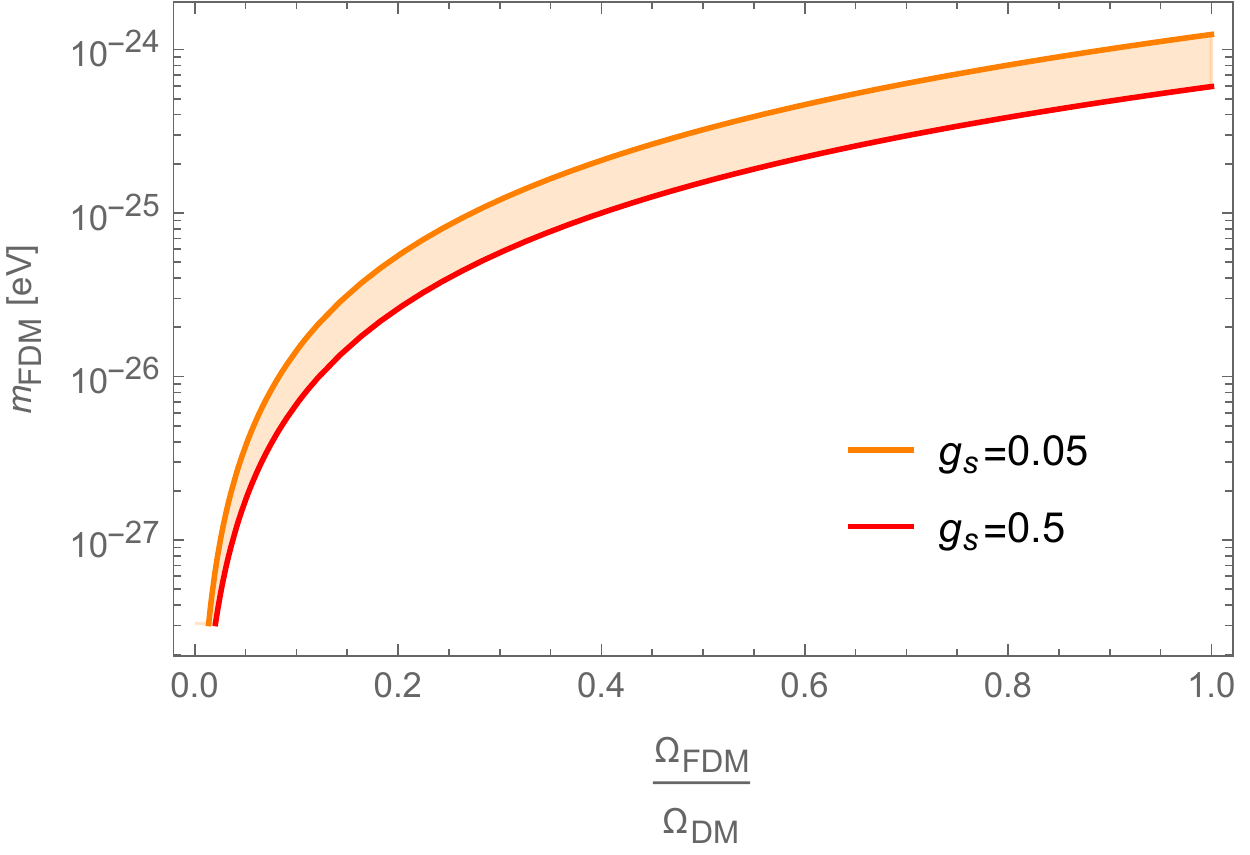}\hspace{6pt}\includegraphics[width=0.49\textwidth]{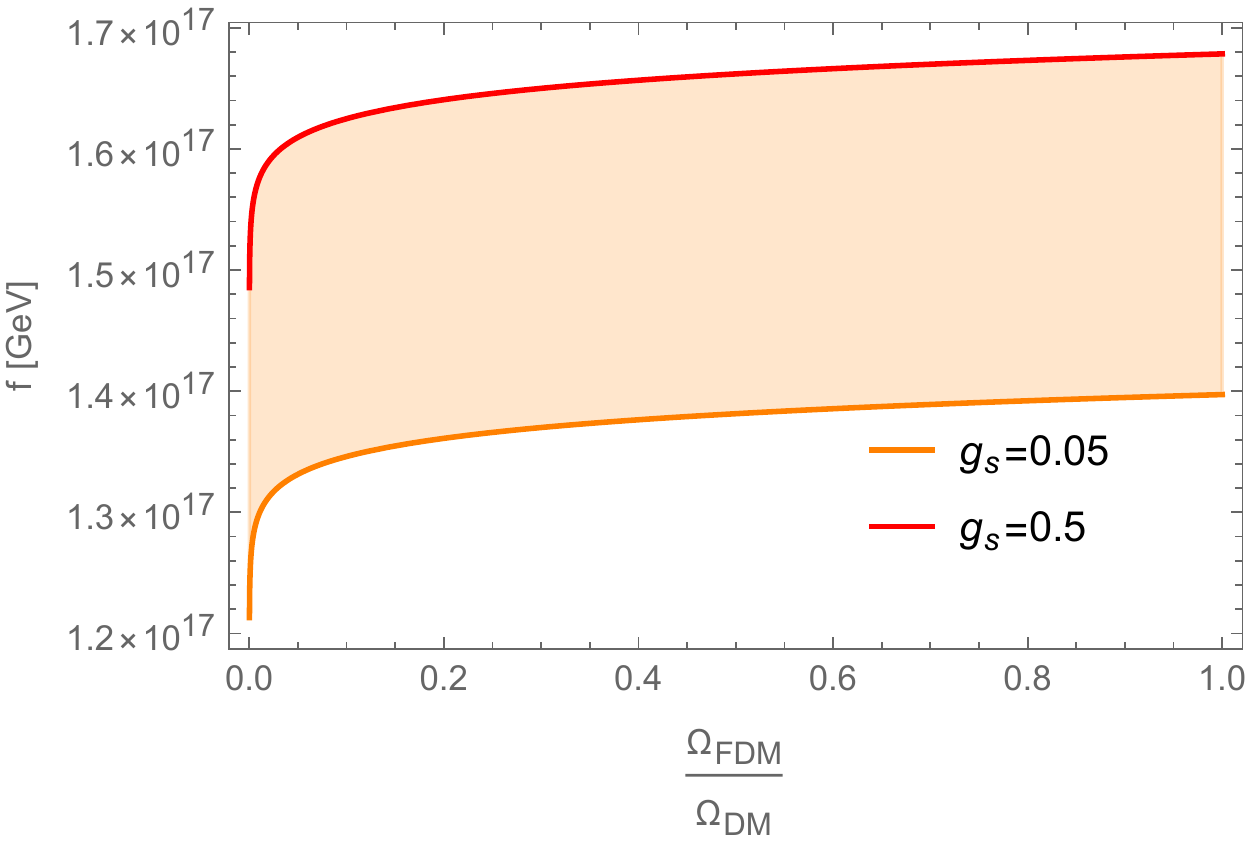}
			
			\caption{Predictions for $C_2$ axion mass (left) and decay constant (right) in LVS (top) and KKLT (bottom) scenarios. The axion receives a mass through ED1 instanton effects coming from an ED1 brane wrapping the volume 2-cycle. }\label{Fig:C2ED1} 
		\end{center}
	\end{figure} 
    Assuming stabilisation of the $B_2$ axion at $\langle b \rangle=0$, the $C_2$ axion decay constant is given by~\cite{Grimm:2004uq}:
	\begin{equation}
       2\pi f = M_P\sqrt{\frac{g_s t_b|\kappa_{b--}|}{2\mc{V}}}\coma
    \end{equation}
    where $\kappa_{b--}\leq0$. The action for the ED1 wrapped around $t_b$ is
    \begin{equation}
    \label{eq:SED1}
        S_{ED1}= \frac{2\pi\, t_b}{\sqrt{g_s}}\fstop
    \end{equation}
    Hence the WGC relation becomes
    \begin{equation}
    \label{eq:C2ED1Sf}
        S f =  \sqrt{\frac{t_b^3|\kappa_{b--}|}{2\mc{V}}}  M_P\quad  \xrightarrow[t_b\gg 1]{\text{LVS}}\quad \sqrt{\frac{3|\kappa_{b--}|}{\kappa_{bbb}}}  M_P\coma
    \end{equation}
    where on the right side we took the LVS limit. The scalar potential for $C_2$ arising from the aforementioned corrections to $\mathcal{K}$ and $W$ is suppressed compared to the LVS terms and scales like
    \be 
    \delta V \simeq - \frac{C\, W_0^2 }{g_s^{1/2}\mc{V}^{11/3}}e^{-2 \,\pi \gamma \frac{\mc{V}^{1/3}}{\sqrt{g_s}}}\cos\left(2\pi c\right)\coma
    \ee
    where $\gamma=\frac{3^{1/3}}{2^{1/6}\kappa_{bbb}^{1/3}}$ and we assumed LVS stabilization for $\tau_b$, $\tau_s$ and $d_s$. Finally, the axion mass is given by
    \be 
    m_c^2\simeq\frac{C\, W_0^2 }{|\kappa_{b--}| g_s^{3/2}\mc{V}^{3}}e^{-2 \,\pi \gamma \frac{\mc{V}^{1/3}}{\sqrt{g_s}}}\,M_P^2\fstop
    \ee
    For simplicity from now on we fix $\kappa_{b--}=-1$ and $C=1$ as their tuning does not really affect our final predictions. We let $W_0$ and $\kappa_{bbb}$ vary in $W_0\in[1,10^2]$ and $\kappa_{bbb}\in\{1,\dots,10\}$ while we set $g_s=\ln^{-1}(\mc{V})$ according to LVS prescription. Our results are shown in Fig. \ref{Fig:C2ED1} where we see that the $C_2$ axion coming from an ED1 brane wrapping the volume 2-cycle can actually represent a good FDM candidate. Just as in the previous cases the decay constant is not sensitive to the variation of the microscopical parameters, showing a constant value $f\sim 10^{16}$ GeV. Instead, in this case the variation of the mass is more pronounced. We have in fact that the $C_2$ axion can represent a considerable percentage of DM if it gets a mass $m\sim[7\cdot10^{-22},10^{-19}]$ eV corresponding to volumes of about $\mc{V}\sim [10^{4},10^{5}]$ and string couplings $g_s\sim[0.08,0.1]$.

    \paragraph{Pure ED1 effects in KKLT} Here we consider the simplest case where $h^{1,1}_+=1$. The overall volume, the $C_2$ axion decay constant and the ED1 instanton action coincide with those listed in the previous section if we neglect the blow-up field contributions. The correction to the superpotential is given by:
    \be
        W=W_0 + A_b e^{-a_b T_b}\fstop
    \ee
    Assuming again that $\langle b \rangle=0$, the $C_2$ scalar potential arising from ED1 corrections to $\mathcal{K}$ and $W$ is suppressed compared to the KKLT AdS scale and reads:
    \be 
    \delta V \simeq \frac{C\,a_b^2 A_b^2 g_s }{ \tau_b^2}e^{-2a_b\tau_b-2\pi\sqrt{2 \frac{\tau_b}{g_s \kappa_b}}} \cos\left(2\pi c\right) \fstop
    \ee
    The axion mass is given by:
     \be 
    m_c^2\simeq\frac{C\,a_b^2 A_b^2 }{ \tau_b \kappa_{b--}}e^{-2a_b\tau_b-2\pi\sqrt{2\frac{\tau_b}{g_s \kappa_b}}} M_P^2\fstop
    \ee
    For simplicity from now on we fix $|\kappa_{b--}|=\kappa_{bbb}=1$, $C=A_b=1$, $a_b=0.1$ as their tuning does not significantly affect our final predictions. We let $W_0$ and $g_s$ vary in $W_0\in[10^{-12},10^{-2}]$ and $g_s\in[0.05,0.5]$ while we set $\tau_b=-\ln(W_0)/a_b$ according to the KKLT prescription. Our results are shown in Fig. \ref{Fig:C2ED1} where we see that the $C_2$ axion can be extremely light in the KKLT scenario, actually too light to represent FDM. In this case both the decay constant and the mass are not very sensitive to the variation of the microscopical parameters. The decay constant is $f\sim 10^{17}$ GeV while the mass $m\sim 10^{-24}$ eV. Low mass values correspond to $W_0\sim 10^{-10}$, high values to $W_0\sim 10^{-2}$.

	\paragraph{ED3/ED1 effects} If the axions acquire a mass via ED3/ED1 instanton contributions, the superpotential receives leading order non-perturbative corrections given by	\be
	W=W_0+A_s\,e^{-a_sT_s}+A_b\,e^{-a_bT_b}+C e^{-a_b \left(T_b+i G\right)}\fstop
	\ee
	These corrections tend to make the volume $C_4$ axion and the $C_2$ axion degenerate in mass. After LVS and $b$ axion stabilisation, which we assume to take place at $\langle b \rangle=0$, we are left with two ultralight axion candidates, namely $d_b$ and $c$. The field space metric associated to these fields is diagonal
	\be 
	\mc{L}_{kin}=\frac{1}{2}\left[\frac{3}{2\tau_b^2} (\partial d_b)^2 -\frac{6 \kappa_b}{\tau_b} (\partial c)^2\right]\coma
	\ee 
	while their scalar potential is given by
	\be 
	V_F\supset \frac{a_b \kappa W_0}{2\tau_b^2}e^{-a_b\tau_b}\left(A_b \cos(a_bd_b) +C \cos\left[a_b\left(d_b+c\right)\right]\right) \fstop
	\ee
	In terms of the canonically normalised fields it becomes  
	\be
	\label{eq:C2pot} 
	V_F\supset \frac{a_b \kappa W_0}{2\tau_b^2}e^{-a_b\tau_b}\left(A_b \cos\left(\frac{\hat{d}_b}{g_{d_b}}\right) +C \cos\left[\left(\frac{\hat{d}_b}{g_{d_b}}+\frac{\hat{c}}{g_{c}}\right)\right]\right)\coma
	\ee
	where
	\be
	g_{d_b}=\sqrt{\frac{3}{2}} \frac{1}{a_b\tau_b}\,, \qquad  g_c=  \frac{1}{a_b}\sqrt{\frac{6 |\kappa_b|}{\tau_b}}\fstop
	\ee
		\begin{figure}
		\begin{center}
			\includegraphics[width=0.49\textwidth]{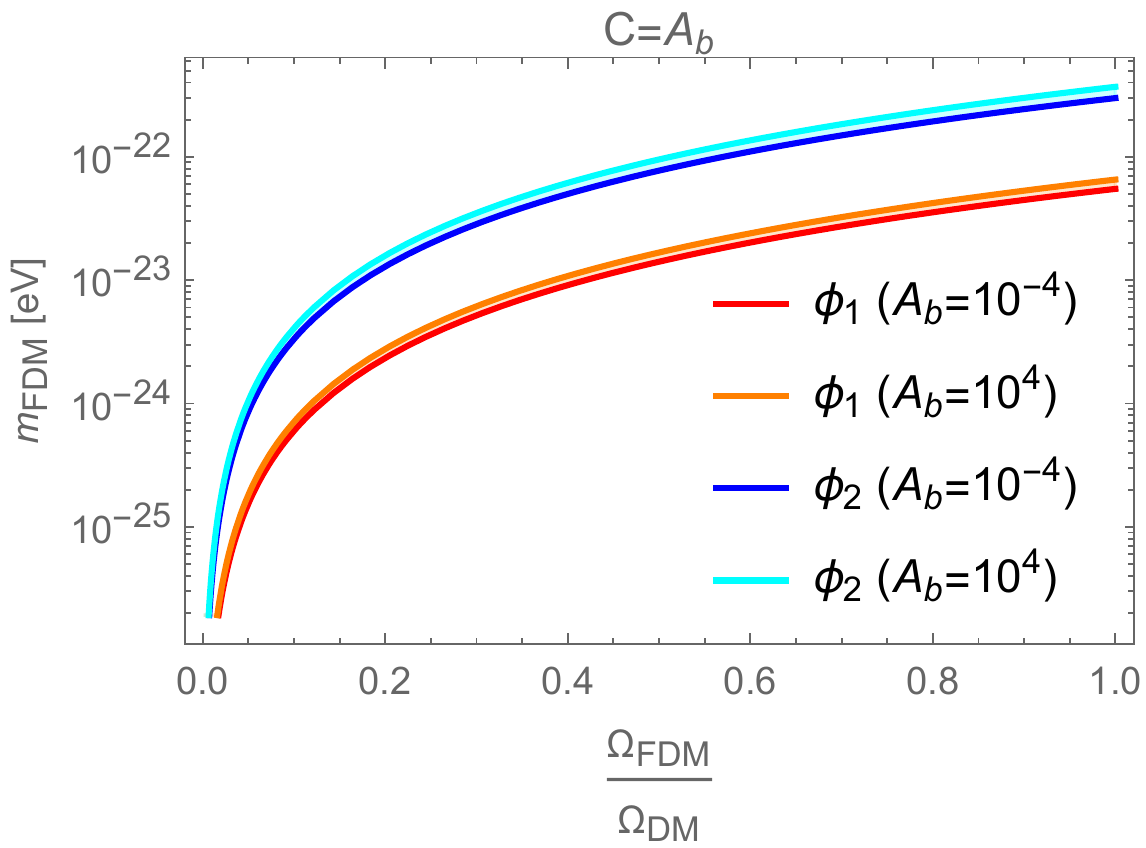}
			\includegraphics[width=0.49\textwidth]{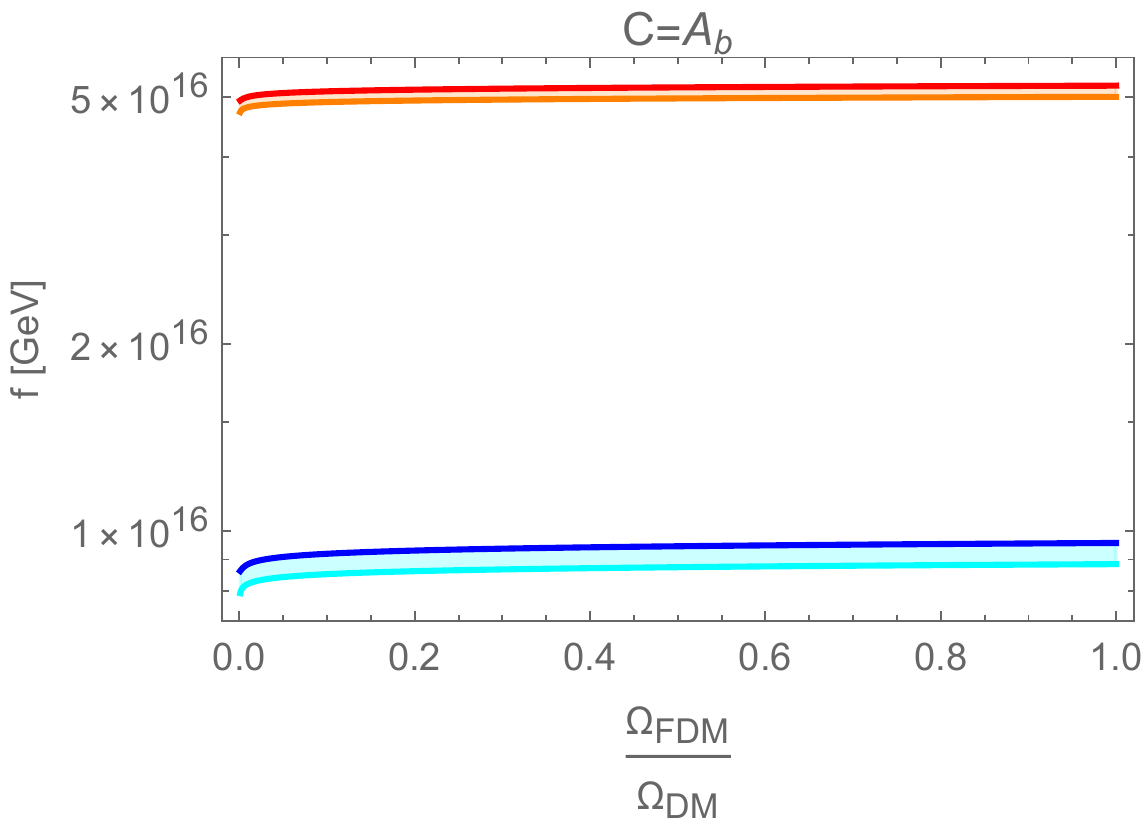}\\[5pt]
			\includegraphics[width=0.49\textwidth]{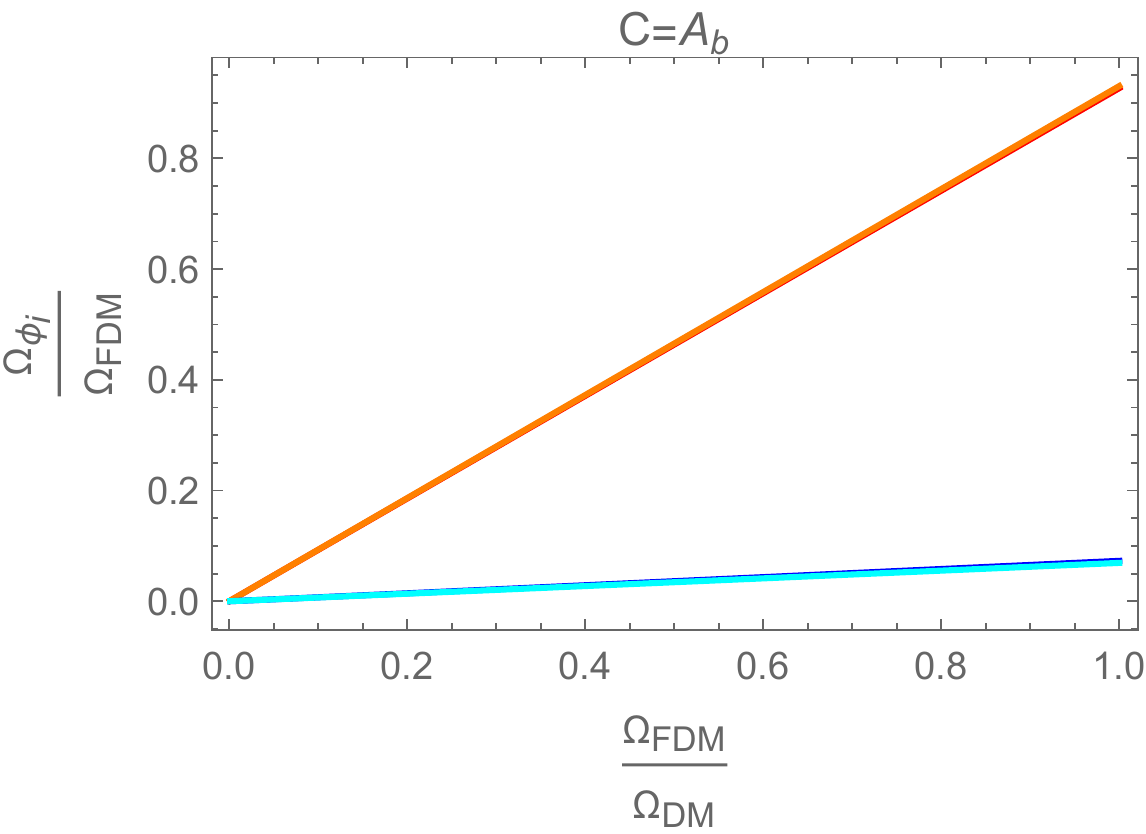}
			\includegraphics[width=0.49\textwidth]{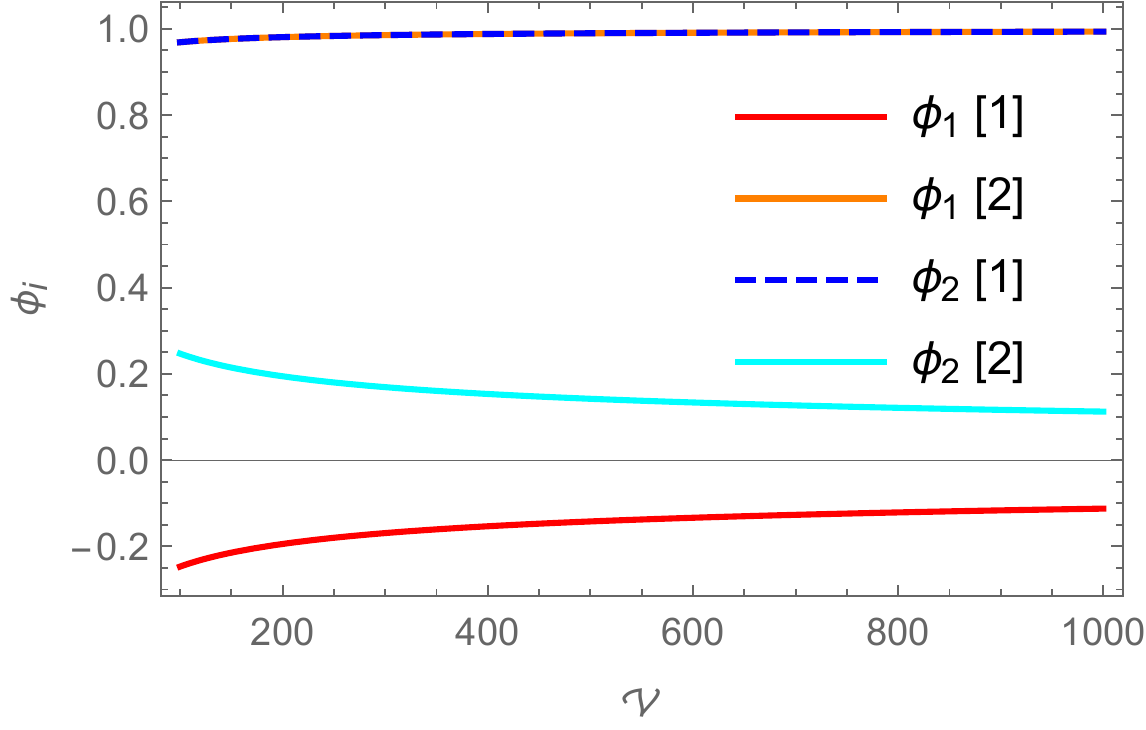}

		\end{center}
		\caption{\label{fig:massC2}Results about $C_2$ and $C_4$ axionic FDM from ED3/ED1-bound state instanton effects wrapping the overall volume cycle in the Swiss-cheese geometry. The results are given in terms of the mass matrix eigenvectors $\phi_1$ and $\phi_2$, Eq. (\ref{eq:evec}). Top: axion masses (left) and decay constants (right) as a function of the total ultralight axionic dark matter fraction. Bottom-left: relative abundance of the axionic DM particles. Bottom-right: eigenvectors components as a function of the overall volume $\mc{V}$. We see that also at small volumes the eigenvectors of the mass matrix $\phi_1$ and $\phi_2$ are mainly given by the $C_2$ and $C_4$ axion respectively.}
	\end{figure}
	As we will show below, we cannot identify $g_{d_b}$ and $g_c$ with the decay constants, as the physical fields are given by the mass matrix eigenvectors that may not be aligned with $d_b$ and $c$. Let us consider for simplicity the case where $A_b=C$. The minimum of the scalar potential is given by
	\be 
	\frac{\hat{d}_b}{g_{d_b}}=(2k+1)\pi \,,\qquad \frac{\hat{c}}{g_c}=2m\pi\,, \qquad m,k \in \Bbb Z
	\ee
	so that the mass matrix in a neighbourhood of the minimum becomes 
	\be 
	M=\Lambda \bar{M}=\Lambda \begin{pmatrix}
		\frac{2}{g_{d_b}^2} & \frac{1}{g_{d_b}g_c}\\
		\frac{1}{g_{d_b}g_c}    & \frac{1}{g_c^2}\\
	\end{pmatrix} \quad \mbox{where}\quad \Lambda= \frac{a_b A_b \kappa W_0}{2\tau_b^2}e^{-a_b\tau_b}\fstop
	\ee
	The eigenvalues, $\lambda_{i}$, and eigenvectors, $\phi_{i}$, of $\bar{M}$ are 
	\be
	\lambda_1=\frac{g_{d_b}^2+2 g_c^2-\sqrt{g_{d_b}^4+4g_c^4}}{2g_{d_b}^2 g_c^2}\coma \quad \lambda_2=\frac{g_{d_b}^2+2 g_c^2+\sqrt{g_{d_b}^4+4g_c^4}}{2g_{d_b}^2 g_c^2}
	\ee
	\be 
	\label{eq:evec}
	\phi_1=\frac{1}{|g_{-}|}\begin{pmatrix}
		\frac{2 g_c^2-g_{d_b}^2-\sqrt{g_c^4+4g_{d_b}^4}}{2g_c g_{d_b}}\\
		1\\
	\end{pmatrix} \coma \quad 
	\phi_2= \frac{1}{|g_{+}|}\begin{pmatrix}
		\frac{2 g_c^2-g_{d_b}^2+\sqrt{g_c^4+4g_{d_b}^4}}{2g_c g_{d_b}}\\
		1\\
	\end{pmatrix}\,
	\ee
	where $|g_{\pm}|=1+\left(\frac{2 g_c^2-g_{d_b}^2\pm\sqrt{g_c^4+4g_{d_b}^4}}{2g_c g_{d_b}}\right)^2$ is just a normalisation factor so that $|\phi_i|=1$. 
	Using these results, we can write the decay constants and the masses of the physical axions as
	\be 
	f_{\phi_i}=\frac{1}{\sqrt{\lambda_i}}\coma m_{\phi_i}^2=\Lambda \lambda_i\fstop
	\ee
	Here we note, that in the limit of large $\tau_b\sim {\cal V}^{2/3}$ we find that the lightest axion $\phi_1$ has $f_{\phi_1}\sim g_c \sim \sqrt{g_s/\tau_b}$ and thus $S_{ED3}f_{\phi_1}\sim \sqrt{g_s\tau_b}$ confirming with our summary in Table~\ref{tab:closedaxions}. This implies a violation of certain strong forms of the WGC.

	Also in this case, we find that FDM particles naturally arise from string compactification only if the overall volume of the extra dimensions is small. For simplicity, in this section we fix $W_0=1$ while we let $A_b$ vary in $A_b\in [10^{-4},10^{4}]$. The overall volumes which are compatible with having 100\% of ultralight axionic DM are $\mc{V}\in 200\div 300$. The results related to this setup are shown in Fig.~\ref{fig:massC2}. Despite the relation $\phi_1\equiv c$ and $\phi_2\equiv \hat{d}_b$ only holds at $\mc{V}\rightarrow \infty$, the eigenvectors $\phi_1$ and $\phi_2$ are mainly given by the $C_2$ and $C_4$ axion respectively. Although the shape of the potential in Eq.~\eqref{eq:C2pot} may suggest some mass degeneracy, the hierarchy in the mass scales and in the abundances of the two fields is apparent. While the two decay constants are comparable and their values lie in the expected range $\sim 10^{16}$ GeV, the $C_2$ axion is much lighter and more represented than the $C_4$. The reason why in this context the lighter axion can represent a higher fraction of DM is that the two fields acquire a mass through instanton corrections that have a different nature. In this way they do not share the same dependence of the mass and the decay constant on the instanton action. The $C_2$ axion field, that would represent the prominent FDM candidate in this setup, exhibits a mass that is lighter than the original FDM estimate, $m_c<10^{-22}$ eV. In this section we are relying again on LVS moduli stabilisation, hence the same energy scales that we have shown in the case of $C_4$ axions in the Swiss-cheese geometry remain valid.

	\subsection{FDM from Thraxions: KKLT \& LVS?}\label{sec:thraxions}
	In the KKLT scenario~\cite{Kachru:2003aw}, it is difficult to realise ultralight axions. In this case, axions get stabilised at the same energy level as their moduli partners by the same non-perturbative effect to $W$. This is a consequence of the fact that KKLT AdS vacuum is supersymmetric. Their masses then are generically of the same order as the gravitino mass. Therefore, axions coming from KKLT moduli stabilisation behave just like the axionic partner of the small-cycle volume moduli in LVS, they are too heavy to be FDM candidates.
	
	However, there is a way out: we could have a viable FDM candidate if the underlying internal manifold admits the presence of \textit{thraxions}~\cite{Hebecker:2018yxs}. Thraxions, or throat-axions, are a recently discovered class of ultralight axionic modes living in warped throats of the CY, near a conifold transition locus in moduli space. They occupy a special corner of the axion landscape as their mass is exponentially suppressed by powers of the warp factor $\omega\sim e^{-S/3}$ of the throat. At the level of complex structure moduli stabilisation via fluxes of~\cite{Giddings:2001yu}, their squared mass scales as~\cite{Hebecker:2018yxs}
	\begin{equation}\label{eq:mthrax}
		\frac{m^2}{ M_P^2}\sim  \frac{4\,g_s\,e^{-2 S}}{\sqrt{3}\,S^{3/2}\,\mathcal{V}^{2/3} M^2}   \coma
	\end{equation} 
	where $\mathcal{V}$ is the volume of the bulk CY and $M$ is a flux quantum coming from the integral of a 3-form field strength $F_3$ over the $\mathcal{A}$-type 3-cycle of the deformed side of the conifold transition. 
	Note that here, compared to the axions studied so far, the dependence on the instanton action $S$ is enhanced by a factor of $2$, resulting in a bigger suppression of the mass. In principle, we should consider also the possibly present effects of ED1 instantons coming from ED1-branes wrapping the 2-cycle, which contribute an action
	\begin{equation}
		S_{ED1} \sim \sqrt{\frac{K M}{g_s}}\coma
	\end{equation}
	where $K$ is another flux quantum defined as the integral of the  3-form field strength $H_3$ over the $\mathcal{B}$-type 3-cycle.
	The effective instanton action generating the thraxion potential reads
	\begin{equation}
		S_{\text{eff}} \sim \frac{2 \pi K}{g_s M}\fstop
	\end{equation}
	Note, that $\omega\ll 1$ is ensured when $K>g_s M$.
	The ED1-brane instanton effects come with a shorter periodicity. Yet, they can remain subdominant in the thraxion scalar potential while satisfying the WGC in its mild version. We should therefore require ED1-contributions to be suppressed compared to the flux-backreaction induced thraxion scalar potential scale. This can be achieved by requiring the following hierarchy among fluxes:
	\begin{equation}\label{eq:MKrel}
		M\gtrsim \sqrt{\frac{K }{M g_s}}\fstop
	\end{equation}
	In this way, we are satisfying a milder version of the WGC. The effective decay constant reads
	\begin{equation}\label{eq:fthrax}
	f_{\text{eff}}\sim \frac{3 \sqrt{g_s} M}{2 \mathcal{V}^{1/3}}   M_P\coma
	\end{equation}
	which is enhanced by a factor $M$ compared to the standard $f$ (cf.~\cite{Hebecker:2018yxs}). Hence, the WGC relation reads
	\be\label{eq:thraxionWGC}
	S_{\text{eff}}f_{\text{eff}} \sim \frac{3\pi K }{\sqrt{g_s}\,\mathcal{V}^{1/3}} M_P\fstop
	\ee
	We can turn Eq.~\eqref{eq:thraxionWGC} into an upper bound on $S_{\text{eff}}f_{\text{eff}}$ by using the relation~\eqref{eq:MKrel} among the flux numbers. This takes the form displayed in Table~\ref{tab:closedaxions}, namely
	\be
	S_{\text{eff}}f_{\text{eff}} \lesssim \frac{3\pi M^3\sqrt{g_s} }{\mathcal{V}^{1/3}} M_P\fstop
	\ee
	
	The presence of no-scale breaking terms, which are necessary to stabilise the K\"ahler moduli sector, \textit{generically} induces cross terms between the thraxion and the moduli in the total potential~\cite{Carta:2021uwv}. These new terms generate a mass for the thraxion which scales as $ e^{- S} $. Hence, the mass loses the double suppression, and the thraxion potentially becomes slightly heavier than in Eq.~\eqref{eq:mthrax}. The mass squared now reads
	\begin{equation}\label{eq:mthrax_single}
	\begin{split}
    \frac{m^2}{ M_P^2}&\sim  \frac{4\,g_s\,e^{- S}}{3^{5/4}\,S^{3/4}\,\mathcal{V}^{2/3} M^2} \frac{|W_0|}{\mathcal{V}^{4/3}} \quad \mbox{ for KKLT stabilisation}\coma\\
     \frac{m^2}{ M_P^2}&\sim  \frac{4\,g_s\,e^{- S}}{3^{5/4}\,S^{3/4}\,\mathcal{V}^{2/3} M^2} \frac{\ln{\mathcal{V}}}{\mathcal{V}^3} \,\,\quad \mbox{ for LVS stabilisation}
    \coma
	\end{split}
	\end{equation} 
	where we distinguished between KKLT and LVS moduli stabilisation procedures. The decay constant remains the same as in \eqref{eq:fthrax}, as it is dominated by the physics in the UV.
	However, for the KKLT case we should impose the consistency condition that $\omega^3>W_0^2$~\cite{Carta:2021uwv}. This relation comes from requiring that gaugino condensation effects in the bulk CY do not become comparable with background fluxes at the IR end of the throat~\cite{Baumann:2010sx}. Since for FDM the scalar potential should scale as $\sim10^{-100}$ in Planck units, we find an upper bound for $|W_0|$, namely $|W_0|<10^{-50}$. Even if we might be able to engineer such values in the landscape, their presence is highly suppressed by the statistics of the flux vacua distribution. Thus we prefer to keep the discussion general and conclude that it is very unlikely that thraxions in KKLT can behave as FDM.\footnote{Notice that also in the cases where the six-fold warp factor suppression can be restored, the value of $|W_0|$ would anyway remain too low to be fully trusted.}
	
		\begin{figure}
	    \centering
	   	\includegraphics[width=0.49\linewidth]{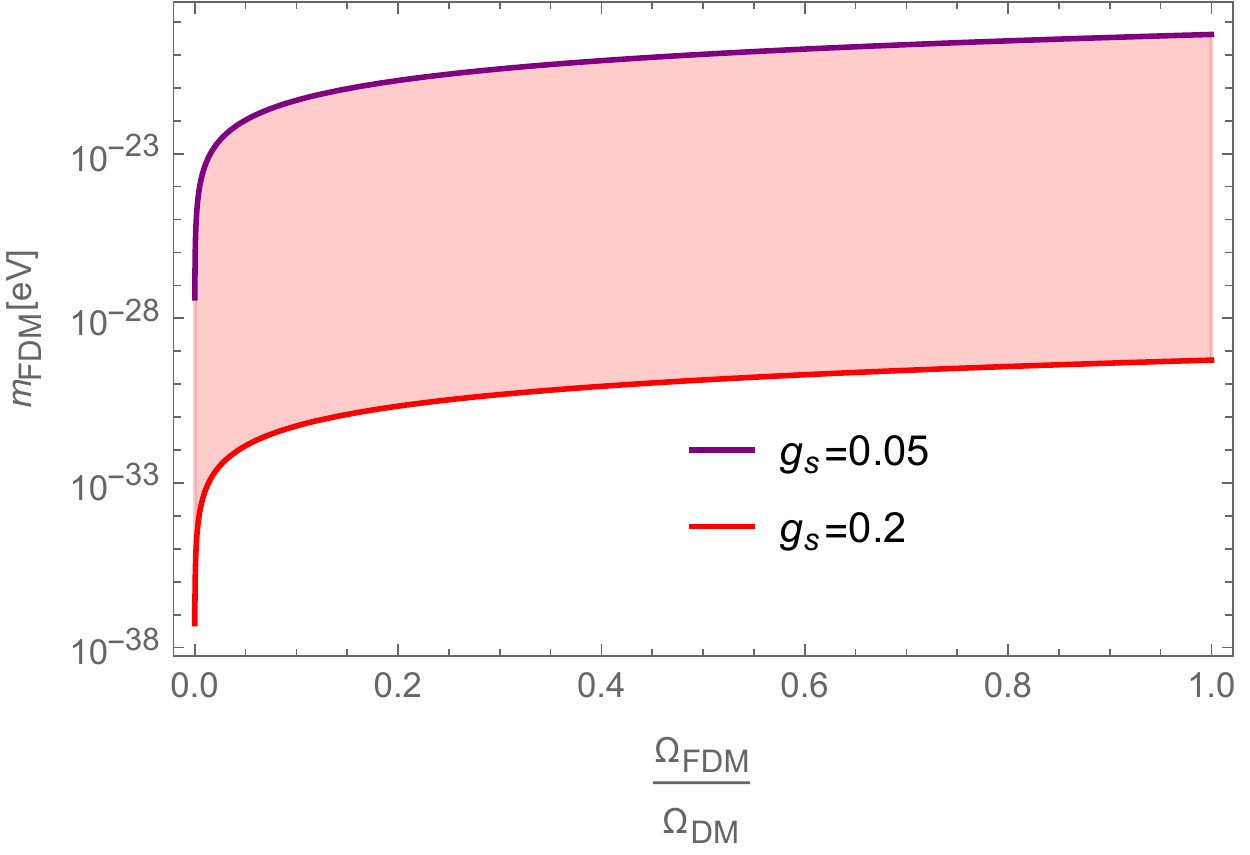}
	   	\includegraphics[width=0.49\linewidth]{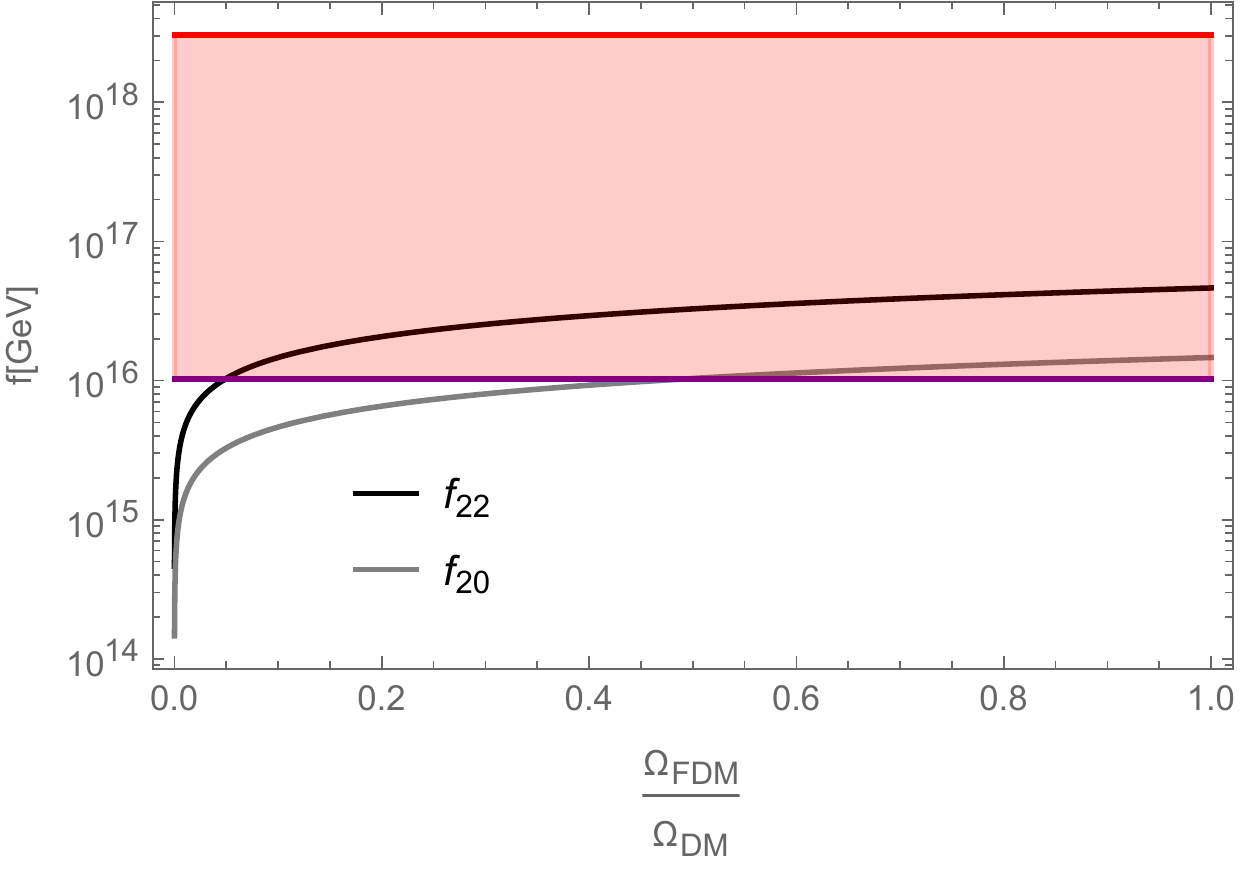}
	    \caption{Predictions for the thraxion mass and decay constant as functions of the FDM abundance. All plots are drawn fixing $M=10$: increasing $M$ makes the thraxions lighter. We then choose a conservative approach which allows us to have a wider phenomenology, given that the flux distribution in the landscape is an open field of investigation.
	    For LVS, the volume scales as $\mathcal{V}\sim e^{1/g_s}$, hence allowing us to deal with only one free parameter (as we consider $W_0\sim\mathcal{O}(1)$).}
	    \label{Fig:MassFPercentage:ThraxS}
	\end{figure}
	
	We display the results for thraxions as FDM candidates in LVS in Fig.~\ref{Fig:MassFPercentage:ThraxS}. First, we point out that we allowed the parameters to vary between the biggest and smallest values compatible with a consistent compactification, regardless of FDM astrophysical constraint. Then, it is indeed remarkable that for a certain parameter space we cover the FDM window. 
	Hence, for the LVS case the thraxion is a viable candidate. The main difference with the other harmonic axions is that now larger values for the total volume $\mathcal{V}$ are preferred: in Fig.~\ref{Fig:MassFPercentage:ThraxS} we are plotting $10^2\leq\mathcal{V}\leq 5\times10^8$, where the upper bound corresponds to the purple line. The fact that thraxions should rely on large volumes of the extra dimensions to lie in the FDM range may turn out to be a drawback. As we will discuss in Sec.~\ref{sec:overall_predictions}, large values of the CY volume may be statistically less represented in the landscape of string vacua.
	
	As explained in~\cite{Carta:2021uwv}, in certain geometries it can happen that the cross terms with the K\"ahler moduli vanish. Hence, the mass scales substantially again as in Eq.~\eqref{eq:mthrax}. We checked also these setups and we found that there is no appreciable difference with the results given in Fig.~\ref{Fig:MassFPercentage:ThraxS} for the single-suppressed mass.

	We are now able to estimate the mass of the warped KK modes living inside the warped-throat systems hosting the thraxion. Indeed, they will be heavier than the thraxion, as their masses scales linearly with the warp factor $\omega$ as
	\begin{equation}
		\frac{m_{w,KK}}{M_P}\sim \frac{\omega}{R}\sim \frac{\omega}{\mathcal{V}^{1/6}\sqrt{\alpha'}}\coma
	\end{equation}
	where $R$ is the throat radius which can be rewritten in terms of the bulk CY volume and the parameter $\alpha'$. The KK masses change drastically from the double to the single suppression case, as we shall discuss below.
	We can express $m_{w,KK}$ in terms of the variables of our setup as 
	\begin{subequations}
		\begin{align}
		&m_{w,KK}^{(s,\text{LVS})}\simeq 2\times 10^{-3}\, g_s^{3/4}e^{5/(9 g_s)}\left(\frac{m}{10^{-22}\mbox{ eV}}\right)^{2/3}K^{1/4}M^{5/12}\mbox{ eV}\coma\\
		&m_{w,KK}^{(d)}\simeq 304\, g_s^{7/12}\left(\frac{10^4}{\mathcal{V}}\right)^{5/9}\left(\frac{m}{10^{-22}\mbox{ eV}}\right)^{1/3}K^{1/4}M^{1/12}\mbox{ GeV}\coma
		\end{align}
		\end{subequations}
	where the index $s$ stand for the single suppressed case. We can give a rough estimate of $m_{w,KK}$ for $10^{-22}\mbox{ eV}\leq m\leq 10^{-19}\mbox{ eV}$ by plugging the other parameters accordingly. Hence, we find
	\begin{subequations}
	\begin{align}
		&  50\mbox{ eV} \lesssim m_{w,KK}^{(s,\text{LVS})}\lesssim  300 \mbox{ eV}\coma\\
		&  0.4\mbox{ GeV} \lesssim m_{w,KK}^{(d,\text{LVS})}\lesssim  8 \mbox{ GeV}\fstop \label{eq:mwkk_s}
	\end{align}
	\end{subequations}
	Note that we expect these modes, which live at the IR ends of the thraxion-carrying multi-throat, to be nearly completely sequestered. Hence, their interactions with standard model particles are suppressed.
	At this point we would like to discuss an intriguing possibility regarding the warped KK modes arising from the single-suppressed case.  With the scaling found above, a warped KK mode might behave as standard CDM. Therefore, in the single-suppression case we may envision a scenario where the thraxion represents part of the total DM abundance as FDM, while the warped KK mode may constitute the rest. We leave this possibility for future work. 
	
	In this setup, the bulk energy scales strongly depend on the moduli stabilisation prescription that we use. In LVS we have that the bulk KK scale ranges in $M_{KK}^{bulk}\in 10^{12}\div 10^{17}$ GeV while the gravitino mass is $m_{3/2}\in 10^{9}\div 10^{16}$ GeV. 
	The constraints on inflation coming from isocurvature perturbations bounds can be shown to be comparable to those related to $C_4$ and $C_2$ axions, implying low inflationary scale and undetectable tensor modes.
	
	Finally, we must point out that the results above rely on the internal manifold to be (almost) CY. This is true when the throats in the multi-throat system are all symmetrical and host one thraxion only: in this particular case the thraxion minimises at vanishing vacuum energy. If this symmetry is not met by the system, the thraxion will not necessarily minimise at zero, and thus it could break the CY condition. Moreover, the single-suppressed terms introduced by K\"ahler moduli stabilisation induce an additional shift on the thraxion vacuum which pushes it further away from the vanishing VEV. This tends to increase the amount of CY breaking and could lead also to a non-supersymmetric vacuum. The fact that vacua at non-zero thraxion VEV break the CY condition implies that the use of the effective 4D supergravity action derived by compactifying type IIB string theory on CY orientifolds is questionable in this situation. 
	However, we could be entitled to keep using the results based on the CY-derived 4D EFT if the CY breaking does not change the EFT (too) drastically. This could happen for instance if the thraxion VEV is sufficiently small, so that the manifold is `close to' the original conformal CY and the CY-based 4D supergravity approach still gives at least the qualitatively right behaviour. Alternatively, the CY-breaking effect of a non-vanishing thraxion VEV may turn out to be largely `decoupled' from the bulk CY (leaving the largest part of the Laplacian eigenvalue spectrum qualitatively unchanged compared to the actual CY) and stays sequestered in the throats.

\section{Overall predictions and comparison with experimental constraints}\label{sec:overall_predictions}
\begin{figure}
\begin{center}
\includegraphics[width=0.9\textwidth]{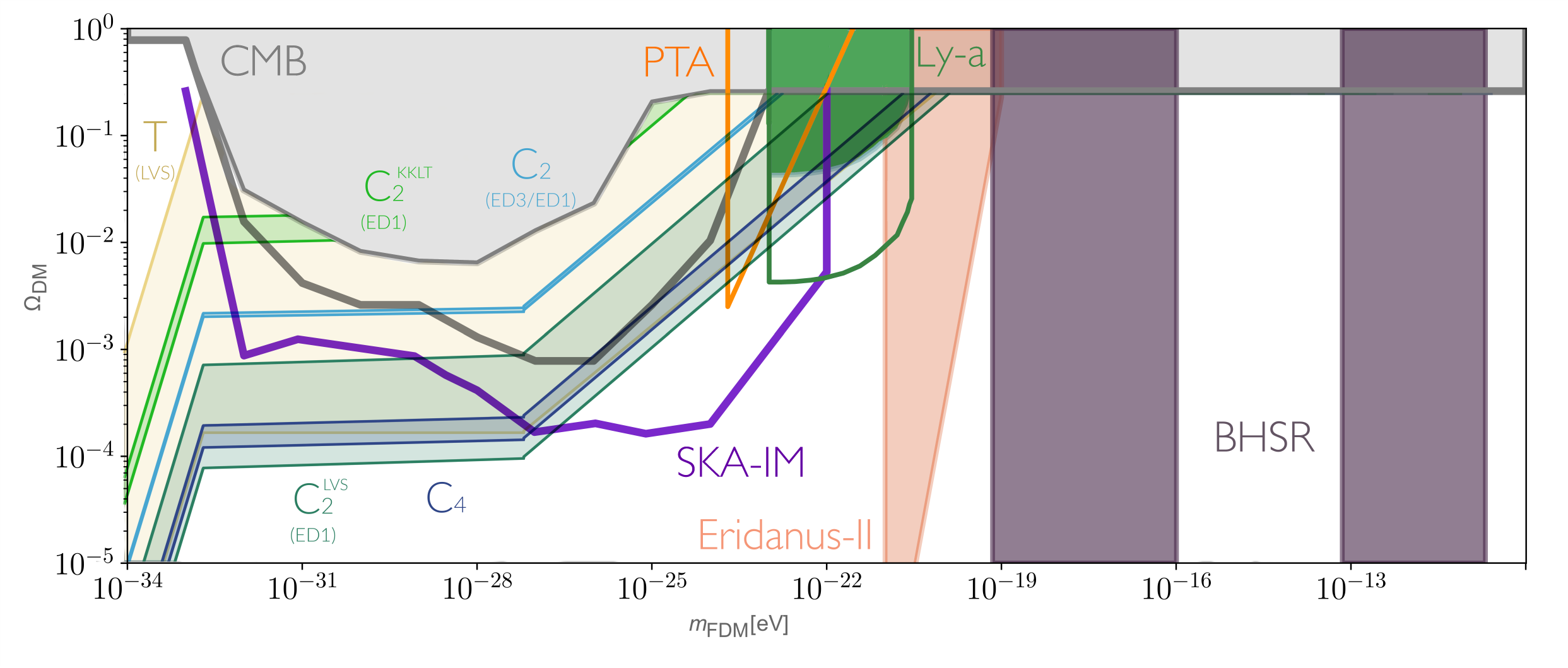}\vspace{5pt}
\end{center}
\caption{\label{fig:final_plot_bounds} Mass and total DM abundance predictions for large cycles $C_4$ axions (blue stripe), $C_2$ axions (light blue stripe for ED3/ED1 effects, dark/light green stripe for pure ED1 effects in LVS/KKLT), and thraxions in LVS (sand stripe) stabilisation. These results are compared to the current experimental bounds coming from CMB (solid grey area), Lyman-$\alpha$ forest (solid red), Eridanus II (solid pink area) observations and with theoretical predictions based on Black Hole Superradiance (solid purple area). Future experimental bounds coming from CMB (grey), Lyman-$\alpha$ forest detection (red), Square Kilometre Array (brown) and  Pulsar Timing Arrays (orange) are identified with solid lines. The reported experimental bounds were adapted from the recent review on ultralight bosonic dark matter~\cite{Marsh:2021lqg}. We refer the reader to that text and to the references therein for more details and extended bibliography. Note that axions moderately evading the WGC (thraxions and $C_2$ axions) are those representing the lightest FDM candidates.}
\end{figure}

In what follows we wrap up all the results coming from the previous sections and we compare our findings with current and future experimental constraints.
As already mentioned, empirical bounds coming from Lyman-$\alpha$ forest, black hole superradiance and ultra-faint dwarf galaxies that are DM dominated put strong constraints on the vanilla FDM model, ruling out a non-negligible area of the parameter space~\cite{Marsh:2018zyw,Chan:2021ukg,Jones:2021mrs,Nadler:2020prv,Zu:2020whs,Nebrin_2019,Maleki:2020sqn,Marsh:2021lqg}. We sum up these bounds together with our results in Fig.~\ref{fig:final_plot_bounds}.
We show the contributions to DM of our light axionic candidates in the mass spectrum $[10^{-34},10^{-10}]$ eV. The dark matter abundance in Eq.~\eqref{eq:DMabundance} applies only to axions in the mass range $m \gtrsim 10^{-28}\text{ eV}$, i.e. to axions which oscillate before matter-radiation equality. The abundance of the axions oscillating after equality ($10^{-33}\text{ eV}\lesssim m\lesssim 10^{-28}\text{ eV}$) and of those that have not yet begun to oscillate ($m\lesssim 10^{-33}\text{ eV}$) is taken from \cite{Marsh:2015xka}.

Our analysis was able to provide some sharp relations between the mass and the abundance of ultralight ALPs coming from type IIB string theory.
We found that non-negligible fractions of DM can only be given by $C_2$ and $C_4$ ALPs or thraxions under the following conditions:
\begin{itemize}
\item{$C_4$: 4-form axions can be good FDM candidates in LVS stabilisation only if the ALPs are related to cycles parametrising the overall volume. The overall extra-dimensions volume needs to be small $\mc{V}\in 10^2 \div 10^4$ and $g_s \sim 0.2$. We considered for simplicity the case where the ALP mass is given by non-perturbative corrections coming from ED3 instanton and gaugino condensation on a stack of $N\leq 10$ branes. Results coming from higher numbers of branes do not show any significant difference and are highly constrained by Eridanus-II and black hole superradiance bounds. These particles can represent $\sim$10\% of DM when their mass is $m\sim 10^{-22}$ eV.}
\item{$C_2$: they can represent FDM in the LVS stabilisation setup when there is non-vanishing intersection between the harmonics $C_2$ and the volume cycle in the extra dimensions. In LVS, if the $C_2$ axions acquire a mass through ED3/ED1 bound state instantons, these particles can represent nearly 50\% of DM when their mass is around $10^{-23}$ eV. In this case the overall extra-dimension volume needs to be small $\mc{V}\sim \mc{O}(10^2)$. If, on the other hand, these axions gain mass due to pure ED1 effects, in LVS they can represent 20\% of DM if their mass $\sim10^{-21}$ eV (for volumes $\mc{V}\sim 10^4\div 10^5$), while in KKLT they can represent up to 100\% of DM for masses $m\sim 10^{-25}\div10^{-24}$ eV (for volumes $\mc{V}\sim 10^2\div 10^3$). Therefore we can conclude that $C_2$ axions in KKLT are too light to be FDM.}
\item{Thraxions: these particles can be FDM candidates in LVS only. Here the allowed parameter region is wider compared to the previous cases. The CY volume can vary between $\mc{V}\in 10^2\div 10^8$ and thus $g_s\in 0.05\div 0.2$. These ALPs can represent 20\% of DM if $m\sim 10^{-21}$ eV and 100\% of DM when $m\in 10^{-25}\div 10^{-23}$ eV.}
\end{itemize}
Scaling the WGC relation up or down amounts to shifting a given axion abundance band up or down.\footnote{Consider an axion satisfying  $S f = \alpha M_P$ with $0<\alpha < \infty$. Given the axion mass in Eq. \ref{eq:axionmassgen}, we see that 
\[S = -2 \log(m) + \mc{O}(\log^2 (m), \log (\alpha)) \coma\]
and the axion DM abundance in Eq.~\eqref{eq:DMabundance} satisfies \[\frac{\Omega_{\phi}h^2}{0.112} \propto m^{1/2} f^2 = m^{1/2} \frac{\alpha^2}{S^2} \sim \frac{m^{1/2}}{(\log (m))^2} \times \alpha^2 (1 + \mc{O}(\log^2 (m), \log (\alpha))\fstop
\]} Generically, this implies that the bands coming from string axions satisfying but not saturating the WGC constraint will place below the $C_4$ constraint band in Fig.~\ref{fig:final_plot_bounds} (which also means, most stringy axions except the ones considered in this work will give negligible FDM abundance).

\label{sec:PredExpConstr}

Given the variety of possible ultralight axionic DM candidates, it is natural to ask whether some of them are more probable than others. Recent works have been analysing the relation between the distribution of string vacua, the axion masses and the decay constants~\cite{Broeckel_2021, Mehta:2020kwu}. Though far beyond the scope of this paper, we try to provide a very short description of how the number of vacua varies across our FDM candidates. 
In LVS, the relation between the overall volume and the string coupling leads to the following differential relation
\be
d\mc{V}\simeq-\frac{e^{1/g_s}}{g_s^2}dg_s \fstop
\ee
Given that the distribution of $g_s$ was shown to be uniform \cite{Broeckel_2021,Blanco-Pillado:2020wjn} we can write $dg_s\sim dN$, $N$ being the number of flux vacua, so that
\be 
dN\sim d\left(\ln \mc{V}\right)^{-1}\fstop
\ee
Instead, in KKLT the relation between the tree-level superpotential and the overall volume (considering a single K\"ahler modulus for simplicity) leads to
\be
d\mc{V}\sim -\frac{3}{2 a}\frac{\mc{V}^{1/3}}{W_0}dW_0\fstop
\ee
The $W_0$ distribution is assumed to be uniformly distributed in the complex plane so that $d|W_0|^2\sim |W_0| d|W_0|\sim dN$ for standard values of $W_0$ \cite{Denef:2004ze} while it scales as $|W_0|\sim e^{-1/g_s}$ for exponentially suppressed values of $W_0$ \cite{Demirtas_2020,Demirtas_2020b,Alvarez-Garcia:2020pxd}. This implies that in KKLT
\begin{subequations}
\begin{align}
&dN\sim d\left[e^{-2a\mc{V}^{2/3}}\right] \qquad \mbox{for not too small $|W_0|$}\coma\nonumber\\
&dN\sim d\left[\mc{V}^{-2/3}\right] \qquad\quad \mbox{for exponentially small $|W_0|$\fstop}\nonumber
\end{align}
\end{subequations}
The relation between the overall volume and the axion mass for large cycles $C_4$ axions, $C_2$ axions and thraxions in KKLT scenario scales as $m\sim e^{-\frac{a}{2}\mc{V}^{\alpha}}$, $\alpha=\frac{1}{3},\,\frac{2}{3}$. Instead, for thraxions in  LVS it reads $m\sim \frac{c}{\mc{V}^{11/6}}$, $c\in \Bbb R^+$. This implies that the relation between the number of vacua and the mass of the ALP is given by:
\begin{subequations}
\begin{align}
&dN \sim d\left[\ln\left(\frac{2}{a} \ln(m^{-1})\right)\right]^{-1} \;\qquad\, \mbox{for $C_2/C_4$ axions in LVS}\coma \nonumber\\[5pt]
&dN \sim d\left[\ln(m^{-1})\right]^{-1} \qquad\qquad\qquad\,  \mbox{for thraxions in LVS}\coma \nonumber\\[10pt]
& dN \sim d\left[m^4\right] \qquad\qquad\qquad\qquad\quad\,\,\, \mbox{for thraxions in KKLT}\coma\nonumber\\
& dN \sim d\left[\frac{2}{a} \ln(m^{-1})\right]^{-1} \qquad\qquad\quad\, \mbox{for thraxions in KKLT $(W_0\sim e^{-1/g_s})$}\fstop\nonumber
\end{align}
\end{subequations}
where we listed only those results corresponding to viable FDM candidates.
We can conclude that ALPs relying on LVS stabilisation do not show a strongly preferred mass value, given that here the number of vacua distribute at most logarithmic with respect to the thraxion mass. On the contrary, thraxions living in the KKLT setup show a polynomial distribution for fairly large values of $W_0$, stating that higher thraxion masses are more likely to appear in the string landscape. This distribution then flattens out towards a logarithmic distribution for exponentially suppressed $W_0$ values.

We would like to stress that our results provide scaling relationships for the simple setups analysed here. A more complete and general treatment of the problem as e.g. the number of moduli increases, also considering different geometries, is well beyond the scope of this paper. Nonetheless, we would like to give a hint about why we believe our results do not substantially change as the complexity of the extra dimensions increases. Thraxion fields depend on the CY geometry only via the overall volume, therefore changing the compactification manifold do not significantly affect their result. On the other hand, $C_4$ axions can be good FDM candidates if and only if they are the axion partners of K\"ahler moduli parametrising the overall volume $\vo$ so that they nearly saturate the WCG bound. Although it is not possible to write the most generic volume of a CY in terms of 4-cycles (the change from 2-cycle variables to 4-cycle volumes enforced by the O7 orientifold action is in general not feasible analytically), the number of moduli entering the volume with a positive sign must be finite. Furthermore, the K\"ahler cone conditions tend to create a hierarchy between the volumes of the 2-cycles, thus reducing the number of very large cycles. Moreover, the presence of many moduli will have to lower the value of $Sf$, as they increase the value of the total volume~\cite{Demirtas:2018akl}. It is therefore quite reasonable to think that as the complexity of the extra dimensions increases, the $C_4$ axions are naturally moved towards lower masses, away from the desired value to represent FDM that was shown to be exponentially sensitive to $\vo$. Similar arguments also apply to the case of $C_2$ coupled to $C_4$ through ED3/ED1 instanton interactions. In fact, this effect tends to make $C_4$ and $C_2$ axions almost degenerate in mass.

\section{Conclusions}
In this work we systematically dissect the long-standing lore that string axions can represent viable FDM candidates. We focus on the string axiverse coming from type IIB string theory compactified down to 4D on a CY orientifold with O3/O7-planes. After studying the properties of the whole axionic spectrum, we restrict the discussion to those axions that can represent good FDM candidates. In simple setups without alignment, tuned parameters and other non-trivial dynamical effects, we find that this request is closely related to the WGC for axions and implies that FDM saturates the bound $Sf \sim M_P $. The best candidates turn out to be $C_2$, $C_4$ closed string axions and thraxions. 

LVS stabilisation naturally gives rise to ultralight $C_4$ and $C_2$ axions. Indeed, being the LVS vacuum non-supersymmetric, these axions can be many orders of magnitude lighter than their volume modulus partners. On the contrary, KKLT stabilisation can only give rise to ultralight $C_2$ axions in presence of ED1 instanton corrections but an accurate computation reveals that these particles are too light to be FDM. Thraxions are axionic modes which stay ultralight regardless of the moduli stabilisation prescription chosen, given that their mass scaling is mostly dominated by the warp factor of the multi-throat systems they live in.

Our results show that string axions can exist in the FDM window allowed by experiments, but this translates into requiring specific properties of the compactification. As mentioned before, for this aim LVS is the preferred stabilisation procedure. For the harmonic zero mode $C_2$ and $C_4$ axions to fit the FDM window, the results suggest that the CY volume should be `smallish' (with respect to LVS standard volumes).
The masses and decay constants are basically insensitive to all the other microscopical parameters, making our predictions quite sharp. We also checked the scenario where more $C_4$ ultralight axions are present by considering a fibred CY. While in general cases heavier axions represent considerably higher DM fractions, in case of isotropic compactifications if we choose similar internal parameters for all the axions, i.e. same rank of gauge group and prefactor coming from complex structure moduli stabilisation, we end up having multiple FDM particles. In this specific case, the relative abundance of the FDM particles is determined by their $Sf$ value. Axions that come closer to saturating the WGC bound will represent higher percentages of DM.

For the $C_2$ axions, the situation is more involved. After checking many possibilities which can give rise to ultralight masses for these modes, we find that the only viable FDM scenarios are the case of a pure ED1 or an ED3/ED1-bound state instanton wrapping the cycle supporting the CY volume. In the former case we have that a FDM $C_2$ axion is compatible with volumes $\mc V\sim 10^4$, implying that an eventual DM contribution coming from the volume axion would be suppressed, this particle being parametrically lighter (potentially constituting dark radiation). In the case of ED3/ED1 effects, $C_2$ axions can be ultralight only in presence of very light $C_4$ axions. Due to different instanton properties, this setup allows for a moderate mass hierarchy such that here the heavier particle, i.e. the $C_4$ axion, constitutes the subdominant FDM fraction in the DM halo.

Then, we analyse the predictions for the masses and decay constants as a function of the DM abundance for the thraxions. These axionic modes allow for a wider range of masses, making them easier to fit the FDM window. We study both K\"ahler moduli stabilisation scenarios  (KKLT and LVS), as well as the two possible regimes arising there: i) the thraxion mass keeps its double-suppression from warping even after K\"ahler moduli stabilisation; or ii) it receives corrections from K\"ahler moduli stabilisation which cut the power of the warp factors suppressing the thraxion mass by half. Surprisingly, our results for thraxion FDM partially decouple from these details, but show that only in LVS thraxions can behave as FDM. The most prominent requirement is that in LVS the volume of the bulk CY should be rather big, as opposite to the cases discussed previously.
A few caveats are in order concerning our results for thraxion FDM. For once, the complete 4D EFT of thraxions is still being developed. Moreover, while warped throats are ubiquitous in CY manifolds, this may not be the case for thraxions, as e.g. for the recently constructed landscape of O3/O7-orientifolds of CICYs thraxion appear only in a fraction of them~\cite{Carta:2020ohw,Carta:2021uwv}. Hence, while they appear to span a large portion of the parameter space in Fig.~\ref{fig:final_plot_bounds}, we leave questions as to their generality for the future.

Finally, we compare our results with current astrophysical and experimental bounds. For each scenario analysed, we discuss the relation between our predictions and the exclusion bands. Moreover, we provide a preliminary discussion of the vacuum distribution for the mass of such axions in the string landscape. The results show that our FDM candidates from string theory have a very flat mass distribution for almost all cases studied. It is particularly exciting that our predictions show overlap with the regions in reach of future experiments. Hence, if at some point axions were to be found at these mass scales, we may be able to learn from the data about the type of axion detected, as well as its couplings, and potentially even something about their underlying microscopic theory.

Given this comparison, we wish to comment in passing on a further observation. Take a final look at the FDM abundance $\Omega_{\rm FDM} \sim \sqrt{m_\phi} f^2 \theta^2 \sim e^{-S/4} f^{3/2} \theta^2$. From this expression we see that for all axions with $f>H$ during inflation, which get populated via the misalignment mechanism $\langle\theta^2\rangle \sim \pi^2/3$, generically heavier axions acquiring their mass from instantons roughly saturating the WGC bound $S f \lesssim M_P$ dominate the DM content. An exception arises if e.g. two different axions acquire masses such that the heavier of the two acquires its mass from an instanton which does not saturate the WGC ($Sf < M_P$), while the instanton giving mass to the lighter axion saturates it ($Sf \simeq M_P$). A simple example for the generic case would be e.g. the two $C_4$ axions from the large and the blow-up 4-cycle of a 2-moduli LVS compactification, while the exception is seen e.g. for the case of the $C_4$-$C_2$ 2-axion system arising from the ED3-ED1 bound state instanton on the volume 4-cycle. From this it becomes clear that for the generic case the heavier axion states would completely dominate the dark matter content. An eventual detection of a sizable FDM fraction would therefore imply one of two possible predictions for the high-scale setup of a UV model: i) all the heavier WGC saturating axion states have $m>H$ during inflation, and there is a desert of axion states between the FDM mass scale and the inflationary $H$. ii) Avoiding the desert requires either fast decay of the heavy $m<H$ axion states significantly before BBN, or an anthropically selected very small heavy axion misalignment angle. Hence, a detection of FDM would put serious constraints on the structure of the allowed UV completion.

With all our caveats having been stated, in the end axions may yet turn out to be the missing link towards testing string theory.

\acknowledgments
\noindent We are particularly indebted to Arthur Hebecker for useful discussions and initial collaboration in this project.
We would like to thank David J. E. Marsh, Federico Carta, Alessandro Mininno, Andreas Schachner and Gary Shiu for useful discussions. N.R. is supported by the Deutsche Forschungsgemeinschaft under Germany's Excellence Strategy - EXC 2121 `Quantum Universe' - 390833306. V.G. and A.W. are supported by the ERC Consolidator Grant STRINGFLATION under the HORIZON 2020 grant agreement no. 647995.

	\appendix
	
	\section{Closed string axions: $S f$ computations}
	\label{sec:closed_examples}
	\subsection{ $C_0$ axion}
	\noindent
	The $C_0$ axion is part of the axio-dilaton field ${\cal S}=\frac{i}{g_s}+C_0$, and its periodicity is defined as $C_0\equiv C_0+1$. The decay constant can be read from the kinetic part of the 4D Lagrangian arising from the K\"ahler potential:
	\be
	\mathcal{K}\supset -\ln({\cal S}-\ol{\cal S})+\cdots\fstop
	\ee
	This implies 
	\be
	{\cal L}=\mathcal{K}_{{\cal S}\ol{\cal S}} |\partial {\cal S}|^2+\cdots = -\frac{1}{({\cal S}-\ol{\cal S})^2}
	(\partial C_0)^2+\cdots = \frac{g_s^2}{4} (\partial C_0)^2+\cdots\fstop
	\ee
	From the conventions given in Eq.~\eqref{eq:AxionLagr}, this means that
	$
	2\pi f=g_s/\sqrt{2}
	$.
	Based on analyticity and periodicity, the instanton contribution (if present) to the superpotential is $\sim\exp(2\pi i{\cal S})$, such that the instanton action reads $S=2\pi/g_s$. Thus,
	\be
	Sf=\frac{1}{\sqrt{2}}\,.
	\ee

	\subsection{$B_2$ axion}
	\noindent Let us consider the ${\cal N}=1$ description of a CY geometry in which K\"ahler moduli are encoded in 2-cycle superfields, with the real part being the $B_2$ axion $b$. Notice that the results derived in what follows do not directly apply to an orientifold of type IIB with D3/D7 branes since we are using the wrong ${\cal N}=1$ part of the original ${\cal N}=2$ SUSY of the CY model. However, the calculation for the two moduli $t_1$, $t_2$ gives the correct value for $Sf$ in a type IIB ${\cal N}=2$ model which is `ready' for the geometric projection associated with $t_1\leftrightarrow t_2$. This includes the restriction to the combined $t_1/t_2$ instanton which will survive the projection. Thus, since $Sf$ does not depend on which SUSY will eventually survive but merely characterises the real axion $b_-$, we can trust our result also for the orientifolded D3/D7 case.
		
	\paragraph{Single modulus}
	In the simplest case with one 2-cycle, the overall CY volume is given by
	\be
	{\cal V}=\frac{1}{6}\kappa_{111}v^3\coma
	\ee
	where $t=iv+b$, and $b\equiv b+1$.
	The K\"ahler potential reads
	\be
	\mathcal{K}\supset -3\ln(t-\ol{t})+\cdots \fstop
	\ee
	The structure of the exponential terms in the non-perturbative corrections is $\sim \exp(2\pi i t)$, such that the only difference with the $C_0$ axion case is the famous no-scale prefactor $3$. Thus,
	\be
	Sf=\sqrt{\frac{3}{2}}\,.
	\ee
	
	\paragraph{Two moduli}
	Now let us generalise to the case of two moduli $t_1$, $t_2$. The standard form of the volume is given by
	\be
	{\cal V}=\frac{1}{6}\kappa_{ijk} v^i v^j v^k\fstop
	\ee
	We require that an orientifolding with $\mathbb{Z}_2$ action $t_1\leftrightarrow t_2$ is possible. This imposes symmetry constraints on the triple intersection numbers $\kappa_{ijk}$ such that the volume becomes
	\be
	{\cal V}=\frac{1}{6}\left(\kappa_{111}\left[(v^1)^3+(v^2)^3\right]+3\kappa_{112}\left[(v^1)^2v^2+v^1(v^2)^2\right]\right)\,.
	\ee
	Changing variables to $t^\pm\equiv t^1\pm t^2$ gives
	\bea
	{\cal V}&=&\frac{1}{24}\left[(\kappa_{111}+3\kappa_{112})(v^+)^3
	+3(\kappa_{111}-\kappa_{112})(v^+)(v^-)^2\right]\nonumber
	\\
	&\equiv&\frac{1}{24}\left[\kappa_{+++}(v^+)^3+3\kappa_{+--}v^+(v^-)^2
	\right]\fstop
	\eea
	We are interested only in the kinetic term for $b^-$, at the locus where $v^-=0$. For this, we need
	\be
	\mathcal{K}_{--}=-\frac{\partial }{\partial t^-}\,\frac{\partial}{\partial \ol{t}^-}\ln{\cal V}(v_{\pm}) \qquad \mbox{with} \qquad v_\pm=-i(t_\pm-\ol{t}_\pm)/2\fstop
	\ee
	This leads to
	\be
	\mathcal{K}_{--}= -\frac{1}{4}\frac{\partial^2}{\partial (v^-)^2}\ln{\cal V}(v_\pm)=-\frac{1}{4}\,\frac{6\kappa_{+--}}{\kappa_{+++}(v^+)^2}\fstop
	\ee
	Since $\kappa_{+++}$ must be positive for positive volume, we learn that $\kappa_{+--}$ is negative. 
	
	The leading instanton for $b^-$ is the product of the instantons coupling to $t^1$ and $t^2$. This is enforced by the $\mathbb{Z}_2$ symmetry. The action of this double instanton is $2\pi(v^1+v^2)=2\pi v^+$ and the corresponding phase factor is $\exp(2\pi i (b^1+b^2))=\exp(2\pi i b^+)$.
	
	Now everything looks very similar to the $C_0$ axion case discussed before. One difference is the factor
	\be
	\sqrt{\frac{6|\kappa_{+--}|}{\kappa_{+++}}}\coma
	\ee
	affecting $f$. The other is the replacement $g_s\to 1/v^+$, but this factor drops out in the end anyway. Thus, since we originally had $Sf=1/\sqrt{2}$, we now arrive at
	\be
	Sf=\sqrt{\frac{3|\kappa_{+--}|}{\kappa_{+++}}}\fstop
	\ee
	The crucial question is how large the ratio
	\be
	\label{eq:bound_kappa}
	\frac{|\kappa_{+--}|}{\kappa_{+++}}=\frac{\kappa_{112}-\kappa_{111}}{3\kappa_{112}+\kappa_{111}} 
	\ee
	can become. If $\kappa_{111}$ and $\kappa_{112}$ are non-negative, then the maximal value of $1/3$ is attained for $\kappa_{111}/\kappa_{112}=0$, leading to $Sf=1$. As $\kappa_{111}/\kappa_{112}$ grows, $Sf$ falls.

	\subsection{ $C_2$ axion}
	
	\noindent
	In the case of IIB with D3/D7 branes, the $b_-$ axion is paired with the corresponding $c_-$ axion coming from $C_2$. The value of $Sf$ for the latter is most easily inferred by noting that, first, the 10D kinetic term changes according to
	\be
	(\partial B_2)^2/g_s^2\quad \to\quad (\partial C_2)^2\coma
	\ee
	and, second, the tension changes between the fundamental string and the euclidean D1 brane as
	\be
	1/(2\pi\alpha')\quad\to \quad 1/(2\pi\alpha' g_s)\fstop
	\ee
	Thus, $f\to f g_s$ and $S\to S/g_s\,$, leading to
	\be
	Sf=\sqrt{\frac{3|\kappa_{+--}|}{\kappa_{+++}}}\leq 1\fstop
	\ee
	where the upper bound in this simple case comes from the discussion around Eq.~\eqref{eq:bound_kappa}.

	\subsection{ $C_4$ axions}
	\paragraph{Single modulus}
	\noindent
	For a single (or one dominant) K\"ahler modulus in type IIB with D3/D7 branes one has
	\be
	\mathcal{K}\supset -3\ln(T+\ol{T})+\cdots \coma\quad T=\tau+id\coma\quad d\equiv d+1\fstop
	\ee
	The non-perturbative term in $W$ is $\sim\exp(-2\pi T)$, such that everything is analogous to the $B_2$ axion without orientifolding:
	\be
	Sf=\sqrt{\frac{3}{2}}\,.
	\ee
	\paragraph{Fibred geometry}
	In the simplest fibred geometry, e.g. K3 over $S^2$, one has
	\be
	\mathcal{K}\supset -2\ln{\cal V}+\cdots=-2\ln(T_1\sqrt{T_2})+\cdots \,.
	\ee
	Relative to the $C_0$ axion, one reads off factors $2$ and $1$ in $f^2$. Thus, one finds for the fibre $T_1$:
	\be
	Sf=1\,,
	\ee
	and for the base $T_2$:
	\be
	Sf=\frac{1}{\sqrt{2}}\,.
	\ee
	
	\section{Open string axions calculations}
	\label{sec:open_example}
	\noindent In the following, we focus on an open string complex scalar matter field $C=|C|e^{i\sigma}$ which lives on a collapsed cycle. The general form of the K\"ahler potential and superpotential which describe the theory for the shrinked cycle near the singularity are given by~\cite{Conlon:2008wa}
	\be
	\label{eq:sequesteredkahlerpotential}
	\mathcal{K}=-2\ln\left(\mathcal{V}+\frac{\hat{\xi}}{2} \right)+\lambda_{seq}\frac{\tau_{seq}^2}{\mathcal{V}}+\mathcal{K}_{matter}
	\ee 
	\be
	\label{eq:sequestered superpotential}
	W=W_0+\sum_{i=1}^{h_{1,1}}A_i\,e^{-a_iT_i} + W_{matter}\coma
	\ee
	where $W_{matter}$ and $\mathcal{K}_{matter}$ are related to the matter sector contributions depending on the field $C$. In the presence of more than one matter field the general form of $\mathcal{K}_{matter}$ is given by~\cite{Conlon:2006tj}
	\be
	\mathcal{K}_{matter}=\mathcal{K}_\gamma(T_i, \bar{T}_i) C^{\gamma}\bar{C}^{\bar{\gamma}}\fstop
	\ee   
	In order to understand  the properties of the ultralight axion candidate, $\sigma$, we have to study the moduli stabilisation procedure in the sequestering scenario \cite{Cicoli:2013cha}. The leading order contribution to the scalar potential, after dilaton and the complex moduli fields stabilisation, comes from the D-term which takes the following form:
	\begin{equation}
		\label{eq:leading D terms }
		V_{D_{D3}}=\frac{1}{Re(f_{seq})}\Bigl(q_{C}\frac{\partial \mathcal{K}}{\partial |C|}|C| - \xi_{seq}\Bigr)^2 \coma
	\end{equation}
	where $q_C$ is the charge of the matter field,  $f_{seq}$ is the gauge kinetic function related to the U(1) symmetry while $\xi_{seq}=-\frac{q_{seq}}{4\pi}\frac{\partial \mathcal{K}}{\partial T_{seq} }$  and $\tau_{seq}$ is the cycle on which the $U(1)$ charge is located. Working near the singularity, $T_{seq}=\tau_{seq}+i d_{seq}$, we have that $\tau_{seq}\rightarrow0$ and the gauge kinetic function 
	$$ f_{seq}=S+q_{seq}\,T_{seq} \fstop$$
	Since we want to find an axion, that is a pseudo Nambu-Goldstone field $\sigma$ with translational symmetry, we want the following conditions to be satisfied
	\begin{itemize}
		\item a Peccei-Quinn mechanism related to the breakdown of the U(1) symmetry of the potential related to $|C|$, i.e. $\langle |C| \rangle\neq0$.
		\item a minimum for the scalar potential which provides an extremely small value of $\langle \tau_{seq} \rangle \ll1$ in order to support the collapsed cycle assumption.
	\end{itemize}
	
	\noindent Working with the canonically normalised matter field, $|\hat{C}|$, the D-terms becomes:
	\be
	\label{eq:leading order d term}
	V_{D_{D3}}=\frac{1}{Re(f_{seq})}\Bigl(q_C |\hat{C}|^2 + \frac{q_{seq}}{8\pi} \frac{\partial \mathcal{K}}{\partial \tau_{seq}} \Bigr)^2\,.
	\ee
	Since the D-term has the same volume dependence as the flux generated F-term potential used to fix both the dilaton and the complex moduli, we have to set it to zero in order to have a consistent stabilisation procedure and preserve supersymmetry at this order in the expansion in inverse powers of $\mathcal{V}$. This implies
	\begin{equation}
		\label{eq:openaxiondecayshrinked}
		|\hat{C}|^2=\frac{q_{seq}}{8\pi\,q_C}\frac{\partial \mathcal{K}}{\partial \tau_{seq}}\sim \frac{\partial \mathcal{K}}{\partial \tau_{seq}}\fstop
	\end{equation}
	This relation fixes one direction in the $(|C|,\tau_{seq})$ plane which corresponds to the supersymmetric partner of the axion which is eaten up by the relative anomalous U(1) gauge boson in the process of anomaly cancellation. The axion which becomes the longitudinal component of the massive gauge boson is a combination of an open and a closed string axion. The mass of the Abelian gauge boson is given by \cite{Cicoli:2013cha}
	\begin{equation}
		\label{eq:Gaugebosonmass}
		m^2_{U(1)}\simeq g_{seq}^2 \Bigl[ (f_{\sigma})^2+(f_{d_{seq}})^2 \Bigr]\coma
	\end{equation}
	where $g_{seq}$ is the gauge kinetic coupling of the theory living on the sequestered cycle.\\
	If we focus on the $U(1)$ charged complex scalar field $C$ living on a D3-brane at a singularity ($\tau_{seq}\ll1$), we will have that the open axion decay constant will be
	\begin{equation}
		\label{eq:tauhidCrelation}
		f_{\sigma}^2=|\hat{C}|^2=\frac{q_{seq}\,\lambda_{seq}}{8\pi\,q_C}\frac{ \tau_{seq}}{\mathcal{V}}M_P^2\sim\frac{\tau_{seq}M_P^2}{\mathcal{V}}\ll M_s^2\,.
	\end{equation}
	On the other hand, the decay constant associated to the closed string axion related to the sequestered cycle is just $f_{d_{seq}}=\frac{\partial^2 \mathcal{K}}{\partial \tau_{seq}\partial\tau_{seq}}=\frac{\lambda_{seq}}{\vo}$. We see that, in this case the open string axion is eaten up by the gauge boson while the open string axion is still a dynamical field. the same process applies to anomalous U(1) on D7-brane stacks in the geometric regime. In that case, the open string axion is the degree of freedom which is eaten up and we are left with just the closed string axion. Coming back  to the sequestered scenario, after D-term stabilisation, we can still consider the open string axion $\sigma$ as a flat direction, while the moduli $\tau_{seq}$ and $d_{seq}$ are fixed and the gauge boson acquires a mass of the order of the string scale, namely
	\be
	M_{U(1)}\sim \frac{M_P}{\sqrt{\vo}}\sim M_s\fstop
	\ee
	The matter field $|C|$ acquires a mass through sub-leading soft terms which look like \cite{Cicoli:2013cha}
	\be
	V_F(|\hat{C}|)=r_2 \frac{|\hat{C}|^2}{\mathcal{V}^{\alpha_2}}+r_3 \frac{|\hat{C}|^3}{\mathcal{V}^{\alpha_3}}+ \left(\frac{r_4}{\mathcal{V}^{\alpha_4}}-\frac{\gamma_4}{\mathcal{V}}\right)|\hat{C}|^4\fstop
	\ee
	The terms proportional to $c_i$ come from the expansion of the scalar potential in powers of $|\hat{C}|$, while the one depending on $\gamma_4$ comes from the breaking of the no-scale structure by $\tau_{seq}$. If $r_2>0$, the matter field has a vanishing VEV and, thanks to the D-term stabilisation condition $\langle\tau_{seq}\rangle=0$. If instead $|C|$ shows a tachyonic mass  from supersymmetry breaking, i.e. $r_2<0$, then, depending on the signs of the different coefficients, $|\hat{C}|$ can develop a non-vanishing VEV. One may think that, given the relation between $\langle|\hat{C}|\rangle$ and $\langle \tau_{seq}\rangle$, the collapsed cycles get stabilised at values larger than the string scale, resolving in this way the singularity. It was shown that in models with just fractional D3-branes, the cycle is still sequestered, being $\alpha_3=2$, $\alpha_4=1$ and either $\alpha_2=3$ or $\alpha_2=4$ depending on the moduli dependence of the K\"ahler metric from matter field. The stabilisation of the matter field gives
	\begin{equation}
		\langle|\hat{C}|\rangle=\frac{2r_2}{3r_3} \frac{1}{\mathcal{V}^{\alpha_2-2}}\,\coma \quad  \langle \tau_{seq}\rangle=\frac{p}{\mathcal{V}^{2\alpha_2-5}}\,\coma
	\end{equation}
	where $p=\frac{32\pi\, q_C\, r_2^2}{9  q_{seq}\,\lambda_{seq}\,r_3^2}$ depends on soft terms and on the terms breaking the no-scale structure. We see that for both values of $\alpha_2$ we are still in sequestered scenario as
	\be
	\begin{array}{ll}
		f_{\sigma}\propto \frac{1}{\mathcal{V}}\coma\quad \tau_{seq}\propto \frac{1}{\mathcal{V}} \qquad &\qquad \mbox{when}\;\; \alpha_2=3\\[10pt]
		f_{\sigma}\propto \frac{1}{\mathcal{V}^2}\coma\quad \tau_{seq}\propto \frac{1}{\mathcal{V}^3} \qquad &\qquad \mbox{when}\;\; \alpha_2=4\fstop\\[10pt]
	\end{array}
	\ee
	At this level of approximation $\sigma$ is still a flat direction. This field receives a mass through hidden sector strong dynamics effects as described in the main text.

	\section{Additional corrections for $C_4$ axions}
	\noindent In order to understand whether there can be some constructions leading to a $C_4$ FDM candidate with mass around $10^{-22}$ eV representing $\sim 100\%$ of DM, we examined several different setups that we list below. 
	\subsection{Non-vanishing 2-form fluxes}
	\noindent Let us consider the fibred geometry described in section \ref{sec:LVSC4} as, having two K\"ahler moduli, it is more flexible compared to the Swiss-cheese case. The overall volume is given by:
	\be
	\label{eq:Vgaugeflux}
	\vo =\frac{2}{3}\,t_2^3+t_1\,t_2^2 =\frac{\sqrt{\tau_1}\,\tau_2}{2}-\frac{\tau_1^{3/2}}{3}\fstop
	\ee
	Let us turn on gauge fluxes as
	\be
	\mathcal{F}= m^{i}\,\omega_i+\dots \qquad \mbox{where}\qquad m^i=2\,\pi\, n_i\,; \quad n_i\in \mathbb{Z}
	\ee
	where  $\omega_j$ are orientifold-even 2-forms, $i=1,\dots,h^{1,1}_+$.
	The presence of non-trivial gauge fluxes $\mathcal{F}$ can induce a $U(1)$-charge $q_{i}$ for
	the i-th K\"ahler modulus together with a flux-dependent correction to the gauge kinetic function $f_j$ of the form
	\begin{equation}
		q_{i}=\int_{CY}\mathcal{F}\wedge\omega_i\wedge \omega_j
	\end{equation}
	\begin{equation}
		f_i=T_i-h_i(\mathcal{F})\,S
	\end{equation}
	where
	\begin{equation}
		h_i(\mathcal{F})=\int_{\rm CY}\mathcal{F}\wedge\mathcal{F}\wedge \omega_i=k_{ijk}\,m_j\,m_k\,.
	\end{equation}
	and $k_{ijk}$ are the intersection numbers. Considering the simple fibred geometry of  Eq. (\ref{eq:Vgaugeflux}) with just two divisors, the K\"ahler form is $J=t_1 \omega_1+t_2 \omega_2$ while the intersection numbers are
	\begin{equation}
		k_{122}=2\,;\qquad k_{222}=4\,.
	\end{equation}
	In this setup, we consider the most general flux form, $\mathcal{F}=m_1\,\omega_1+m_2\,\omega_2 $ 
	and we compute the corrections to the gauge kinetic couplings and the induced charges
	\begin{equation}
		\begin{array}{ll}
			h_1(\mathcal{F})=2\, m_2^2\coma\\[10pt]
			h_2(\mathcal{F})=4\, m_2\coma\,(m^2+m^1)
		\end{array}
	\end{equation}
	\begin{equation}
		\begin{array}{ll}
			q_{22}=2\, m_1+4\, m_2\coma\\[10pt]
			q_{21}=2\, m_2\fstop
		\end{array}
	\end{equation}
	The non-perturbative corrections to the superpotential induced by non-vanishing worldvolume gauge fluxes are given by:
	\be
	\begin{array}{lll}
		W_{\rm n.p.}&=A_i\, e^{-a_i f_{\,i}}\\[10pt]
		&=A_1\, e^{-a_1\left( T_1 + \frac{2}{g_s} \, m_2^2\right)}+A_2\, e^{-a_2\left( T_2 + \frac{4 f_2}{g_s}\,(m_2+m_1)\right)}\fstop
	\end{array}  
	\ee
	We see that gauge fluxes can induce an extra suppression in the axion mass. On the other hand, being interested in the perturbative regime of the theory, we need $g_s\ll 1$. This implies that the contributions to $W$ coming from gauge fluxes will induce an $\exp(-\mc{O}(10))$ correction that can produce considerably lighter FDM candidates. Nevertheless, the correction coming from 2-form fluxes cannot disrupt the predictions given in the main text and its precise contribution is model dependent. For this reason, in the body of this paper we treat the simplest case neglecting gauge flux effects.
	
	\subsection{Ample divisors}
	\noindent We focus again on the fibred geometry discussed above. In this section we consider the case where the fibred CY contains an ample divisor of the form $\tau_D= \tau_f+ \tau_b$ so that the superpotential receives non-perturbative corrections of the form~\cite{Bobkov:2010rf}:
	\be
	W=W_0+A e^{-a\left( T_f +  T_b\right)}\fstop
	\ee
	The leading order contributions to the F-term scalar potential are given by:
	\be
	V_F\supset \Lambda_2\cos\left(\frac{\phi_f}{f_{f_{mix}}}+\frac{\phi_b}{f_{b_{mix}}}\right)\coma
	\ee
	where 
	\be
	\Lambda_2\simeq \frac{ 4\kappa\, a A W_0 (\tau_f+\tau_b)}{\mc{V}^2} e^{-a (\tau_f+\tau_b)}\fstop
	\ee
	The eigenvalues of the mass matrix are:
	\be
	m_\lambda^2=\left\{0, \left(\frac{1}{f_{f_{mix}}^2}+\frac{1}{f_{b_{mix}}^2}\right)\Lambda_2 \right\} \coma
	\ee
	so that the effective decay constant of the massive axion is then given by 
	\be
	\bar{f}=\frac{f_{f_{mix}} f_{b_{mix}}}{\sqrt{f_{f_{mix}}^2+f_{b_{mix}}^2}}\fstop
	\ee
	A numerical inspection of this setup reveals that the  natural amount of DM  having mass $\sim 10^{-22}$ eV is around $1\%$, while the most likely values of the mass and of the decay constant are $\sim 10^{-19}$ eV and  $10^{15}$ GeV respectively. We can therefore conclude that the presence of ample divisors does not affect the predictions given in the main text. 
	
	\subsection{Poly-instantons}
	\noindent For completeness, we now check the possibility of getting FDM through poly-instanton corrections. For instance let us consider  let us consider the following corrections to the super potential given by an Euclidean D3-instanton wrapping the cycle $\tau_p$ yielding to non-perturbative corrections to the gauge kinetic functions of the condensing gauge group on $\tau_i$~\cite{Blumenhagen:2008ji}:
	\be
	W=W_0+A e^{-a_i T_i + C e^{- 2\pi T_p}} \simeq W_0+A e^{-a_i T_i }+A\,C  e^{-a_i T_i - 2\pi T_p}\fstop
	\ee
	Let us assume that $T_i$ is a blow up cycle which can be stabilised in the usual LVS fashion and that the real part of $T_p$ is stabilised through $g_s$ loops. In this way there is no explicit relation between the overall volume stabilisation and the VEV of $\tau_p$. The axion $d_p$ receives mass contributions only through these n.p. corrections. Then the potential related to $d_p$ will scale as
	\be
	V(d_p)\sim V_{\rm LVS} e^{ - 2\pi \langle\tau_p\rangle}\sim\mc{O}(\vo^{-3-p})\coma
	\ee
	where the value of $p$ depend on the geometry of the cycle $T_p$ and is usually of order unity. The decay constant of $d_p$ also depends on the geometry, for instance is $\tau_p$ is a rigid blow-up cycle or the fibre modulus, we have
	\be
	f_{d_p}\sim
	\left\{
	\begin{array}{ll}
		\frac{M_P}{2\pi}\frac{1}{\sqrt{\vo}}\qquad\qquad & \mbox{blow up}\\
		\frac{M_P}{2\pi \tau_p}\qquad\qquad & \mbox{fibre}\\
	\end{array}
	\right.
	\ee 
	then we see that in the first case, in order to satisfy the condition on the decay constant, we have to deal with an extremely small overall volume, in which case it is not possible to get the desired tiny mass for the axion $d_p$. In the second case there can be a chance of getting extra mass suppression with respect to the results presented in the main text. Nevertheless, given that this setup is more model dependent and we cannot provide sharp predictions for the exact mass of the FDM candidate, we consider the examples provided in the main text as the most general predictions.

	\section{Anharmonicity and isocurvature bounds}
	\label{sec:anharm}
	\noindent All the results presented in the main text assume that ALP self-interaction can be neglected. This is valid for small misalignment angles $\theta_{mi}\lesssim 1$. In the most general case, assuming that the PQ symmetry is broken before inflation, $f>>H_{I}$, we can have an enhanced axion density which depends on the form factor function $F(\theta)$ as follows \cite{Visinelli:2009zm}:
	
	\be
	\label{eq:DMabundanceF}
	\frac{\Omega_{\theta}h^2}{0.112}\simeq 1.4 \times \left(\frac{m}{10^{-22} \mbox{eV}}\right)^{1/2}\left(\frac{f}{10^{17}\mbox{GeV}}\right)^2
	\theta_{mi}^2\, F(\theta_{mi})\coma
	\ee
	
	\noindent where $F(\theta_{mi})\rightarrow 1$ for $\theta\sim 1$ and $F(\theta_{mi})\rightarrow \infty$ for $\theta\sim \pm \pi$. This function has been found to be given by \cite{Visinelli:2017imh}
	\begin{equation}
		F(\theta_{mi})=\left[\ln\left(\frac{e}{1-\theta_{mi}^4/\pi^4}\right)\right]\fstop
	\end{equation}
	Let us now focus for simplicity on $C_4$ ALPs in Swiss-cheese geometry. We can estimate the misalignment angle value that correspond to a FDM particles with mass $\sim 10^{-22}$ eV representing 100\% of DM without fine-tuning any of the microscopical parameters $W_0=1$ and $A_i=1$. This is given by a value that is extremely near to the maximum of the axion potential, namely  $\theta_{mi}\simeq 0.99 \pi$. For an extended treatment of the phenomenology arising in this last regime, see~\cite{Reig:2021ipa}.\\

	Such a large value for $\theta_{mi}$ may lead to the over-closure of the universe through the domain wall problems.
	In order to check whether different vacua, separated by domain walls, are populated in space, we need to compare quantum fluctuations and the classical initial field displacement. Domain walls problem can be avoided if 
	\be
	\Delta \theta_{in}\gg H_I/(2\pi f)\qquad \mbox{where}\qquad \Delta\theta_{in}\simeq 10^{-2}\pi \coma
	\ee
	which for $C_4$ in Swiss-cheese geometry implies $H_I\ll  0.1 f\sim 10^{15}\,\mbox{GeV}$ that does not significantly impact on model building.  
	
	Indeed, the most stringent inflationary constraint related to FDM models comes from the experimental boundaries on isocurvature perturbation. This looks like:
	\be 
	\Delta_{\mc{S}}^2= \Delta_{\mc{R}}^2 \frac{\beta_{iso}}{1-\beta_{iso}} < 5.6\times 10^{-11} \coma
	\ee
	where the scalar power spectrum $\Delta_{\mc{R}}^2$ and the isocurvature parameter $\beta_{iso}$ have been constrained to be $\Delta_{\mc{R}}^2\simeq 2\times 10^{-9}$ and  $\beta_{iso}\lesssim 2.6\times 10^{-2}$ at a pivot scale  $k^{*}= 0.05 \mbox{Mpc}^{-1}$ \cite{Akrami:2018odb}.
	When the PQ symmetry is broken before inflation, the isocurvature perturbations produced by the axion field are given by \cite{Kitajima:2014xla,Visinelli:2017imh}: 
	\be 
	\Delta_{S}^2=\left(\frac{H_I}{\pi\theta_{mi} f}\right)^2 \left(1+ \frac{\theta_{mi}}{2}\frac{F'(\theta_{mi})}{F(\theta_{mi})}\right)^2\fstop
	\ee 
	Given the experimental constraint on $\Delta_{\mc{S}}^2$, the previous equation induces an upper bound on the inflationary scale. This bound is strongly related to the assumption that the ALP is decoupled from the inflaton dynamics. In string theory this is not always the case as the field space turns out to be curved and the shape of the scalar potential is highly non-trivial. Nevertheless, whatever kind of coupling, both kinetic or in the potential, heavily depends on the inflationary model under study. For this reason, we decide to focus on the simplest assumption. 
	
	Using the bound on the inflationary scale, we can derive a rough estimate of the bound on the tensor-to-scalar ratio, $r$ that can be expressed as
	\be 
	r=\frac{\Delta_t^2}{\Delta_\mc{R}^2}\simeq\frac{2}{\Delta_s^2\pi^2}\frac{H_I^2}{M_P^2}\,.
	\ee
	\begin{table}[h!]
		\centering
		\begin{tabular}{c|ccc}
			& \multicolumn{3}{l}{$C_4$ \textbf{Swiss-cheese geometry}}\\
			&    $H_I$ [GeV]     & $r$   & $m$  [eV]  \\[3pt]
			\hline
			harmonic		& $<5\cdot 10^{11}\, $ & $<4\cdot 10^{-6}$ & $10^{-20}$       \\[3pt]
			anharmonic 	& $<5\cdot 10^{10}\, $  & $<10^{-7}$  & $10^{-22}$   \\[3pt]
			\hline
		\end{tabular}\\[15pt]
		
		\begin{tabular}{cc|cccc}
			&& \multicolumn{4}{l}{$C_4$ \textbf{fibred geometry}}\\
			&&    $H_I$ [GeV]     & $r$   &    $m_{d_b}$[eV] & $m_{d_f}$ [eV]     \\[3pt]
			\hline
			\multirow{2}{*}{isotropic}           & harmonic	& $<2\cdot 10^{11}\, $& $< 8\cdot 10^{-7}$    &$9.8\cdot 10^{-20}$ & $1.4\cdot 10^{-20}$ \\[3pt]
			& anharmonic& $<3\cdot 10^{10}\, $ & $<2 \cdot 10^{-8}$ & $9.3\cdot 10^{-23}$ & $1.3\cdot 10^{-22}$\\[3pt]
			\multirow{2}{*}{anisotropic $q=100$} & harmonic & $<4\cdot 10^{9}\, $ & $<3\cdot 10^{-10}$   &$2.3\cdot 10^{-20}$ & $\sim 0$    \\[3pt]
			& anharmonic 	& $<6\cdot 10^{8}\, $  & $<7 \cdot 10^{-12}$  &$2.1\cdot 10^{-22}$ & $\sim 0$     \\[3pt]
			\multirow{2}{*}{anisotropic $q=0.01$}& harmonic	&  $<3\cdot 10^{9}\, $ & $<1\cdot 10^{-10}$    & $\sim 0$ &$7.8\cdot 10^{-20}$   \\[3pt]
			& anharmonic 	& $<5\cdot 10^{8}\, $  & $<4 \cdot 10^{-12}$ & $\sim 0$ &$7.4\cdot 10^{-22}$      \\[3pt]
			\hline
		\end{tabular}\\[15pt]

		\begin{tabular}{cc|ccc}
			&            & \multicolumn{3}{l}{\textbf{Thraxions}}\\
			&            &    $H_I$ [GeV]       & $r$         &    $m$[eV] \\[3pt]
			\hline
			\multirow{2}{*}{unlifted}& harmonic	  & $<8\cdot 10^{11}\,$ & $< 10^{-5}$         &$1.1\cdot 10^{-21}$ \\[3pt]
			& anharmonic & $<1\cdot 10^{11}\,$ & $<3 \cdot 10^{-7}$ &$ 10^{-23}$ \\[3pt]
			\multirow{2}{*}{lifted}  & harmonic   & $<8\cdot 10^{11}\,$ & $< 10^{-5}$         &$1.2\cdot 10^{-21}$ \\[3pt]
			& anharmonic & $<1\cdot 10^{11}\,$ & $<3 \cdot 10^{-7}$ &$ 10^{-23}$ \\[3pt]
			\hline
		\end{tabular}
		
		\caption{Boundaries on the inflationary scale $H_I$ and the tensor-to-scalar ratio $r$ coming from isocurvature perturbations
			constraint. We consider both the harmonic approximation (quadratic potential) that is valid for small misalignment angles and the full anharmonic setup (cosine potential) where the $\theta_{mi}$ is tuned to $\theta_{mi}=0.99\pi$. Top: $C_4$ FDM from CY Swiss-cheese geometry. We fix $W_0=1$, $A_s=1$ and $a_s=1$. Center: $C_4$ FDM from CY fibred geometry.  We fix $A_b=A_f=1$ and $W_0=1$. We consider both isotropic, $q=f_{d_b}/f_{d_f}=\sqrt{2}$, and anisotropic compactification $q=100,\, 0.01$. Bottom: Thraxion FDM considering both the stabilisation in absence (unlifted) and in presence (lifted) of moduli coupling.   }		\label{tab:recapiso}
	\end{table}
	
	In Table \ref{tab:recapiso} we derive the constraints on the inflationary scale and the tensor-to-scalar ratio coming from the setups discussed in the main text. We consider both the harmonic approximation and the tuned initial misalignment angle case with $\theta_{mi}=0.99\pi$. In general, we observe that tensor modes that are produced during inflation will be undetectable and, as expected, a large misalignment angle is compatible with lower axion masses. On the other hand, this tuned initial condition implies stronger constraints on $H_I$ and $r$. 
	
	In this section we saw that the requirement of having a FDM particle with the standard FDM mass coming from $C_4$ axions leads to heavy fine-tuning of either the microscopical parameters, or the misalignment angle. Since from a statistical perspective it is not clear how to justify the tuning on $\theta_{mi}$ and the parameters tuning may lead the EFT out of the controlled regime, we decided to focus on the most likely cases where the ALP is heavier and the axion self-interactions can be ignored at leading order.

\bibliographystyle{JHEP}
\bibliography{refs}

\end{document}